\documentclass[11pt,twoside,a4paper]{article}
\usepackage{graphicx,color}
\usepackage{times}
\usepackage{booktabs}
\usepackage{cite}
\usepackage{epsfig}
\usepackage{feynmp}
\usepackage{color}
\usepackage{a4wide}
\usepackage{xspace}
\usepackage{amsmath, amssymb}
\usepackage[small,bf]{caption} 
\usepackage[colorlinks=true, pdftex, pdftitle={Timelike Dipole-Antenna Showers with Massive Fermions}, pdfauthor={A. Gehrmann-De Ridder, M. Ritzmann, and Peter Zeiler Skands}, pdfsubject={Monte Carlo generators}, pdfkeywords={Monte Carlo, generators, QCD, Parton Showers, HEP, VINCIA}]{hyperref}\usepackage{subfig}
\usepackage{enumerate}

\newcommand{\mrm}[1]{\mathrm{#1}}

\newcommand{\eqRef}[1]{equation~\eqref{#1}\xspace}

\newcommand{\eqsRef}[1]{equations~\eqref{#1}\xspace}

\newcommand{\secRef}[1]{section~\ref{#1}\xspace}

\newcommand{\secsRef}[1]{sections~\ref{#1}\xspace}
\newcommand{\SecsRef}[1]{Sections~\ref{#1}\xspace}
\newcommand{\tabRef}[1]{table~\ref{#1}\xspace}

\newcommand{\figRef}[1]{figure~\ref{#1}\xspace}
\newcommand{\FigRef}[1]{Figure~\ref{#1}\xspace}
\newcommand{\tabsRef}[1]{tables~\ref{#1}\xspace}

\newcommand{\figsRef}[1]{figures~\ref{#1}\xspace}

\newcommand{\ant}{\ensuremath{a}}
\newcommand{\pT}[1][]{\ensuremath{p_{\perp #1}}\xspace}
\newcommand{\mqqbar}{\ensuremath{m_{q \bar{q}}}}
\newcommand{\QE}[1][]{\ensuremath{Q_{E #1}}}
\newcommand{\Qe}[1][]{\QE{#1}}

\newcommand{\pAri}{\ensuremath{P_{\text{Ari}}}}
\newcommand{\TeV}{\,\mbox{Te\kern-0.2exV}\xspace}
\newcommand{\GeV}{\,\mbox{Ge\kern-0.2exV}\xspace}
\newcommand{\MeV}{\,\mbox{Me\kern-0.2exV}\xspace}
\newcommand{\keV}{\,\mbox{ke\kern-0.2exV}\xspace}
\newcommand{\eV}{\,\mbox{e\kern-0.2exV}\xspace}
\newcommand{\qbar}{\ensuremath{\bar{q}}}
\renewcommand{\d}[1]{\ensuremath{\mrm{d}#1}\hspace*{0.2em} }

\newcommand{\PS}{\ensuremath{\Phi}}
\newcommand{\dPS}[2][]{\d{\PS_{#2}^{#1}}}

\newcommand{ \CFhat }{\ensuremath{\widehat{C}_F}}
\newcommand{ \TRhat }{\ensuremath{\widehat{T}_R}}

\newcommand{\s}{s}
\newcommand{\y}{y}

\newcommand{\sij}{\s_{i j}}
\newcommand{\sik}{\s_{i k}}
\newcommand{\sjk}{\s_{j k}}
\newcommand{\yij}{\y_{ij}}
\newcommand{\yik}{\y_{ik}}
\newcommand{\yjk}{\y_{jk}}

\newcommand{ \tvec }[1]{ \mathbf{ #1 } }

\newcommand{\Ecm}{m_{IK}}
\newcommand{\Gdet}{\ensuremath{\Delta_3}}
\newcommand{\Gdetij}{\Delta_{ij}}

\newcommand{\Gdetjk}{\Delta_{jk}}

\newcommand{ \mq }{ m_q }
\newcommand{ \mqb }{ m_{\bar{q}} }
\newcommand{ \mqbar }{ \mqb }
\newcommand{ \mqp }{ m_{q'}}
\newcommand{ \muq }{ \mu_q }
\newcommand{ \muqb }{ \mu_{\bar{q}}}
\newcommand{ \muqp }{ \mu_{q'} }

\newcommand{ \Ctil }{ \widetilde{C} }
\newcommand{ \Mtil }{ \widetilde{M} }
\newcommand{ \Ct }{ \Ctil }

\newcommand{ \evo }{\mathcal{S} }

\newcommand{\aint}{\mathcal{A}}
\newcommand{\mant}{m_{IK}}
\newcommand{\qstart}{ \ensuremath{Q_{\text{start}}} }
\newcommand{\qemit}{ \ensuremath{Q_{\text{emit}}} }
\newcommand{\qemin}{ \ensuremath{Q_{E\text{min}}} }
\newcommand{\qstop}{ \ensuremath{Q_{\text{stop}}} }
\newcommand{\colijk}{\mathcal{C}_{j/IK}}

\newcommand{\colijktrialemit}{\ensuremath{C_\text{trial-emit}}}
\newcommand{\alphaStrial}{\hat{\alpha}_S}

\newcommand{ \paccept }{ \ensuremath{ P_\text{accept} } }

\newcommand{ \me }{ \mathcal{M} }
\newcommand{ \abs }[1]{ \left| #1 \right| }
\newcommand{ \order }[1]{ \mathcal{O} \left( #1 \right) }

\newcommand{\Qprime}{\Sigma}

\newcommand{ \msbar }{ \ensuremath{ \overline{MS} } }

\newcommand{ \asy }{ \chi }

\newcommand{\Ar}{{ARIADNE}\xspace}

\newcommand{\Mg}{{MADGRAPH}\xspace}
\newcommand{\Py}{{PYTHIA}\xspace}

\newcommand{\Vc}{{VINCIA}\xspace}

\begin{document}

\begin{minipage}{\textwidth}
\flushright
CERN-TH-2011-133\\
MCNET-11-19
\end{minipage}
\vskip5mm
\begin{center}
{\Large\bf Timelike Dipole-Antenna Showers with Massive Fermions}
\end{center}
\vskip5mm
\begin{center}
{\large A.~Gehrmann-De~Ridder$^1$, M.~Ritzmann$^1$, P.~Skands$^2$}
\end{center}
\begin{center}
\parbox{0.9\textwidth}{
$^1$: Institute for Theoretical Physics, ETH, CH-8093 Zurich, Switzerland\\
$^2$: Theoretical Physics, CERN, CH-1211 Geneva 23, Switzerland
}
\end{center}
\vskip5mm
\begin{center}
\parbox{0.84\textwidth}{
\textbf{Abstract \  --- } 
We present a complete formalism for final-state (timelike) 
dipole-antenna showers including fermion masses, but neglecting
polarization and finite-width effects. 
We make several comparisons of tree-level
expansions of this shower algorithm to fixed-order
matrix elements for hadronic $Z$ decays, up to and including $Z\to 6$
partons, to which the algorithm can be consistently matched over all of phase
space. We also compare to analytical resummations at the NLL
level. The shower algorithm has been implemented in the
publicly available VINCIA plugin to the PYTHIA~8 event generator, which
enables us to compare to experimental data at the fully hadronized level. We
therefore also include  comparisons to selected observables in 
$b$-tagged $Z$ decays.
}
\end{center}

\section{Introduction}

The large phase space opened up by the LHC is rekindling 
interest in the collider phenomenology of heavy
coloured particles. Appreciable samples of top quarks with large Lorentz 
boosts are becoming accessible for the first time; 
energetic top and bottom quarks are sought as decay products
of new-physics or Higgs particles; and coloured new-physics particles may also 
themselves give off radiation, though at suppressed rates close to
threshold. 

Indeed, for massive particles produced near threshold, 
most of the radiation produced  results from the violent
deceleration of the \emph{incoming} (massless) colour charges, i.e
initial state radiation is dominant. The
effects of multiple soft emissions from the massive partons themselves
are then largely unimportant for the description of the event as a
whole and only become relevant to define precisely the mass of the
produced particle
\cite{Skands:2007zg,Fleming:2007qr,Buckley:2011ms}. 

However, for the
production of boosted heavy quarks or other coloured particles, 
either directly, via decay, or through gluon splitting processes, 
multiple emissions cannot be neglected. Mass corrections will generate 
differences in the shape of the evolving jet, and in the energy loss
of the evolving particle. These effects must be taken systematically into
account if we are to rely on physics models of these phenomena to
distinguish ``signal'' from ``background'' production sources.

On the theory side, calculations of observables involving massive
particles present a unique set of challenges. 
The introduction of an additional scale in the problem, for each non-zero mass, 
leads to an increased number of terms in amplitudes, to modifications to
the pole structure caused by the massive propagators and to 
more complicated phase-space boundaries and kinematics.
The presence of massive final state particles shrinks the size of the
phase space available for additional QCD emissions, both in fixed-order calculations
and in parton showers. 

The modifications to the pole structure imply different infrared
limits in the massive case. In particular, QCD radiation from massive
particles can lead to soft divergences but  
cannot lead to strict collinear divergences, since the mass is acting
as an infrared regulator. 
Traditional Monte Carlo (MC) shower descriptions, which rely on the relative dominance of
collinear-enhanced terms, therefore become intrinsically less
accurate when non-zero masses are involved. Though it is possible to 
systematically improve shower descriptions to take into account
universal mass effects (as, e.g., in \cite{Norrbin:2000uu}), 
one would still expect a
relatively larger uncertainty from non-universal and/or subleading
terms than in the massless case, simply because the leading singular
behaviour itself is less strong. 
Consequently, corrections from higher-order matrix elements, generally 
referred to as ``matrix element matching'', may
be relatively more important. 

In this paper, we shall attempt to address a relevant subset of 
these challenges, in the specific context of matched time-like dipole-antenna showers
\cite{Giele:2007di,Giele:2011cb}. 
We restrict our attention to unpolarized stable massive particles, deferring a
detailed treatment of helicity dependence, as discussed recently by 
\cite{Larkoski:2011fd} and finite-width effects to a future
study. Still, our approach has some advantages. To the best of our knowledge,
this is the first time a rigorous and systematic approach to mass effects has 
been incorporated  in an antenna-based shower Monte Carlo code
\cite{Lonnblad:1992tz,Winter:2007ye,Giele:2007di}. We
also generalize the fixed-order antenna functions derived in
\cite{GehrmannDeRidder:2005hi,GehrmannDeRidder:2005aw,GehrmannDeRidder:2005cm,GehrmannDeRidder:2009fz,Abelof:2011jv}
to include variations in their non-singular behaviour and  
extend the unitarity-based  matching formalism presented in
\cite{Giele:2011cb} to include tree-level matrix elements with up to four
additional massive partons beyond the Born level. 
Due to the unitary 
nature of the matching corrections, this prescription can be used also
in the soft and (quasi-)collinear regions and hence we expect the subleading
properties of the resulting shower to be improved. This is a feature which is
not possible with other approaches to multileg matching, such as
MENLOPS~\cite{Hamilton:2010wh},
CKKW~\cite{Catani:2001cc,Krauss:2002up} and related 
approaches~\cite{Lonnblad:2001iq,Mrenna:2003if,Lavesson:2007uu},   
or MLM~(see \cite{Alwall:2007fs} for a description). 
The speed of the resulting matched calculations is
also greatly improved as compared to the existing approaches, 
as discussed in \cite{LopezVillarejo:2011ap}.

Corrections at the next-to-leading order level 
have not yet been included in this work. We therefore do not attempt to distinguish
rigorously between different possible mass definitions, such as
``constituent'' vs  ``pole'' vs ``running'' masses
\cite{Fleming:2007qr,Buckley:2011ms,Abazov:2011pt}. 
For the purpose of our studies here, we treat parton masses simply
as effective parameters, to be determined from data. It has been
argued that this should be comparable to using a 
perturbative mass definition evaluated at a scale of the order of the
infrared shower cutoff~\cite{Hoang:2008xm}, though the corresponding 
scheme is only defined numerically by the shower algorithm. We expect
that NLO matching for massive fermions will be able to provide some
further insight into this question, but that is beyond the scope of
the work presented here.

This paper is organized as follows. 
In \secRef{sec:formalism}, we discuss the factorization, kinematics,
and infrared limits of a single $2\to 3$ splitting involving massive
partons. In \secRef{sec:showerformalism}, we
introduce the additional ingredients required to turn this into a
framework for parton showering, including a discussion of trial
functions and veto algorithm steps. The generalization of our
evolution variables to the massive case and the treatment of $g\to
q\bar{q}$ splittings in the shower are also addressed. 
\SecsRef{sec:fixed-order}, \ref{sec:resummation}, and
\ref{sec:lep} then present comparisons to fixed-order matrix elements,
to analytical resummations, and to $b$-tagged experimental data,
respectively. We round off with conclusions and an outlook in
\secRef{sec:conclusion}.

We note
that the work reported has been made publicly available as a plug-in to the
\Py~8 event generator \cite{Sjostrand:2007gs}, starting from
\Vc\ version 1.026 \cite{VinciaPage}. 


\section{Massive Phase-Space Factorization and Massive 
Dipole-Antennae}
\label{sec:formalism}
The dipole-antenna formalism
\cite{Gustafson:1987rq,Kosower:1997zr,Kosower:2003bh,GehrmannDeRidder:2005cm} 
is constructed from two basic
ingredients: 
\begin{enumerate}[1)]
\item
an exact momentum-conserving and Lorentz-invariant
phase-space factorization based on $2\to 3$ mappings between on-shell
partons 
\item
a set of antenna functions that, 
combined with the phase-space factorization, capture the leading
singular behaviour of gauge field theory amplitudes
\cite{Kosower:1997zr}.
\end{enumerate}
\noindent
We return to how these are
implemented in the shower context in  \secRef{sec:showerformalism}. In
this section, we focus on a single ``elementary'' $2\to 3$ branching,
e.g., as it would appear during a single step in a shower algorithm, and/or
in the context of an antenna-based NLO
calculation. 
We here focus on the generalizations necessary in the
 massive case, with details on the massless treatment available in
 \cite{Giele:2007di,Giele:2011cb,GehrmannDeRidder:2005cm}. 
We begin by giving some conventions concerning the notations we use, in
\secRef{sec:notation},
then turn to the phase-space factorization in
\secsRef{sec:phase-space}--\ref{sec:phasespacemap} and finally discuss 
  the structure of the antenna functions in
  \secsRef{sec:polestructure}--\ref{sec:massiveantennae}. 

\subsection{Notation and conventions \label{sec:notation}}
Given momenta $p_a, p_b$ of massive particles $a$ and $b$ with
masses $m_a$ and $m_{b}$,  
it is convenient to use the notation, 
\begin{equation}
s_{ab} = 2\, p_a \cdot p_b = (p_a + p_b)^2 - m_a^2 - m_b^2~,
\label{eq:sab}
\end{equation}
which we adopt throughout this paper. 
With this notation, the relation
expressing the conservation of the total centre-of-mass (CM) energy 
in a massive $2\to 3$ branching, $IK\to ijk$, becomes
\begin{equation}
\mant^2 \ = \ (p_I+p_K)^2 \ = \ s_{IK} + m_I^2 + m_K^2 \ = \ s_{ij} + s_{jk} + s_{ik} +
m_i^2 + m_j^2 + m_k^2~.
\label{eq:Econs}
\end{equation}
The momenta involved in this branching are either called parent or
pre-branching momenta for $I,K$ and  daughter or post-branching
momenta for $i,j,k$.  
Hence we may express, the dot product of two daughter momenta $i,k$ 
as,
 $s_{ik}= \mant^2 - s_{ij} - s_{jk} - m_i^2 -
m_j^2 - m_k^2$. 

We shall also work with scaled invariants, normalized to the CM energy
of the dipole-antenna given by $\mant^2$ 
\begin{equation}
y_{ab} = \frac{s_{ab}}{\mant^2}~,\label{eq:y}
\end{equation}
and scaled mass values,
\begin{equation}
\mu_a = \frac{m_a}{\mant}~.
\end{equation}

We denote dipole-antenna functions by the symbol $a$ (for antenna) and
represent the partons participating in the branching process $IK \to
ijk$ by subscripts $a_{j/IK}$. The normalization of $a$ is 
such that $|M_{n+1}|^2 \sim a |M_{n}|^2$ in the relevant 
soft/collinear limits, to be elaborated upon in the sections below. 
We also define a corresponding colour- and coupling-stripped dipole-antenna function
$\bar{a}_{j/IK}$. The relation between $\ant$ and $\bar{\ant}$  
is\footnote{Note that compared to the equivalent relation between 
these two antenna functions 
presented in \cite{Giele:2011cb}, the normalization 
to the phase-space factor is not present here anymore. We prefer 
to keep the phase-space factor normalization outside the dipole-antenna functions 
which are defined from matrix elements squared only.}
\begin{equation}
\label{eq:aabar}
	a_{j/IK} = \frac{\alpha_s}{4\pi}  \colijk \bar{a}_{j/IK}
\end{equation} 
where $\colijk$ denotes the appropriate colour factor for the
branching, as follows:
$\CFhat$ for $q\bar{q} \to qg\bar{q}$, 
$C_A$ for $gg\to ggg$, either $\CFhat$ or $C_A$ for $qg\to qgg$, 
and $\TRhat$ for the splitting of a gluon into a quark-antiquark pair. 
In the conventions for the colour factors we use~\cite{Giele:2011cb}, 
we have $C_A = N_C$, $\CFhat = 2C_F = N_C - 1/N_C$, $\TRhat = 2T_R = 1$ which makes the 
difference between $\CFhat$ and $C_A$ explicitly colour-subleading.

Note furthermore that the lower index of  the
dipole-antenna function $a_{j/IK}$ fully specifies the partons involved in the
branching process, and hence those functions are uniquely
determined. In particular, the radiators $IK$ are also uniquely
identified, and our antenna functions 
are therefore related to so-called ``sub-antenna'' functions 
in the context of fixed-order subtraction
\cite{GehrmannDeRidder:2005cm}. 

\subsection{Phase-space factorization
\label{sec:phase-space}
\label{sec:phasespacemappings}
}

To define the dipole-antenna phase space 
characterizing the massive $2 \to 3$ branching process $I,K \to i,j,k$,
we consider the exact
factorization of the $n+1$-particle phase space $ \dPS{n+1} $
into a $n$-particle phase space $ \d \Phi_{n}$  
and a dipole-antenna phase space given by $\dPS{3} / \dPS{2}$ 
\begin{equation}
	\dPS{n+1} \left( p_1\dotsc, p_i, p_j, p_k, \dotsc
          p_{n+1}; q\right) = 
\dPS{n} \left(p_1 \dotsc, p_I, p_K, \dotsc p_{n+1}; q\right) \frac{ \dPS{3} \left( p_i, p_j, p_k \right) }{ \dPS{2} \left( p_I, p_K \right) }~.
	\label{eq:phasespacefactorization}
\end{equation}
In this equation, $\dPS{n}$ corresponds to the phase space for $n$
outgoing particles with momenta $p_1, ..p_n$ and  masses $m_1.. m_n$, with
total four-momentum $q^{\mu}$.
In $\dPS{n}$ only the parent momenta $p_I$, $p_K$ appear.
The relation between $p_I$, $p_K$ and 
$p_i$, $p_j$, $p_k$, typically called ``momentum mapping'' in fixed-order 
subtraction contexts and ``recoil strategy'' in parton-shower ones, 
 will be discussed in \secRef{sec:phasespacemap} below.

The dipole antenna phase space $\frac{\dPS{3}^{ijk}}{\dPS{2}^{IK}}$
is proportional to the three-particle phase space.
It involves only the pre- and post-branching momenta 
$p_{I},p_{K}$ and $p_i,p_j,p_k$
respectively and is given by \cite {Giele:2007di,Giele:2011cb,GehrmannDeRidder:2009fz}, 
\begin{equation}
\frac{\dPS{3}^{ijk}}{\dPS{2}^{IK}}	 = \frac{1}{16 \pi^2}
        \frac{1}{\sqrt{\lambda(\Ecm^2, m_I^2, m_K^2)}} \
		\d \s_{ij} \d \s_{jk} \frac{ \d \phi}{2 \pi}~
	\label{eq:dipantps}~,
\end{equation}
where the  K\"allen function $\lambda$ is given by
\begin{equation}
\label{eq:Kaellen}
\lambda( a, b, c ) = 
	a^2 + b^2 + c^2 -2 (ab+bc+ac)~,
\end{equation}
$\phi$ parametrizes rotations around the 
$\tvec{P}_I$-$\tvec{P}_K$-axis in the centre-of-mass frame.
As long as we restrict ourselves to unpolarized processes which we do
in this paper, all
emission probabilities are independent of $\phi$. The factor 
$\d \phi/(2 \pi)$ will therefore be suppressed in the following.

The boundaries of the three-particle phase space with general masses
follow from momentum conservation and the on-shell conditions.
They are given by

\begin{equation}
	2 m_i m_j = \s_{ij}^- \leq \s_{ij} \leq \s_{ij}^+ = (\Ecm - m_k)^2 - m_i^2 - m_j^2 
\end{equation}
\begin{equation}
\begin{split}
	\s_{jk}^\pm (\s_{ij}) 	= 	\frac{1}{ 2 ( \s_{ij} + m_i^2
          + m_j^2 ) } & \Bigg[
								\left(	\s_{ij} + 2 m_j^2 \right)
								 \left( [ \s_{ij}^+  + 2 m_k (\Ecm - m_k) ] - \s_{ij} \right)
						 \\
					& 	
								\pm \sqrt{ \s_{ij}^2 - \left(\s_{ij}^-\right)^2 }
									 \sqrt{ \s_{ij}^+ - \s_{ij} }
									 \sqrt{ \s_{ij}^+ + 4 \Ecm m_k  - \s_{ij}} 
							\Bigg].
\end{split}
\end{equation}
Equivalently, these  boundaries characterizing 
the physical phase space for the daughter partons $i,j,k$ 
are determined by requiring the positivity of the Gram
determinant $\Gdet$ defined as, 
\begin{equation}
	\Gdet 	= \frac{1}{4} \left( \sij \sik \sjk - \sij^2 m_k^2 - \sik^2 m_j^2 - \sjk^2 m_i^2 + 4 m_i^2 m_j^2 m_k^2 \right).
\end{equation}

\subsection{Phase-space mappings \label{sec:phasespacemap}}

To specify the phase-space factorization in
\eqRef{eq:phasespacefactorization}, a momentum-conserving 
mapping, or ``recoil strategy'' in the parton shower language, 
that relates the three on-shell daughter momenta, $p_i$,
$p_j$, and $p_k$  to the two on-shell parent momenta $p_I$ and $p_K$
is needed.  
In a dipole-antenna approach, the radiators $I$ and $K$ can be 
both emitter or recoiler and the radiation 
emitted between them 
is shared smoothly and symmetrically amongst them. 
The mapping between daughter and parent momenta presented below will 
reflect this fundamental property. 
Independent of whether the momenta involved are massive or not, 
the on-shell and momentum-conserving $2 \to 3$ mapping is not unique 
except on the boundaries of the phase space. 
Instead, there is a 
one-parameter family of such mappings. 

Generalizing the analysis in \cite{Kosower:1997zr} to the case of
non-vanishing particle masses we start by relating the momenta 
of the daughter particles to those of the parent momenta as follows,
\begin{eqnarray}
	p_I &=&  x  p_i + r p_j + z p_k\\[1.5mm]
	p_K &=& (1-x) p_i + (1-r) p_j + (1-z) p_k
\end{eqnarray}
with the on-shell conditions 
\begin{equation}
p_i^2 = m_i^2~~~,~~~p_j^2=m_j^2~~~,~~~p_k^2 = m_k^2~~~,~~~p_K^2 =
m_K^2~~~,~~~p_I^2 = m_I^2~. 
\end{equation}
We use these on-shell conditions to re-express the parameters $x$ and $z$ 
in terms of the single free parameter $r$ and obtain 
\begin{eqnarray}
		x	& = &  \frac{1}{2 \left( 4 \Gdet + \Ecm^2 \left(
                  \s_{ik}^2 - ( \s_{ik}^- )^2 \right) \right) } \ 
\Big[ \Qprime^2 \left( \s_{ik}^2 - ( \s_{ik}^- )^2 + 4 \Gdetij
\right)
\label{eq:xofr}
\nonumber\\[1mm]
		&& \hspace*{0.3\textwidth}+ R \left( \s_{ij}^+ + 2 \Ecm m_k - \s_{ij} \right)
				+ 8 r \left( \Gdet - \Ecm^2 \Gdetij \right) \Big]
			\\[3mm]
		z	& =& \frac{1}{2 \left( 4 \Gdet + \Ecm^2 \left(
                  \s_{ik}^2 - ( \s_{ik}^- )^2 \right) \right) } \ 
\Big[ \Qprime^2 \left( \s_{ik}^2 - ( \s_{ik}^- )^2 + 4 \Gdetjk \right)\nonumber
\\[1mm]
			& & \hspace*{0.3\textwidth} 
				- R \left( \s_{jk}^+ + 2 \Ecm m_i - \s_{jk} \right)
				+ 8 r \left( \Gdet - \Ecm^2 \Gdetjk \right) \Big] ~
\label{eq:zofr}
\end{eqnarray}
where we have defined
\begin{eqnarray}
	\label{eq:R2}
	R^2 			&=& 16 \Gdet \left[ \Ecm^2 r (1-r) - (1-r) m_I^2 - r m_K^2 \right]
						+ \left[ \s_{ik}^2 - ( \s_{ik}^- )^2 \right] \left[ \s_{IK}^2 - ( \s_{IK}^- )^2 \right] \\[1.5mm]
	\s_{IK}^-		&=& 2 \, m_I \, m_K \\[1.5mm]
	\Qprime^2 		&=& \Ecm^2 + m_I^2 - m_K^2 \\[1.5mm]
	\Gdetij	&=& (-1) \cdot \det
				\begin{pmatrix}
					\frac{\sij}{2}	&	\frac{\sjk}{2}	\\[1.5mm]
					\frac{\sik}{2}	&	m_k^2
				\end{pmatrix}
			= \frac{1}{4} \left( \sjk \sik - 2 \sij m_k^2 \right)	\\[1.5mm]
	\Gdetjk	&=& (-1) \cdot \det
				\begin{pmatrix}
					m_i^2		&	\frac{\sij}{2}	\\[1.5mm]
					\frac{\sik}{2}	&	\frac{\sjk}{2}
				\end{pmatrix}
			= \frac{1}{4} \left( \sij \sik - 2 \sjk m_i^2 \right)~.
\end{eqnarray}

These equations characterize our 1-parameter family of massive mappings.
The parameters $z$ and $x$ are related to each other 
with the replacements $i \leftrightarrow k$ and $R \rightarrow -R$.
Contrary to the massless case, however, $R^2>0$, (which corresponds 
to real momentum fractions $x$ and $z$),
is not true for arbitrary values of the momentum fraction $r$.

We note that the 
massive dipole mapping  of \cite{Catani:2002hc} 
corresponding to a dipole made of the partons $i,j$ and $k$ which
play respectively the roles of 
emitter ($i$), emittee ($j$) and spectator ($k$), 
is obtained as a
special case, by setting $r = x$ in the above formula. In this case we have,
\begin{equation}
	x 	= 	\frac{ \Qprime^2 }{ 2 \Ecm^2 } 
			+ \frac{ \left( \s_{ij}^+ + 2 \Ecm m_k - \s_{ij} \right) \sqrt{ \s_{I K}^2 - ( \s_{IK}^-)^2 } }
					{ 2 \Ecm^2 \sqrt{ \lambda(
                                            \Ecm^2, (p_i + p_j)^2,
                                            m_k^2 ) } }
\label{eq:dipole}
\end{equation}
where, in the rest frame of  $p_i + p_j$, 
we can rewrite the K\"all\'en function using
\begin{equation}
\lambda( \Ecm^2, (p_i + p_j)^2, m_k^2 ) = 4 (p_i+p_j)^2 E_k^2
\tvec{v}_k^2
\end{equation}
where $E_k$ denotes the energy of the parton $k$, $\tvec{v}_k$ its three-velocity. 
As in the massless case, the massive dipole mapping is asymmetric 
under the interchange of the particles $i$ and $k$,
but symmetric under the interchange of $i$ and $j$. 
While it is appropriate to 
use this mapping for a shower based on CS-dipoles (CS:Catani-Seymour) 
which distinguishes 
between emitter $i$, emittee $j$ and spectator $k$, 
it would clearly be inappropriate to use it in a dipole-antenna shower 
like \Vc where the roles of $i$ and $k$ are interchangeable. The
dipole mapping is mentioned for comparison only.

To get a geometrical picture of the mapping used in \Vc,
it is convenient to express the 
free parameter $r$ in the massive mapping family presented above in terms of 
the angle $\psi$ between the daughter parton $p_i$ and the parent parton $p_I$
(see, e.g., \cite{Lonnblad:1992tz,Giele:2007di}).
To this end, we write down the 4-product $p_i \cdot p_I$ as:
\begin{eqnarray}
\label{eq:pidotpI}
  p_i \cdot p_I	&= & E_i E_I - |p_i||p_I|\cos\psi	\nonumber \\[2mm]
    & = & \frac{1}{4 \left( 4 \Gdet + \Ecm^2 ( \s_{ik}^2 - ( \s_{ik}^-
      )^2 \right) }
  \Bigg[ \Qprime^2 \left( \s_{ik}^2 - ( \s_{ik}^- )^2 \right) 	
          \left( \s_{jk}^+ + 2 \Ecm m_i - \s_{jk} \right) \nonumber \\[1mm]
				& &	\qquad\qquad + 8 r \left( \s_{jk}^+ + 2 \Ecm m_i - \s_{jk} \right) \Gdet
						- R \left( \s_{ik}^2 - ( \s_{ik}^- )^2 + 4 \Gdetjk \right) 
					\Bigg]~.
\end{eqnarray}

Looking at equations
\eqref{eq:R2} and \eqsRef{eq:pidotpI}, we see that if 
we approach the boundaries of the phase space 
(for example if we consider a soft emission $p_j \to 0$ 
or if we take the (quasi-)collinear limit for $p_i$ and $p_j$), 
the Gram-determinant $\Gdet$ tends to zero and  
the dependence of $p_i \cdot p_I$ on 
the free parameter $r$ drops out.\footnote{
The sole exception to this occurs for $\s_{ik} = \s_{ik}^-$ where
the angle $\psi$ does depend on the functional form of $r$. 
} 


Inside the parton shower \Vc, to define appropriately the $2\to
3$ branching, we need to fix the mapping. In other words, we need to fix 
the functional form of the free parameter $r$.
If all particles are massless, the default mapping used is given by 
\cite{Kosower:1997zr}
\begin{equation}
\label{eq:defaultmapping}
r = \frac{\s_{jk}}{ \s_{ij} + \s_{jk} }~.
\end{equation}
This mapping has the properties that the interchange 
 $i \leftrightarrow k $ corresponds to $ r \leftrightarrow (1-r) $ 
and the momentum fractions are 
restricted to $x \geq 1$, $ 0 \leq r \leq 1 $ and $ z \leq 0 $, 
where $r=0$ corresponds to the collinear limit $p_j || p_k$ 
and $r=1$ corresponds to the collinear limit $p_i || p_j$.

In the massive case, we shall consider the following mapping,
\begin{eqnarray}
\label{eq:massivemapping}
	r 	& = & 	r^- + \frac{ \s_{jk} - \s_{jk}^- }{ \s_{ij} - \s_{ij}^- + \s_{jk} - \s_{jk}^- } \left( r^+ - r^- \right) \nonumber\\[2mm]
		&= & \frac{ \Qprime^2 }{ 2 \Ecm^2 } + \frac{ \sqrt{ \s_{IK}^2 - ( \s_{IK}^- )^2 } }{ 2 \Ecm^2 }
			\frac{ \s_{jk} - \s_{jk}^- - ( \s_{ij} - \s_{ij}^- ) }{ \s_{ij} - \s_{ij}^- + \s_{jk} - \s_{jk}^- }
\end{eqnarray}
where the condition
\begin{equation}
	r^- 	=	 \frac{ \Qprime^2 - \sqrt{ \s_{I K}^2 - ( \s_{I K}^- )^2 }}{ 2 \Ecm^2 }
		\leq 	r
		\leq r^+ 
		=  	\frac{ \Qprime^2 + \sqrt{ \s_{I K}^2 - ( \s_{I K}^- )^2 }}{ 2 \Ecm^2 }
\end{equation} 
ensures that $R^2 > 0$.

\begin{figure}[tp]
\centering
\subfloat[$Q \qbar \to Qg\qbar$ ($Q$ massive, $\qbar$ massless)]{
\includegraphics[width = 0.4 \textwidth]{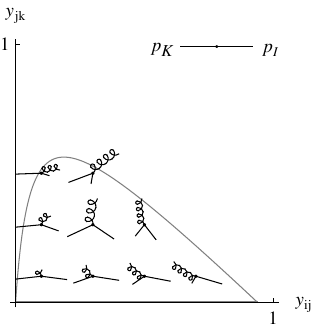}}
\hspace*{1cm}
\subfloat[$q \qbar \to qg \qbar$ ($q$ massless)]{
\includegraphics[width = 0.4 \textwidth]{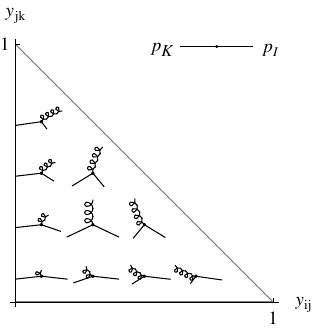}}
\caption{
Dalitz plot of the dipole-antenna phase space for $IK \rightarrow
ijk$ for massive partons (left) as compared to massless ones (right), 
using the scaled invariants $y_{ab}$ defined in
\eqRef{eq:y} as coordinates. The boundary of the physically 
allowed phase space is drawn as a solid grey line.
Insets show the orientation of the $ijk$ momenta corresponding to
the centre of each inset, in the CM frame of the parent partons, with
parents oriented horizontally and 
$\phi$ chosen such that the gluon is radiated ``upwards''. The
mass values used in the left-hand pane are $m_I = m_i = 0.25 \, m_{IK}$, $m_K = m_k = m_j =
0$. 
\label{Fig:mapping}
}
\end{figure}
Apart from reducing to the massless mapping in 
\eqRef{eq:defaultmapping} for vanishing masses, 
this phase-space mapping has the ``swapping'' property that $ i
\leftrightarrow k $ combined with $ I \leftrightarrow K $  
corresponds to $ r \leftrightarrow ( 1 - r ) $. 
For $m_I = m_i$ and $ m_K = m_k $  it also satisfies 
$ x \geq 1 $ and $ z \leq 0$ and can therefore be viewed as a
generalization of the massless mapping given in \eqRef{eq:defaultmapping}. 

In \figRef{Fig:mapping}, the mapping given in
\eqRef{eq:massivemapping}
is illustrated in a Dalitz plot of the three-particle phase space of the 
daughter momenta $p_i, p_j, p_k$. The phase-space boundary is 
marked with a solid grey line. Insets show the orientation 
of the daughter momenta for a branching with $s_{ij}$ and $s_{jk}$ given 
by the centre of the inset, in the CM frame of the parent partons,
with $\phi$ is chosen such that the radiated particle is moving ``upwards''.
A mass configuration characteristic for 
$Q \qbar \to Qg \qbar$, with $m_Q = 0.25 \, \mant$, $m_{\qbar}=0$ (left), is compared
to  the massless case (right). Notice that 
the physically  allowed phase space shrinks considerably in the
massive case, and that the
invariant $s_{ij}$ can only vanish in the soft limit $p_j \to 0$. The
limit $s_{ij} \to 0$ with $j$ hard is not accessible. 
The mass effects on the mapping are most pronounced for configurations 
which are close to the edge of the phase space and far away 
from the soft limit. For the rest of the phase space they are
relatively unimportant. A similar illustration for massless partons 
can be found in \cite{Giele:2011cb}.

\subsection{Pole structure \label{sec:polestructure}}

\begin{figure}
\centering
\includegraphics*[scale=0.9]{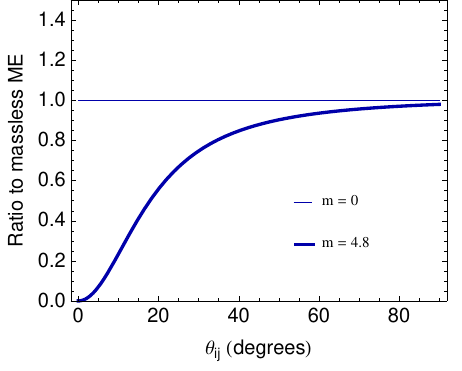}
\caption{Illustration of the dampening of the collinear singularity
for $Z\to Qg\bar{Q}$: squared matrix elements with (thick) and without
  (thin) mass corrections, normalized to the massless case, as a
  function of the opening angle between the quark and the gluon, for
 constant $E_g = 10\,$GeV and $m_Q=4.8\,$GeV.\label{fig:QgNorm}}
\end{figure}

Since masses act as infrared regulators in the collinear region,  
the pole structure of massive amplitudes is  actually simpler (less
divergent) than that of
their massless counterparts. A specific example of this is given in 
\figRef{fig:QgNorm}, in which
we show the ratio of the amplitudes squared for the processes 
$Z\to Qg\bar{Q}$ relative to $Z\to qg\bar{q}$, 
as a function of the $Qg$ opening angle, for
$M_Z=91\,$GeV, $E_g=10\,$GeV and $m_Q=4.8\,$GeV 
($Q$ stands for a massive quark while $q$ stands for a massless
one). The dip in the thick solid line for $\theta_{ij} \to 0$
is generated by the mass-shielding of the collinear enhancements, 
relative to the massless case (thin line). 

However, the calculation of observables with massive final state
particles still involves the treatment  
of potentially large mass-dependent logarithmic terms.
They correspond to collinear divergences which are regulated 
by the quark mass, therefore they become
divergent in the massless limit. 
For observables that are infrared safe in the massless limit, these logarithmic terms
cancel in the final result, but they can still appear at intermediate
steps of the calculation, for example in the separate evaluation of
real and virtual contributions. They are of the form
$\ln (Q^2/m^2)$, where $m$ is the parton mass 
and $Q$ is a characteristic scale  of the hard-scattering process.
These mass-dependent logarithmic terms are 
related to the  {\it quasi-collinear} \cite{Catani:2000ef} limit 
of the matrix element, the definition of which we shall recall below.

In a fixed-order approach, the potentially large logarithmic
contributions induced by mass terms are taken care of in 
the context of subtraction methods \cite{Frixione:1995ms}; terms which
mimic the singular behaviour of real matrix elements are added and
subtracted.  The construction of these terms relies heavily on the 
factorization properties of amplitudes in their soft 
and (quasi-)collinear limits \cite{Catani:2000ef}.
In the antenna framework presented in 
\cite{GehrmannDeRidder:2009fz,Abelof:2011jv} (and in the dipole
formalism \cite{Catani:2002hc} that predates it), the main 
building blocks, massive antenna (dipole) functions and
phase-space factorizations, are therefore constructed so as to 
reproduce exactly  the quasi-collinear and soft behaviours of real
radiation matrix-elements in the corresponding  limits. For cross
sections which are well-behaved in the 
massless limit, the explicit cancellations of the  $\ln(Q^2/m^2)$-terms also  
ensure numerical stability in the limit $m \to 0$.

For some observables which are not infrared safe in their massless
limit, such as ones sensitive to the details of the 
fragmentation process for example, 
the cancellation of the mass-dependent logarithms is incomplete. 
Terms of the form $\alpha_S^{n} \ln^n(Q^2/m^2)$ appear 
in every order of the expansion. 
In the case of a large hierarchy $m \ll Q$, these terms 
jeopardize the convergence of the perturbative series.
It is necessary to resum them to all orders to obtain a meaningful
result, as is done, for example, 
for  the $b$-quark fragmentation process in
\cite{Cacciari:2001cw}, to which we compare 
the massive VINCIA dipole-antenna shower in section
\ref{sec:resummation}. However, in order to construct this shower, we must
first consider the soft and quasi-collinear limits more carefully  
and define how the massless splitting functions and 
soft Eikonal factors are generalized in the presence of massive particles.

The infrared singularity properties of tree-level colour-ordered matrix
elements involving only massless partons have been well studied in
\cite{Catani:2000ef}. 
In the limit where a gluon $j$  is soft with respect to its
neighbouring partons $i$ and $k$, the colour-ordered matrix-elements
squared $|{\cal M}_{n+1}|^2$ for $(n+1)$ partons
factorizes into a universal soft Eikonal factor  $S_{ijk}$ and a colour-ordered
tree-level squared amplitude where gluon $j$ has been removed.
For the squared amplitudes we have,
\begin{equation}
\abs{ \me_{n+1}(1,\cdots,i,j,k,\cdots,n+1) }^2 \ \xrightarrow{j_g \to
  0} \ 
 g_s^2 {\cal C}_{ijk} \ 	S_{ijk}
	\abs{ \me_{n}(1,\cdots,i,k,\cdots,n+1) }^2
\end{equation}
where $g_s^2 = 4\pi\alpha_s$ is the strong coupling, ${\cal
  C}_{ijk}$ is a colour factor that tends to $N_C$ in the
leading-colour limit, and the massless Eikonal factor is given by
\begin{equation}
\label{eq:eikonalmassless}
S_{ijk} = \frac{2 s_{ik}}{s_{ij}s_{jk}}~.
\end{equation}
Similarly when two neighbouring gluons or a quark and a gluon 
become collinear the colour-ordered matrix elements factorize. 
Depending on the nature of the partons involved different collinear
factors are obtained. Partons which are not
colour-connected do not lead to singular behaviours of the colour
ordered matrix-elements squared, hence the soft or
collinear factors only involve the neighbouring particles to which the 
unresolved particle is colour-connected. 

In the massive case, 
essentially the same factorization properties still hold, provided the collinear 
limit is generalized to the quasi-collinear limit (see below). 
For the emission of a soft gluon from massive radiators, the factorization 
of the matrix element into a soft Eikonal factor times a reduced 
matrix element with the soft gluon omitted works in the same way as for 
massless partons. The soft Eikonal factor given in
\eqRef{eq:eikonalmassless} needs however to be 
generalized. 
Written in terms of the parent parton masses 
$m_I$ and $m_K$ and the invariants between the daughter 
partons $i$, $j$ and $k$, the massive soft Eikonal factor reads
\begin{equation}
\label{eq:softeikonal}
{{S}_{ijk}}(m_{I},m_{K})= \frac{2 s_{ik}}{s_{ij}s_{jk}} 
- \frac{2 m_I^2}{s_{ij}^2} - \frac{2 m_K^2}{s_{jk}^2}
\end{equation}
which has two new mass-dependent terms compared to the massless
Eikonal factor defined above.

The quasi-collinear limit of a massive 
parton with momentum $p^\mu$ decaying into two massive partons 
$j$ and $k$ is given by,
\begin{equation}
p_{j}^{\mu} \to z \, p^{\mu}, \,p_{k}^{\mu} \to (1-z) \, p^{\mu},
\end{equation}
\begin{equation}
p^2=m_{(jk)}^{2}.
\end{equation}
with the constraints,
\begin{equation}
p_{j}\cdot p_{k},m_{j},m_{k},m_{jk} \to 0
\end{equation}
at fixed ratios,
\begin{equation}
\frac{m_{j}^2}{p_{j}\cdot p_{k}},
\frac{m_{k}^2}{p_{j}\cdot p_{k}},
\frac{m_{jk}^2}{p_{j}\cdot p_{k}}.
\end{equation}

The key difference between the massless collinear limit and the quasi-collinear
limit is given by the constraint that the on-shell masses squared have to be
kept of the same order as the invariant mass $(p_{j}+p_{k})^2$, 
with the latter becoming small.
In these corresponding quasi-collinear limits, 
the colour-ordered $(m+1)$-parton matrix element 
squared factorizes into a reduced $m$-parton matrix element 
 squared multiplied by quasi-collinear splitting functions, the latter
are generalizations of the
Altarelli-Parisi splitting functions \cite{Altarelli:1977zs}
from which they differ by  mass-dependent terms. In four dimensions, they read
\begin{equation}
\label{eq:splittingmassive}
\begin{split}
P_{qg \to Q} (z, m_q, s_{qg}) &= \frac{1 + (1-z)^2}{z} - \frac{ 2 m_q^2}{s_{qg}}~,\\
P_{q \qbar \to G} (z, m_q, s_{q \qbar} ) &= z^2 + (1-z)^2 + \frac{2 m_q^2}{s_{q \qbar} + 2 m_q^2}~.
\end{split}
\end{equation}
We now turn to a description of the full massive dipole-antenna
functions as implemented in VINCIA.  

\subsection{Massive dipole-antenna functions
\label{sec:massiveantennae}}

In general, the full forms of the 
dipole-antenna functions are obtained by normalizing 
a three-parton tree-level matrix-element squared 
to a corresponding two-parton squared matrix
element, stripped of all couplings and colour factors and normalized 
to reproduce the known collinear splitting functions and soft Eikonal
factors in the corresponding 
unresolved limits. 

In the fixed-order context, specific sets of 
such dipole-antenna functions have been derived for the massless case in 
\cite{GehrmannDeRidder:2005cm,GehrmannDeRidder:2005hi,GehrmannDeRidder:2005aw}
and for the massive one in \cite{GehrmannDeRidder:2009fz, Abelof:2011jv}. 
In principle, there is an infinite set of similar dipole-antenna
functions, differing by non-singular (``finite'') terms and hence having 
the same soft and  (quasi-)collinear limits, which 
could equally well be used to construct the subtraction terms. For
example, consider gluon emission off 
a $q\bar{q}$ antenna. 
To cover the limiting behaviour of this emission process, 
we could use the matrix element for $\gamma^* \to Q g \bar{Q}$
normalized to the one for $\gamma^* \to Q \bar{Q}$, 
\begin{equation}
\bar{a}_{g/q\bar{q}}^{\gamma^* \to Q g \bar{Q}} \left( \Ecm^2, \sij, \sjk, \mq, \mqb \right) \ =
\ 
\frac{2 \s_{ik} }{\s_{ij} \s_{jk} } - \frac{2 \mq^2}{\s_{ij}^2} - \frac{ 2 \mqb^2 }{ \sjk^2 }
+ \frac{1}{s_{IK} + 4 \, \mq \mqbar}
	\left( \frac{\s_{ij}}{\s_{jk}} + \frac{ \s_{jk}}{\s_{ij}} 
\right)~.
\end{equation}
Alternatively, we could use the process $H \to Q g \bar{Q}$, which
gives
 \begin{equation}
\bar{a}_{g/q\bar{q}}^{H \to Q g \bar{Q}} \left( \Ecm^2, \sij, \sjk, \mq, \mqb \right) \ =
\ 
\frac{2 \s_{ik} }{\s_{ij} \s_{jk} } - \frac{2 \mq^2}{\s_{ij}^2} - \frac{ 2 \mqb^2 }{ \sjk^2 }
+ \frac{1}{s_{IK} -2 \, \mq \mqbar}
	\left( \frac{\s_{ij}}{\s_{jk}} + \frac{ \s_{jk}}{\s_{ij}} +2
\right)~.
\end{equation}
In both of these expressions, the denominator factor $s_{IK} + x \mq \mqb$
is proportional to the two-parton matrix element to which the three-particle 
matrix element is normalized. 
These two different constructions 
would give rise to slightly different integrated and unintegrated 
subtraction terms, but the final result would in either case 
be completely independent  of which one is used. 

For a parton shower, however, the behaviour 
of the dipole-antenna functions away from the 
phase-space boundaries is important to determine the 
amount of radiation produced. 
The most obvious example is that of adding a positive 
constant to a dipole-antenna function. This would result in 
a slightly higher rate for hard emissions in the parton shower 
(and consequently smaller Sudakov factors) without 
changing the limiting behaviour of the dipole-antenna function. 
In the context of shower uncertainty evaluations, it is therefore 
useful to generalize the definition of the dipole-antennae, 
to allow for continuous variations of the ambiguous non-singular
terms, as done for the massless case in~\cite{Giele:2011cb}. 
The possibility of varying finite parts in the parton-shower framework 
is a particular and important feature of the \Vc code. 
Other parton showers, whose evolution equations are  based on fixed
kernels, such as the Altarelli-Parisi splitting
functions \cite{Altarelli:1977zs} and/or the Catani-Seymour dipole ones
\cite{Catani:1996jh}, do not 
provide this particular uncertainty measure.

However, the presence of quark masses greatly increases the number 
of possible finite terms that could be added, hence we shall still
place some limitations on the type of terms we will allow for, 
as will be described in detail below.
 As a starting point, we require that we must be able to reproduce 
the dipole-antenna functions which were derived from physical matrix 
elements, such as those given above.
We then choose a generalization of the resulting parametrization, 
in such a way that the finite parts of all the dipole-antenna functions
are parametrized in a similar way.
We consider it preferable to have a rather general parametrization 
of the finite parts of the antenna functions because a change in the 
parametrization itself would require a change in the program code,
whereas changes to individual terms within a given parametrization can
be made without even recompiling the code.

Subsequently, we must decide which values to assign the finite
coefficients by default. Since we chiefly intend to use them for
variations, our philosophy is to set most of them to zero from the
start, allowing only for a few non-zero values to bring the tree-level 
expansion  of the resulting parton shower into reasonable agreement 
with the fixed-order matrix elements for $Z$ decay up to $Z \to
\text{6 partons}$. Note that explicit comparisons to such matrix elements 
are given in \secRef{sec:comparison}.

In the context of our shower model, one must
also require that the  dipole-antenna functions 
be positive definite, since they act as 
branching probability densities. This is the
case for all dipole-antennae considered in this paper.

With the default choices fixed, we also define two antenna-function 
variations which we
consider reasonably extremal, which we call ``MIN'' and ``MAX''.
Our approach here has been to choose the coefficients for the 
MIN set as small as possible without introducing negative values 
for the dipole-antennae and then choosing the MAX coefficients 
such that the difference between the default coefficients and the MIN 
coefficients is at least as big as that between the MAX coefficients 
and the default coefficients.
Note that we have not varied all possible finite coefficients in 
the MIN and MAX sets, but only a small subset of them, so it is
conceivable that some physical variations could fall outside the range
we define here. 
As always, uncertainty evaluations are more of an art than an
exact science. We expect to learn more about the reasonableness of
our choices as we expand to more processes in the future 
and can make explicit comparisons to more matrix elements. 

In the following we shall present the decomposition of the 
dipole-antenna functions used in VINCIA into their singular 
and finite parts. The following antennae are needed: 
gluon emission from a $q\bar{q}$, $qg$, and $gg$
parent antenna, and gluon splitting from a $qg$ or 
$gg$ parent. As mentioned in \secRef{sec:notation}, 
the hard radiators are always uniquely determined, and hence the
antennae we discuss here are equivalent to 
the ones referred to as ``sub-antennae''  in fixed-order contexts
(see, e.g., \cite{GehrmannDeRidder:2005cm}). For each of these five
antenna types, we give four separate sets of ``finite terms'':
DEF (default), MIN, MAX, and GGG, with the latter reproducing the fixed-order
antennae  defined in \cite{GehrmannDeRidder:2005cm,GehrmannDeRidder:2009fz,Abelof:2011jv}. 
Condensed summaries of the corresponding finite-term values are given
in \tabsRef{Tab:CqqEmit}, for gluon emission antennae, and in tables
\ref{Tab:Csplit}, for gluon-splitting ones.

To define the dipole-antenna for  
gluon emission off a massive $Q\bar{Q}$ pair (where the quark and the
antiquark may or may not be of identical flavour), 
we start with the generic form
\begin{multline}
\bar{a}_{g/q\bar{q}} \left( \Ecm^2, \sij, \sjk, \mq, \mqb \right) \ = \\
\ \frac{1}{m_{IK}^2} \left(
\frac{2 y_{ik} }{y_{ij} y_{jk} } - \frac{2 \muq^2}{y_{ij}^2} - \frac{ 2 \muqb^2 }{ \yjk^2 }
+ \frac{1}{1 - \muq^2 - \muqb^2 + x \, \muq \muqb}
	\left( \frac{y_{ij}}{y_{jk}} + \frac{ y_{jk}}{y_{ij}} 
+ F_{g/q\bar{q}}
\right)\right)~,\label{eq:a03}
\end{multline}
where $F_{g/q\bar{q}}$ represents an arbitrary ``finite'' function, i.e.\ a function 
which is regular in all soft and (quasi\mbox{-})collinear limits. 
The antenna function derived from $Z$ decay, as in the fixed order context 
and called $A^{0}_{3}=a_{3}^{0}$ 
corresponds to $F_{g/q\bar{q}}=0$ with $x=4$ and $m_q=m_{\bar{q}}$ as listed in 
\tabRef{Tab:CqqEmit}.

We allow for the following optional terms in
$F_{g/q\bar{q}}$, 
\begin{eqnarray}
F_{g/q\bar{q}}\left(\yij, \yjk, \muq, \muqb \right) & = & C_{00} +
C_{10}(\yij+\yjk) + C_{20}(\yij^2+\yjk^2) + C_{11}\,\yij\yjk \nonumber \\[1mm]
& & \quad + (\muq+\muqb)(M^{10}_{00} + M^{10}_{10}(\yij+\yjk))\nonumber \\[1mm]
& & \quad + (\muq^2+\muqb^2)(M^{20}_{00} + M^{20}_{10}(\yij+\yjk))\nonumber \\[1mm]
& & \quad + \muq\muqb(M^{11}_{00} + M^{11}_{10}(\yij+\yjk))
\end{eqnarray}
with the default values $C_{ab}=M^{cd}_{ab}=0$. 
Note that this form of $F$ explicitly respects charge conjugation symmetry
($i\leftrightarrow k$). In principle, one could allow for terms with
higher powers of masses and/or invariants, but for the  simple purpose
of uncertainty estimates, we believe the form above gives sufficient 
flexibility.\footnote{
The denominator factor $s_{IK} + x \, \mq \mqbar$ in \eqRef{eq:a03} 
has $x=0$ in \Vc, but we retain the possibility to vary it for
uncertainty estimates.}

The default $Q\bar{Q}$ antenna function used in
\Vc\ is thus,  
\begin{equation}
\bar{a}_{g/q\bar{q}}^{\mrm{(def)}} \left( \Ecm^2, \sij, \sjk, \mq,
\mqb \right) = \frac{1}{m_{IK}^2}\left(
\frac{2 \yik }{\yij \yjk } - \frac{2 \muq^2}{\yij^2} - \frac{ 2 \muqb^2 }{ \yjk^2 }
+  \frac{1}{1- \muq^2- \muqb^2} \left( \frac{\y_{ij}}{\y_{jk}} + \frac{ \y_{jk}}{\y_{ij}} \right)
\right)~\label{eq:a03def}
\end{equation}
which corresponds to choosing zero values for the finite part and for $x$ 
in \eqRef{eq:a03}. In the left-hand pane of \tabRef{Tab:CqqEmit}, 
we compare the default values in \Vc\ to the GGG
functions and to a ``MIN'' and ``MAX'' variation that we use for
uncertainty estimates. Note that a non-zero $M$ coefficient is
introduced in the ``MIN'' case, in order to avoid negative regions in
the massive dipole-antenna function.
\begin{table}[tp]
\centering
\parbox[t]{0.3\textwidth}{
\begin{tabular}[t]{l c c c c }
\multicolumn{5}{c}{$\mathbf{q\bar{q}\to qg\bar{q}}$}\\
\toprule
\scriptsize$\bar{a}_{g/q\bar{q}}$\! &	\textbf{Def} & GGG & MIN & MAX \\
\midrule
${\cal C}$ & $\frac83$ & $\frac83$  & $\frac83$ &  $\frac83$ \\
\midrule
 $C_{00}$   &    - &   - &  -6  & 6 \\
 $C_{10}$   &    - &   - &  4.5 & -4.5 \\
 $C_{01}$   &    - &   - & 4.5 & -4.5 \\
\midrule
 $x$   &    - &  4 &  4  & -2 \\
$M^{20}_{00}$\! & - & - & 9 & -\\
\bottomrule
\end{tabular}\\[5mm]
\begin{tabular}[t]{l c c c c }
\multicolumn{5}{c}{$\mathbf{gg\to ggg}$}\\
\toprule
\scriptsize$\bar{a}_{g/q\bar{q}}$\!&		\textbf{Def} & GGG & MIN & MAX \\
\midrule
${\cal C}$ & $3$ & $3$  & $\frac83$ &  $3$ \\
\midrule
 $C_{00}$   &    2 &  $\frac83$ &  -8  &  10 \\
 $C_{10}$   &    - &  -1 &  7.5 & -7.5 \\
 $C_{01}$   &    - &  -1 &  7.5 & -7.5 \\
\bottomrule
\end{tabular}}
\hspace*{1cm}
\begin{tabular}[t]{l c c c c }
\multicolumn{5}{c}{$\mathbf{qg\to qgg}$}\\
\toprule
\scriptsize $\bar{a}_{g/q\bar{q}}$\! & \textbf{Def} & GGG & MIN & MAX \\
\midrule
${\cal C}$ & $3$ & $3$  & $\frac83$ &  $3$ \\
\midrule
 $C_{00}$   &  2 &  2.5 &  -6 & 10 \\
 $C_{10}$   & -1 &  -2 &   6 & -8 \\
 $C_{01}$   &  - &  -0.5 &   7 & -7 \\
\midrule
$M_{00}^{1}$&  - &  1 &  -  &  -\\ 
$M_{00}^{2}$&  - & -3.5 & 8.5 &  - \\ 
$M_{10}^{2}$&  - &  - &  8  &  -\\ 
$M_{01}^{2}$&  - &  - &  8  &  -\\ 
$M_{-11}^{2}$\!\!\!  &  -& -2 & - & -\\ 
$\widetilde{M}_{-1-12}^{2}$\!\!\!&  - & -1 & -& -\\ 
$\widetilde{M}_{-1-11}^{3}$\!\!\!&  - & -1 & -& - \\ 
$\widetilde{M}_{-1-11}^{4}$\!\!\!&  - & 1 & - & -\\ 
\bottomrule
\end{tabular}
\caption{Colour factors and finite parts for four different examples of the
  colour-ordered gluon-emission antenna functions. 
  The \Vc\ default antenna set (\textbf{Def}) is compared to the ``GGG'', MIN,
  and MAX variations. Coefficients which are not listed (or which are
  represented by ``-'') are zero. \label{Tab:CqqEmit}}
\end{table}

\begin{figure}[tp]
\centering
	\includegraphics[width = 0.4 \textwidth]{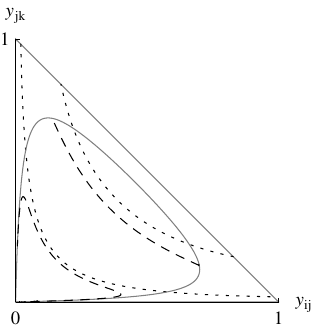}\\
\scriptsize $Q\bar{Q} \to Qg\bar{Q}$
\caption{Dalitz plot showing contours of the
  massive (dashed) and massless (short dashed) 
  gluon-emission dipole-antenna
  function $\bar{a}_{g/q\bar{q}}$, with $m_Q = 0.15 \, m_{IK}$. 
  Contours of constant values of the dipole-antennae are shown 
  for $\bar{a}=5,50$.
  Grey solid lines denote the boundaries of the massless and of 
  the massive phase space respectively.
\label{Fig:a03}}
\end{figure}

The influence of quark masses on the default $Q\bar{Q}$ antenna function is 
illustrated in figure \ref{Fig:a03}. This figure shows contours of constant
values for the antenna $a_{g/q\bar{q}}$ 
in a Dalitz plot of the three-particle phase space for massless quarks (short dashed) 
and for massive quarks (dashed). For massive quarks, the contour lines 
start to avoid the boundaries of the phase space (drawn as a solid grey line) 
for high values of the antenna 
function. This is a direct consequence of the presence of terms of the
form 
$(-\sij^2/m^2)$ in the massive soft Eikonal factor given in \eqRef{eq:softeikonal}.

For gluon emission in a $Qg$ dipole-antenna, we use the generic\footnote{
There is an additional ambiguity which originates from the collinear gluon singularity.
In \Vc, we use identified particles and therefore we have to distribute the collinear 
gluon splitting singularity $p_j \parallel p_k$ onto two dipole-antennae, one 
in which $j$ is a hard radiator and only $k$ can become soft and one in which 
$k$ is a hard radiator and only $j$ can become soft.
When doing so, we could introduce a $j \leftrightarrow k$-asymmetric term
of the form $\asy (\yij - \yik)/[\yjk (1-\muq^2) ]$ to the 
dipole-antenna function which contributes in the collinear limit $j \parallel k$.
To first order, such an asymmetric term cancels out, but it would still 
influence the shower at higher orders.
However, since values different from $\asy=0$ tend to lead to negative
dipole-antenna functions for high quark masses, we have so far not
enabled the option to vary it  in the {\Vc{}} code.
} form
\begin{multline}
\bar{a}_{g/qg}\left( \Ecm^2, \sij, \sjk, \mq \right) = \\
\frac{1}{m_{IK}^2}\left(\frac{2 \yik}{\yij \yjk} - \frac{2 \muq^2}{\yij^2} 
	+ \frac{1}{1-\mu_q^2} \left(
		 \frac{\yjk}{\yij} +
		 \frac{ \yij}{\yjk} \left(
		 1 - \frac{ \yij }{ 1- \muq^2}
		 \right)
	+ \frac{1}{1-\muq^2}F_{g/qg}\right)\right)
	~. \label{eq:d03}
\end{multline}
We allow for the following finite terms, 
\begin{eqnarray}
\label{eq:d03general}
F_{g/qg} & = &  C_{00} + C_{10}\, \yij + C_{01}\, \yjk + C_{20}\,\yij^2 +
C_{02}\,\yjk^2 + C_{11}\,\yij\yjk \nonumber\\[2mm] & &
		\quad + \muq (M^{1}_{00}  + M^{1}_{10}\, \yij +
                M^{1}_{01}\yjk) \nonumber\\[2mm]
 && \quad + \muq^2 \big(M^{2}_{00}  + M^{2}_{10}\, \yij +
                M^{2}_{01}\,\yjk + M^{2}_{-10}\, \frac{1}{\yij} + 
                M^{2}_{-11}\, \frac{\yjk}{\yij} \nonumber\\
 && \qquad + \widetilde{M}^{2}_{-1-12} \frac{\yjk^2}{\yij\yik}
 		+ \widetilde{M}^{3}_{-1-11} \frac{\muq \yjk}{\yij\yik}
		+ \widetilde{M}^{4}_{-1-11} \frac{\muq^2 \yjk}{\yij\yik}
 \big)~.
\end{eqnarray}
In a fixed order context, the corresponding massive $d_{3}^{0}$ 
was derived from the matrix element for the decay of  
a neutralino into a massive gluino and two gluons, 
$\tilde{\chi} \to \tilde{g}gg$ \cite{GehrmannDeRidder:2009fz}, 
using an effective Lagrangian~\cite{GehrmannDeRidder:2005hi}. 
Similarly to the massive dipole-antenna $\bar{a}_{g/q\bar{q}}$, 
the denominator factor $s_{IK}^2=(\Ecm^2 -
\mq^2)^2$ is proportional to the coupling-stripped two-particle matrix
element for neutralino decay into a gluino and a gluon, $\tilde{\chi} \to
\tilde{g}g$. 
It has mass dimension four  
in this case because the neutralino-gluino-gluon coupling has mass
dimension $-2$. 
%
The parametrization in \eqRef{eq:d03general} contains finite terms 
which are proportional to $\mu_q^2/(y_{ij} y_{ik})$.
Although these terms seem out of place in a 
quark-gluon dipole-antenna function,
they are indeed part of the 
fixed-order antenna $d^0_3$ if the quark is massive.
Their appearance is connected to the fact
that the physical matrix element for $\tilde{\chi} \to \tilde{g} g g$
from which $d^0_3$ is extracted
is symmetric under the interchange of the two gluons.
We are using the quark-gluon dipole-antenna function to generate 
emissions in a situation which is decidedly asymmetric between 
the two gluons since one is identified as the hard radiator and 
the other is the emitted particle.
For this reason, the terms proportionate to $\mu_q^2/(y_{ij} y_{ik})$
have been deactivated by default 
in \Vc and are not even considered for the purpose of uncertainty 
estimates.
 
The values for the other coefficients are summarized
in the right-hand pane of \tabRef{Tab:CqqEmit}, with 
the coefficients reproducing the fixed-order $d^0_3$ given in the
``GGG'' column. 

Obviously, there are no (leading-order) mass effects for $gg\to ggg$. We include
the corresponding generic\footnote{
If a term $\asy (\yij - \yik)/[ \yjk (1- \muq^2) ]$ is added to \eqRef{eq:d03}, we
would also need to  
introduce a term  $\asy (\yij - \yik)/\yjk + \asy (\yjk - \yik)/\yij$ in 
\eqRef{eq:f} to ensure that 
$a_{g/qg}(p_i, p_j, p_k) + a_{g/gg}(p_j, p_k, p_l)$ reproduces
$P_{g/gg}$  for $p_j \parallel p_k$.
} form of the dipole-antenna function 
here for completeness,
\begin{equation}
\bar{a}_{g/gg} \left( \Ecm^2, \sij, \sjk \right) = \frac{1}{m^2_{IK}} \left(
\frac{2\yik}{\yjk \yij}
	+
        \frac{\yij (1-y_{ij})}{\yjk}  
        + \frac{\yjk (1-y_{jk})}{\yij} 
	+\sum_{a,b\ge 0}^{a+b\le 2} C_{ab}\, 
	\yij^{a}  \yjk^{b}\right)~, \label{eq:f}
\end{equation}
where the last term represents $F_{g/gg}$, with coefficients as given
in the bottom left-hand pane of \tabRef{Tab:CqqEmit}. The ``GGG''
column tabulates the coefficients of the $f^0_3$ function derived in 
\cite{GehrmannDeRidder:2005aw} using an effective Lagrangian for the
process $H \to ggg$ normalized to $H\to gg$.

\begin{table}[tp]
\centering
\small{\hfill
\begin{tabular}[t]{l r r r r}
\multicolumn{5}{c}{$\mathbf{qg\to q\bar{q}'q'}$ }\\
\toprule
$\bar{a}_{\qbar'/qg}$&		\textbf{Def} & GGG & MIN & MAX \\
\midrule
${\cal C}$ & $n_f$ & $n_f$  & $n_f$ &  $n_f$ \\
\midrule
$C_{00}$  &  0.3 &  0.5 & - & 0.6\\
 $C_{10}$   &    - &   - & -  & -\\
$C_{01}$  &  - & -0.5 & - & -\\ 
\midrule
$\Ct_{01}$&  - & -2.0 & - & -\\ 
$M_{00}^{10}$&  - & -1.0 & - & -\\ 
$M_{00}^{20}$&  - & -0.5 & - & -\\ 
$\widetilde{M}_{00}^{02}$&  -& -2.0 & - & -\\ 
$\widetilde{M}_{00}^{10}$&  - & -4.0 & - &- \\ 
$\widetilde{M}_{00}^{20}$&  - & -2.0 & - & -\\ 
\bottomrule
\end{tabular}\hfill
\begin{tabular}[t]{l r r r r }
\multicolumn{5}{c}{$\mathbf{gg\to g\bar{q}q}$}\\
\toprule
$\bar{a}_{\qbar/gg}$ &		\textbf{Def} & GGG & MIN & MAX \\
\midrule
${\cal C}$ & $n_f$ & $n_f$  & $n_f$ &  $n_f$ \\
\midrule
 $C_{00}$   &  0.3 &   - &  - & 0.6 \\
 $C_{10}$   &    - &   - &  - & - \\
 $C_{01}$   &    - &   - & -  & -\\
\midrule
 $\Ct_{00}$ &    - &-2.0 &  - & - \\ 
 $\Ct_{01}$ &    - & 1.0 &  - & - \\ 
 $M_{00}^2$ &    - & 2.0 &  - & -\\ 
\bottomrule
\end{tabular}\hfill}
\caption{Colour factors and finite parts for four different examples of the
  colour-ordered $qg\to q\bar{q}'q'$ and $gg\to g\bar{q}q$ antenna
  functions. The \Vc\ default antenna set (\textbf{Def}) is compared
  to the ``GGG'', ``MIN", 
  and ``MAX" variations. Coefficients which are not listed (or which are
  represented by ``-'') are zero.}
\label{Tab:Csplit}
\end{table}

We proceed with the dipole-antennae describing the gluon splitting processes.
The gluon-splitting process $Qg\to Q\bar{Q}'Q'$ is described by the
following generic form,

\begin{equation}
\bar{a}_{\bar{q}'/qg} \left( \Ecm^2, \sij, \sjk, \mq, \mqp \right) = 
\frac12\frac{1}{\sjk + 2 \mqp^2}\left(
  \frac{\sik^2 + \sij^2}{ s_{IK}^2}  
+ \frac{ 2\mqp^2 }{\sjk + 2 \mqp^2}\right)
+  \frac{\mant^2}{ s_{IK}^2 } F_{\bar{q}'/qg} ~,
\end{equation}
with finite terms, 
\begin{eqnarray}
F_{\bar{q}'/qg} & = & 
	\frac{\yjk}{\yjk + 2 \muqp^2} \left(
		C_{00} + C_{10}\,\yij + C_{01}\, \yjk + C_{20}\,\yij^2 +
		C_{02}\,\yjk^2 + C_{11}\,\yij\yjk
	\right)
\nonumber\\[2mm] & & \quad 
	+ \frac{\yjk \, \muq}{\yjk + 2 \muqp^2} \left(
		M^{10}_{00}  + M^{10}_{10}\, \yij +M^{10}_{01}\yjk +
		\muq \left( M^{20}_{00}  + M^{20}_{10}\, \yij +M^{20}_{01}\yjk \right)
	\right)  
\nonumber\\& &  
	+ \frac{\muqp^2}{\yjk + 2 \muqp^2} \left(
		\Ctil_{00} + \Ctil_{10}\,\yij + \Ctil_{01}\, \yjk + \Ctil_{20}\,\yij^2 +
		\Ctil_{02}\,\yjk^2 + \Ctil_{11}\,\yij\yjk
	\right)
\nonumber\\[2mm] & & \quad 
	+ \frac{\muqp^2 \, \muq}{\yjk + 2 \muqp^2} \left(
		\Mtil^{10}_{00}  + \Mtil^{10}_{10}\, \yij +\Mtil^{10}_{01}\yjk +
		\muq \left( \Mtil^{20}_{00}  + \Mtil^{20}_{10}\, \yij +\Mtil^{20}_{01}\yjk \right)
	\right)
\nonumber\\[2mm] & & \quad 
	+ \frac{ \muqp^4 }{\yjk + 2 \muqp^2} \left(
		\Mtil^{02}_{00}  + \Mtil^{02}_{10}\, \yij +\Mtil^{02}_{01}\yjk \right)
 ~,
\end{eqnarray}
whose values are listed in the left-hand pane of \tabRef{Tab:Csplit}. 
In a fixed order context, the massive $e_3^0$ quark-gluon antenna 
can be derived from the decay
of a neutralino into a gluino and quark-antiquark, i.e from the
process $\tilde{\chi} \to \tilde{G}Q\bar{Q}$. The resulting
partons, gluino and a quark-antiquark pair can either be massless or massive.
The presence of two different masses is the main reason why the structure 
of the finite term $F_{\bar{q}'/qg}$ is more complicated than for the other 
dipole-antenna functions.
The values of the coefficients which reproduce the sub-antenna
$e_3^0$ for the most general case (all partons massive)  are given 
in the GGG column of \tabRef{Tab:Csplit}.

\begin{figure}
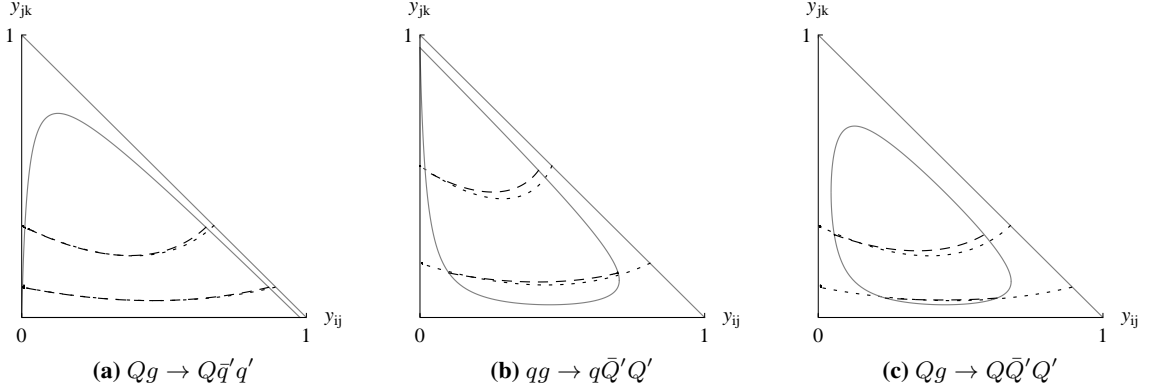

\centering
\subfloat[$Qg\to Q\bar{q}'q'$]{
	\includegraphics[width = 0.28
          \textwidth]{{{e03-15-0}}} 
} 
\hspace*{0.5cm}
\subfloat[$qg\to q\bar{Q}'Q'$]{
	\includegraphics[width = 0.28 \textwidth]{{{e03-0-15}}}
} \hspace*{0.5cm}
\subfloat[$Qg\to Q\bar{Q}'Q'$]{
	\includegraphics[width = 0.28 \textwidth]{{{e03-15-15}}}
}
\caption{Dalitz plot showing contours of the
  massive (dashed) and massless (short dashed) 
  gluon-splitting antenna
  function $\bar{a}_{\bar{q}'/qg}$, 
  for three different combinations of massive and massless partons,
  with $m_Q = 0.15 \, m_{IK}$. Contours are shown for $\bar{a}=1,4$ in
  (a) and (c) and for  $\bar{a}=0.5,2$ in (b).
  Grey solid lines mark the boundary of the phase space.
  Contrary to the gluon emission case, there are no qualitative changes 
  caused by the introduction of quark masses.
\label{fig:e03}}
\end{figure}

Contours of constant value of this function 
with default parameters for the finite part
are illustrated in
\figRef{fig:e03}, for three different mass combinations, shown from
left to right. Dashed contours represent massive functions, short dashed contours
represent massless functions.
Apart from the modifications to the size of the
physical phase space,  the mass effects are only numerically 
important away from the collinear limit.

Finally, the gluon-splitting process $gg \to g\bar{Q}Q$ is described by
the following generic form of its dipole-antenna function
\begin{equation}
\label{eq:g03}
\bar{a}_{\bar{q}/gg} \left( \Ecm^2, \sij, \sjk, \mq \right) = 
\frac12\frac{1}{\sjk + 2 \mq^2}\left(
  \frac{\sik^2 + \sij^2}{ s_{IK}^2}  
+ \frac{ 2\mq^2 }{\sjk + 2 \mq^2}\right)
+  \frac{\mant^2}{ s_{IK}^2 } F_{\bar{q}'/qg} ~,
\end{equation}
with finite terms, 
\begin{eqnarray}
F_{\bar{q}/gg} & = & 
	\frac{\yjk}{\yjk + 2 \muq^2} \left(
		C_{00} + C_{10}\,\yij + C_{01}\, \yjk + C_{20}\,\yij^2 +
		C_{02}\,\yjk^2 + C_{11}\,\yij\yjk
	\right)
\nonumber\\& &  
	+ \frac{\muq^2}{\yjk + 2 \muq^2} \left(
		\Ctil_{00} + \Ctil_{10}\,\yij + \Ctil_{01}\, \yjk + \Ctil_{20}\,\yij^2 +
		\Ctil_{02}\,\yjk^2 + \Ctil_{11}\,\yij\yjk
	\right)
\nonumber\\[2mm] & & \quad 
	+ \frac{ \muq^4 }{\yjk + 2 \muq^2} \left(
		M^{2}_{00}  + M^{2}_{10}\, \yij +M^{2}_{01}\yjk \right)
 ~.
\end{eqnarray}
In this case, the corresponding fixed-order antenna $g_{3}^{0}$   
can be derived from the decay of a
Higgs into a gluon and a massive quark-antiquark pair i.e from $H \to
Q\bar{Q}g$. It can be
obtained from the dipole-antenna analogue given in \eqRef{eq:g03}
by setting the finite parts F according to the right-hand pane of 
\tabRef{Tab:Csplit}.
		
Note that the finite parts of the the two dipole-antennae 
related to gluon splitting $\bar{a}_{\qbar'/qg}$ and $\bar{a}_{\qbar/gg}$ have been 
parametrized in the same way. In the $\bar{a}_{\qbar/gg}$ case,
some simplifications occur though due to the
presence of a massless parton in the final state.

Finally, let us mention that \Vc is not necessarily restricted to
describe processes in the Standard Model with massive fermions in the
final state.
It could equally be used to describe
processes with massive final state particles with different spin-statistics
properties.
Since many models of physics beyond the Standard Model contain 
new heavy coloured particles which can be scalars, 
\Vc needs a default dipole-antenna function for those.
Since the soft Eikonal factor of \eqRef{eq:softeikonal} is 
spin-independent, it can be used as a default  
dipole-antenna in \Vc for those cases:
\begin{equation}
\bar{a}_{\mrm{Eikonal}}(\mant^2,s_{ij},s_{jk},m^2_I,m_K^2) =
\frac{2 s_{ik}}{s_{ij}s_{jk}} - \frac{2 m_{I}^2}{s_{ij}^2} - \frac{2
  m_K^2}{s_{jk}^2}~,\label{eq:eikonal}
\end{equation}
where $m_I=m_i$ and $m_K=m_k$ are the masses of the radiating partons
and $s_{ik}$ can be obtained from the other arguments using
\eqRef{eq:Econs}. The colour factor used is equal to
$\hat{C}_F=2C_F$ if both parents are in the fundamental
representation, $C_A$ if both are adjoints, and options for anything
in between for mixed-parent antennae, with the default being
$\frac12(\hat{C}_F + C_A)$.

\section{The VINCIA Formalism with Massive Particles \label{sec:showerformalism}}

In this section we present the main ingredients 
of our dipole-antenna shower formalism implemented 
in the present version of \Vc.
This formalism was first derived in \cite{Giele:2007di,Giele:2011cb} to
describe QCD radiation off massless partons. We here generalize it to
take quark mass effects into account.
For completeness, some aspects which are carried over from the massless
case without modification 
are also summarized.\footnote{
We  encourage  readers unfamiliar with shower
formulations to consult 
\cite{Buckley:2011ms,Skands:2011pf} for recent pedagogical reviews.}

\Vc is based on the dipole-antenna picture of QCD
radiation~\cite{Gustafson:1987rq}: its fundamental evolution step is a
Lorentz-invariant $2\to 3$ branching process $ I K \to ijk$ 
by which two on-shell parent partons ($I$ and $K$) 
are replaced by three on-shell daughter partons ($i$, $j$,
$k$), conserving four-momentum exactly. Dipole-antenna functions and
phase-space mappings were discussed in \secsRef{sec:phasespacemap} and
\ref{sec:massiveantennae}.

To construct an explicit shower algorithm, one must furthermore introduce 
an evolution variable (a.k.a.\ shower ordering variable or resolution
scale), $Q_E$, suitably generalized to massive particles as will be discussed in
\secRef{sec:ordering}. Together with this variable one also needs to
define a concrete iterative strategy for generating and accepting ``trial
branchings'' according to the Sudakov form factors, 
as described in \secsRef{sec:evolution} -- \ref{sec:trial-splittings}. 

Matching to fixed-order matrix elements is performed using a unitary matching 
scheme, which results in unweighted events matched to full-colour tree-level matrix
elements squared, as explained in \cite{Giele:2011cb}. We give a brief
summary of the main points of this method in \secRef{sec:matching}. 
\Vc also calculates uncertainty estimates for its predictions at a moderate 
speed penalty, 
those are summarized in \secRef{sec:uncertainties}.

\subsection{Ordering \label{sec:ordering}}

In the dipole-antenna formalism, subsequent emissions are naturally
ordered by the nesting of the on-shell $2\to 3$ phase spaces: as more and more
emissions (or $g\to q\bar{q}$ splittings) are added,
each dipole-antenna will, on average, carry a progressively smaller
fraction of the total original centre-of mass energy squared $s$. 
We refer to this as ``no ordering'', but it could equally well be
called ``phase-space-ordering'', since the only constraint implied
 is that energy-momentum inside
the nested $2\to 3$ phase spaces is conserved. 

As demonstrated in \cite{Skands:2009tb,Giele:2011cb}, however, and
further elaborated on for the massive case in our
\secRef{sec:comparison}, a dipole-antenna shower without any
additional constraints would produce far too much
radiation outside the double-logarithmic limit, i.e in the so-called
``hard region''. 
This is essentially due to the  fact that an ``unordered'' shower
approximation can be viewed as a sum of {\it independent dipoles}, while full
QCD, beyond the $2\to 3$ level, has a more complicated multipole
structure, with in particular destructive interference produced by
colour-coherence effects. A reasonable agreement with all-orders QCD
can be restored by enforcing a strict ordering of the emissions in
terms of some ``evolution scale'', $Q_E$. This variable represents a measure of
(inverse) formation time or characteristic wavelength. 

The \Vc formalism can accommodate a whole class of evolution
variables which differ in how they prioritize 
soft emissions relative to collinear ones. 
We refrain from characterizing all possible evolution variables here, 
referring instead to the original 
publication\cite{Giele:2011cb} for details.

The two most important evolution variables implemented in \Vc 
are the following:
\begin{equation}
\begin{array}{c}
\mbox{\underline{(Type 1) Transverse Momentum}}\\[1mm] \displaystyle
 Q_E^2 \ = \ 4p_\perp^2  \ \stackrel{\text{def}}{=}
\ 4\frac{s_{ij}s_{jk}}{\mant^2} \ = \ 4\frac{(2p_i\cdot p_j)(2p_j\cdot
  p_k)}{\mant^2} ~\stackrel{m_j=0}{=} 
\ 4 \frac{(m_{ij}^2 - m_i^2)(m_{jk}^2-m_k^2)}{\mant^2}~,
\end{array}
\end{equation}
\begin{equation}
\begin{array}{c}
\mbox{\underline{\mbox{(Type 2) Dipole Virtuality}}}\\[1mm] 
Q_E^2 \ = \ m_D^2  \ \stackrel{\text{def}}{=}
\ 2\min(s_{ij},s_{jk}) ~\stackrel{m_j=0}{=} \ 2\min(m_{ij}^2-m_i^2,m_{jk}^2-m_k^2)~,
\end{array}
\end{equation}
where we again emphasize that the notation $s_{ij}$ is used for the 
dot product $s_{ij} = 2p_i\cdot p_j$, which differs from the 
invariant mass squared $m_{ij}^2=(p_i+p_j)^2$ when non-zero rest
masses are involved. Note that the rightmost expressions in the above
equations are appropriate only to gluon emission, for which $m_j=0$.

While the imposition of such ordering conditions can 
extend the agreement with higher-order QCD to a
much larger region of phase space outside the double-logarithmic limit 
(see, e.g., \cite{Skands:2009tb,Giele:2011cb}), 
it does have a side effect, which is formally beyond LL:  
in general, there 
will be small corners of the $n$-particle phase space which are not 
accessible through any sequence of strongly ordered branchings
\cite{Andersson:1991he,Skands:2009tb}. Those are called ``dead zones''
, which the shower does not populate at all. These zones correspond to regions
of phase space that are classified by the ordering condition as having
no LL contributions. It is therefore consistent to set these
corresponding contributions to zero at the LL level.

As shown in \cite{Giele:2011cb}, it 
is possible to avoid dead zones without re-introducing the large
overestimates present in the ``no-ordering'' scenario. This can be done 
by allowing unordered branchings to occur  with a
suppressed probability which does not affect the LL accuracy of the shower. 
This is technically achieved by starting from an unordered shower and, instead
of applying the strong-ordering condition as a step function, 
apply a smooth damping factor instead. We label this as an 
``improved'' ordering condition, $P_{\mrm{imp}}$, 
\begin{equation}
\Theta_\mrm{strong-ordering} \ \to \	
  P_\mrm{imp}(\widehat{Q}_E^2, Q_E^2)  
    = \frac{ \widehat{Q}_E^2 }{ \widehat{Q}_E^2 + Q_E^2 }~,
\label{eq:pimp}
\end{equation}
where $Q_E$ is the evolution scale of the current $n \to n+1$ branching.
$\widehat{Q}_E$ is a measure of the scale of the 
$n$-particle configuration, defined as the minimal value of the
evolution scale evaluated over all partons in the $n$-parton parent
configuration. For branchings which are at a much lower scale than 
the last one, i.e.\ the strongly ordered limit $Q_E^2 \ll
\widehat{Q}_E^2$, this factor is unity, 
whereas branchings with the opposite hierarchy $Q_E^2 \gg
\widehat{Q}_E^2$ are strongly suppressed.  
At the point $Q_E^2 = \widehat{Q}_E^2$, 
the suppression factor in \eqRef{eq:pimp} is equal to $1/2$. We refer
to this as  the ``smooth ordering'' condition. 

Note that, since the
$P_{\mrm{imp}}$ factor is everywhere smaller than unity, it can be
applied as a probabilistic veto, which we make use of in the technical
implementation. 

\subsection{The Evolution Algorithm \label{sec:evolution}}

Formally, we can define a unitary evolution 
operator $\evo( \{ p \}_n, \qstart, \qstop )$
which generates the perturbative radiation off an $n$-parton 
state $\{p\}_n$ between the two resolution 
scales $Q_{\text{start}}$ and $Q_{\text{stop}}$ in terms 
of an iterative Markov chain\footnote{
Strictly speaking, strongly ordered showers  depend on 
the scale of the last branching, and are
therefore not completely Markovian. In the context of VINCIA, the
difference is only really relevant in the context of matching to
matrix elements and can therefore be ignored for the discussion of the
pure shower algorithm.} 	
\begin{multline}
\evo( \{ p \}_n, \qstart, \qstop ) = \Delta ( \{ p \}_n, \qstart, \qstop ) \\[1mm]
+\sum_{IK \to ijk} \int_{\qstop}^{\qstart} 
	\frac{ \dPS[ijk]{3} }{ \dPS[IK]{2}}
	a_{j/IK}
	\Delta \left( \{ p \}_n, \qstart, Q^{ijk} \right)
	\evo( \{ p \}_{n+1}^{IK \to ijk}, Q^{ijk}_\text{restart}, \qstop )~,
\label{eq:S}
\end{multline}
where the first line represents the fraction of states that remain
unchanged by the evolution (i.e., the exclusive $n$-parton fraction
at the resolution scale $\qstop$), 
while the second line includes all states that do evolve (i.e., the
inclusive $(n+1)$-parton fraction). To maintain
unitarity, the second line must be minus the derivative of the
first. Analogously to other 
time-dependent decay problems, the solution to this differential
equation is that the Sudakov factor, $\Delta$, must be the exponential of 
minus the integrated tree-level branching probability,
\begin{equation}  
	\Delta ( \{ p \}_n, \qstart, \qemit ) = \prod_{IK \to ijk} \exp \left( -\aint_{j/IK} ( \qstart, \qemit)  \right), 
	\label{eq:sudakov}
\end{equation}
\begin{equation}
\aint_{j/IK}( \qstart, \qemit) = \int_{\qemit}^{\qstart} \frac{ \dPS[ijk]{3}}{\dPS[IK]{2}} 
 			a_{j/IK} \left( p_i, p_j, p_k \right),
\label{eq:dipInt}
\end{equation}
with phase-space measures and dipole-antenna functions as defined 
in \secRef{sec:formalism}. 

The sum  in  \eqRef{eq:S} and the product in \eqref{eq:sudakov} 
run over all possible $2 \to 3$
branchings $IK\to ijk$. The integrals in \eqsRef{eq:S} and
\eqref{eq:dipInt} are performed over the 
range $\qstart > Q^{ijk} > \qstop$, with $Q^{ijk} = \Qe( p_i, p_j, p_k
)$ the evolution variable. 

The starting scale, $\qstart$, represents the
``factorization scale'' of the $n$-parton configuration and 
may be given either by the invariant mass of the evolving dipole-antenna, by 
the restart scale defined by a previous branching, or by some
externally imposed scale, depending on the type of ordering criterion imposed on
the shower evolution.

In a standard shower application, $\qstop$ represents 
the infrared shower cutoff, or hadronization scale, 
with a value of
$\sim$ $1\,$GeV. 
For simplicity, we have assumed here that the definition of 
\qstop in terms of the post-branching 
momenta $p_i$, $p_j$ and $p_k$ is the same as that of \qemit.
In practice, this is not necessarily the case, see the section on
hadronization in \cite{Giele:2011cb}.

The presence of $\evo( \{ p \}^{IK \to ijk}_{n+1},
Q^{ijk}_\text{restart}, \qstop )$ in \eqRef{eq:S} generates the
continued (iterated) evolution of the $(n+1)$-parton state after
branching,
with $Q^{ijk}_{\mrm{restart}}$ normally taken to be equal to
$Q^{ijk}$, for a traditional so-called  
strongly-ordered shower. 

Now we turn to the algorithmic steps themselves. Since the $2\to 3$ branching
phase space is three-dimensional (two independent Lorentz invariants, 
e.g.\ $s_{ij}$ and $s_{jk}$, and 
the azimuthal angle $\phi$ which determines the global orientation), 
three independent random numbers 
must be picked for each step of the algorithm. The first of these is 
the \qemit\ scale,  distributed according $1-\Delta(\qstart,\qemit)$ 
with the ``no-emission" probability $\Delta$ given in \eqRef{eq:sudakov}. 
Next, the second independent Lorentz invariant must be generated. 
It is generated according
to the integrand in \eqRef{eq:dipInt}. From the two Lorentz invariants and the 
azimuthal angle $\phi$ (which is chosen according to a flat probability 
density because the integrand does not depend on it), 
the momenta $p_i$, $p_j$, and $p_k$ can then be
constructed from the parent momenta $p_I$ and $p_K$, based on the chosen kinematics map and the relations
given in \secRef{sec:phasespacemap}. 

The algebra and computational overhead involved in generating the
three branching invariants can be 
simplified considerably by judicious use of the veto algorithm (see
\cite{James:1980yn,Weinzierl:2000wd,Buckley:2011ms} for pedagogical 
reviews). First, so-called ``trial branchings'' are generated, 
using a  simplified form of the integrand in \eqRef{eq:dipInt}. 
A percentage of these trials are 
then rejected, using 
\begin{equation}
\label{eq:paccept}
\paccept = 
	\frac{\ant_{j/IK} \left( \mant^2, \sij, \sjk, \left\{m\right\} \right) }
		{ a_{\mrm{trial}} \left( \mant^2, \sij, \sjk \right) }~,
\end{equation}
to determine whether a given trial branching should be accepted or
not. Symbolically $\{ m \}$ stands for all the masses of the parent 
and daughter partons, $a_{j/IK}$ is the desired integrand in
\eqRef{eq:dipInt}, and $a_\mrm{trial}$ is the simplified trial
function. The veto algorithm
ensures that the final answer 
has no dependence (apart from the speed with which it is obtained) 
on the form of the trial function used, requiring only that this function is
an overestimate of the correct integrand over all of phase space, so
that the accept probability \eqRef{eq:paccept} does not exceed
unity. 

\subsection{Trial Gluon Emissions \label{sec:trials}}

For gluon emission, the trial function used in VINCIA is based on the 
double-pole singular behaviour of the soft Eikonal factor,
\eqRef{eq:softeikonal}, which it coincides with in the soft limit and 
overestimates everywhere else. Using the notation conventions adopted
in \secRef{sec:notation}, it can be adapted straightforwardly from the
massless case, and is given by 
\begin{equation}
\label{eq:trial}
	\ant_\text{trial-emit} 
		 =
		\frac{\alphaStrial}{4 \pi} C_A \frac{2 \mant^2}{s_{ij}s_{jk}}~,
\end{equation}
where the 
overestimates $\alphaStrial \geq \alpha_S$ and 
$\colijktrialemit=C_A \geq \colijk$ can be used to guarantee sufficient
``headroom'' for arbitrary coupling constants and colour factors,
respectively. 
Since the mass corrections to the soft Eikonal factor 
and to the quasi-collinear splitting function are negative (as given 
in \eqsRef{eq:softeikonal} and \eqref{eq:splittingmassive} and
illustrated, e.g., in \figRef{fig:QgNorm}), this function is also
guaranteed to be an overestimate in the massive case. 

\begin{figure}[t]
\centering
\scalebox{1.025}{
\begin{tabular}{c}
\includegraphics*[scale=0.6]{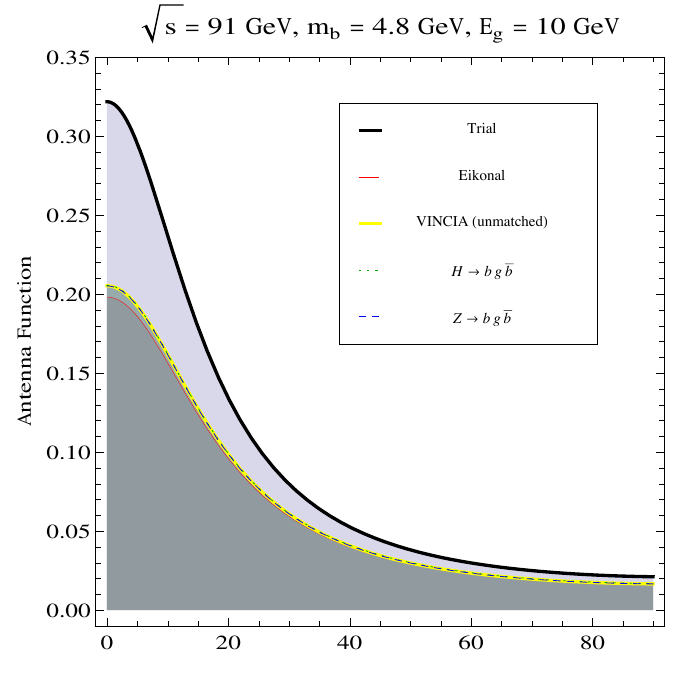}\\[-1.32cm]
\hspace*{-1.03mm}\includegraphics*[scale=0.5983]{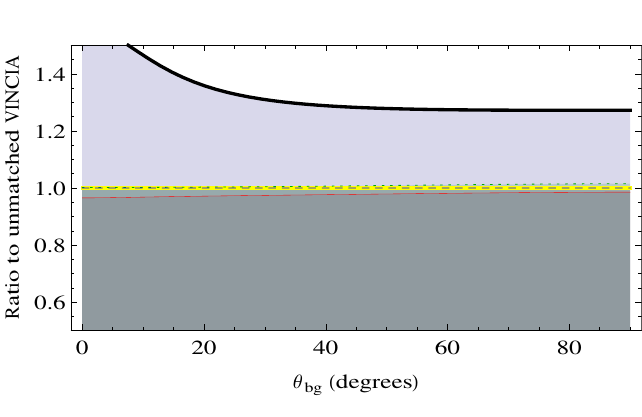}
\end{tabular}}
\scalebox{1.025}{
\begin{tabular}{c}
\includegraphics*[scale=0.6]{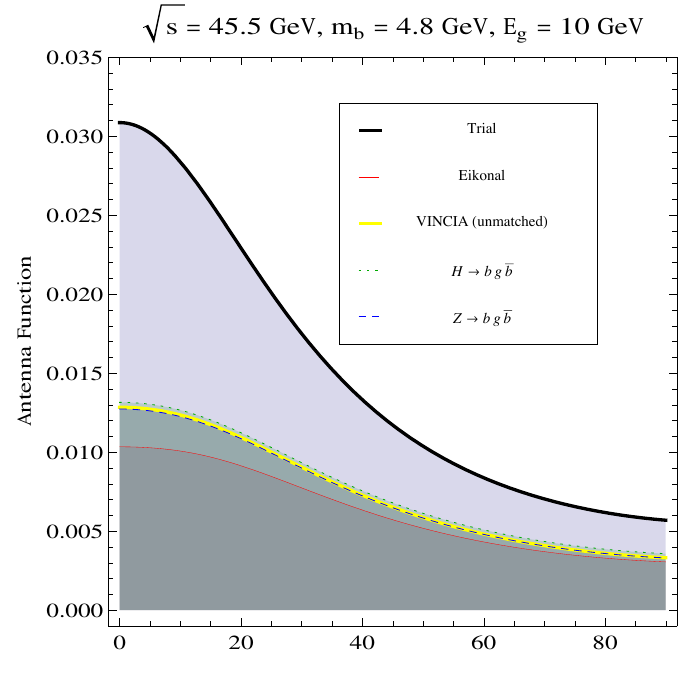}\\[-1.32cm]
\hspace*{1.55mm}\includegraphics*[scale=0.5983]{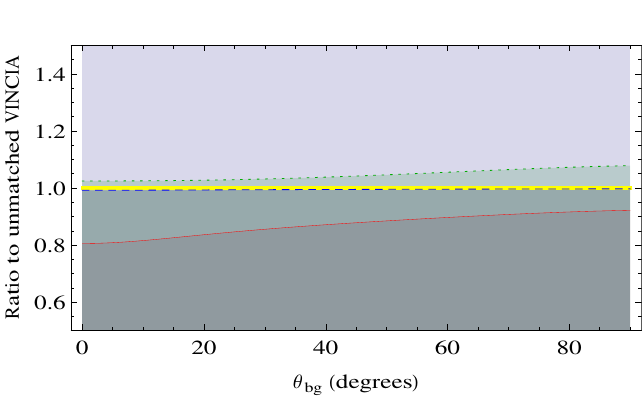}
\end{tabular}}
\caption{Various levels of approximation for $Q\bar{Q}\to
  Qg\bar{Q}$, compared to the \Vc trial and default (unmatched) 
shower functions. Left: the size of the dipole-antenna functions as a
function of emission angle for a gluon energy of $10\,$GeV, for $\sqrt{s}
= 91\,$GeV and $m_Q=4.8\,$GeV. Right: the same for
$\sqrt{s}=45.5\,$GeV. Lower panes show results normalized to the
default \Vc dipole-antenna function.
\label{fig:trial}}
\end{figure}

In \figRef{fig:trial}, we attempt to 
give a more concrete impression of the suitability of
this trial function, for a $b\bar{b} \to bg\bar{b}$ branching in a
dipole-antenna of mass $\sqrt{s}=91\,$GeV (left pane) and $=45.5\,$GeV 
(right pane), with 
$m_b=4.8\,$GeV in both cases and a gluon energy of
$E_g=10\,$GeV. These values were selected so as to yield plots that
can be compared directly to those in
\cite{Norrbin:2000uu}. The
$x$ axes show the gluon emission angle in degrees, going from the collinear
limit (zero angle) at origin to a 90-degree emission angle on the
right-hand edge of the plots. The value
of the gluon-emission trial function defined in 
\eqRef{eq:trial} is shown as a thick solid black line. It is everywhere
larger than all the other curves  and hence
represents an overestimate, as desired. 
The Eikonal
factor defined in \eqRef{eq:softeikonal} is shown as a thin solid line
(red). The default \Vc $q\bar{q}$ antenna function defined in
\eqRef{eq:a03def} is shown as a thick lighter (yellow) curve; it is
slightly larger than the Eikonal factor, but is still everywhere  
smaller than the trial function. For completeness, antenna functions
derived from two different LO matrix elements are also shown (for $H^0$
and $Z^0$ decay, shown with dots and
dashes, respectively). 
Note that the matrix-element curves are closer to the default
\Vc\ antenna function than to the Eikonal factor, in particular in the 
zero-degree region. This is due to
the fact that the default \Vc\ antenna function not 
only reproduces the soft limit, but also the quasi-collinear limit 
of the full matrix elements. 

Note also that the trial function is closest to the physical antenna functions
  in the left-hand pane, where the values of the antenna functions are
  ten times larger than in the right-hand one (notice the factor 10
  difference in the $y$ axis scales). The overall efficiency of
  the trial algorithm, which is dominated by the regions in which the trial
  function is large, is hence quite reasonable. 

The corresponding evolution integral for trial gluon emission, \eqRef{eq:dipInt}, was defined for massless
partons in \cite{Giele:2011cb},  in terms of the variables $Q_E$ and $\zeta$,  
\begin{equation}
\label{eq:trialint}
\aint_\text{trial-emit}(\qstart, \qemit ) 
	= C_A \int_{\qemit^2}^{\qstart^2} \d{Q_E^2} \d{\zeta} 
		\abs{ J } \frac{ \alphaStrial( Q_E^2 ) }{4 \pi} \frac{ 2 }{ \sij( Q_E^2, \zeta) \sjk( Q_E^2, \zeta ) }~.
\end{equation}
We have suppressed a trivial integration over $\phi$ and 
changed variables from $\sij$ and $\sjk$ to $Q_E^2$ and an
arbitrary (linearly independent) phase-space variable $\zeta$, with
$\abs{J}$ the Jacobian of the 
transformation. For the two
$Q_E$ definitions discussed in \secRef{sec:ordering}, a convenient
definition of $\zeta$ is  
\begin{equation}
\zeta = \frac{\sij}{\sij + \sjk}~.
\end{equation}
For both of the evolution variables considered above, the inverse
relations can be written 
\begin{equation}
	s_{ij} = \zeta \tilde{Q}^2  , \quad
	s_{jk} = (1-\zeta) \tilde{Q}^2~,
\end{equation}
with the definition of $\tilde{Q}$ depending on the choice of
evolution variable. For $Q_E = 2p_\perp$, 
\begin{equation}
\tilde{Q}^2 = \frac{Q_E \ \mant}{2\sqrt{\zeta (1-\zeta)} }~,
\end{equation}
while for $Q_E=m_D$,
\begin{equation}
\tilde{Q}^2 = \frac{Q_E^2}{\min( \zeta, 1-\zeta)}~.
\end{equation}
Using these relations, a set of $(Q_E,\zeta)$ values can be translated
unambiguously back to the original phase-space invariants
$(s_{ij},s_{jk})$. We therefore emphasize that, although the $\zeta$ variable  
plays a role analogous to the $z$ fraction of traditional parton
showers, it here serves merely to (re)parametrize phase space;
there is no explicit dependence on this choice \cite{Giele:2011cb}.  

Since the massive phase space is contained within the massless one,
the massless phase space generator developed in \cite{Giele:2011cb}
can be recycled for massive momenta. In this case, the points 
which do not correspond to physical massive momenta are rejected. 
We do have to take into account, however, that the two-particle phase
space $\rm d\Phi_2(p_I,p_K)$, which normalizes the  
the dipole-antenna branching phase space, depends on the masses of the parent 
partons $I$ and $K$ and is not anymore just given by the invariant
mass of the parent partons as in the massless case.
Instead,  we have
\begin{equation}
\label{eq:phasespacevolume}
\sqrt{ \lambda( 1, \mu_I^2, \mu_K^2) }
\frac{ \dPS{3} (p_i, p_j, p_k) }{ \dPS{2}( p_I, p_K ) }= 
	\left. \frac{ \dPS{3} (p_i, p_j, p_k) }{ \dPS{2}( p_I, p_K )} \right|_{m_I = m_K = 0}
\end{equation}
where the K\"all\'en function $\lambda$ is defined in
\eqRef{eq:Kaellen} 
and the reduced masses are defined as $\mu= m/\mant$, 
with $m_{IK}^2=(p_I+p_K)^2$. 
In the massive case, the relevant evolution integral can therefore be
written as the massless evolution integral 
modified by a massive phase space factor~,
\begin{equation}
\aint_\text{trial-emit} \to \frac{ \left. \aint_\text{trial-emit} (\mant^2, \qstart, \qemit) \right|_{m_I = m_K = 0}}
	{ \sqrt{ \lambda (1, \mu_I^2, \mu_K^2 ) } }~.
\end{equation}
We take this into account by applying the K\"all\'en factor as a
multiplicative pre-factor on the trial emission probabilities. With
this replacement, the treatment of trial gluon emission derived in
\cite{Giele:2011cb} in the massless case, can be carried over to the
massive one without further modifications. 

\subsection{Trial Gluon Splittings \label{sec:trial-splittings}}

For gluon-splitting antennae, i.e.\ those 
containing a $g\to q\bar{q}$ branching, we again make use of the
leading singularity structure of the underlying process. This improves
on the treatment in \cite{Giele:2011cb}, in which \eqRef{eq:trial} was used
for both gluon emission and gluon splitting. 
Irrespective of whether a gluon splits into a massive or into a massless 
quark-antiquark pair, the process is characterised by an $s$-channel gluon 
propagator with a singularity structure given by the factor 
$1/m_{q \bar{q}}^2$. The effect of
non-zero quark masses is only to restrict the emission phase
space in such a way that  the $m_{q\bar{q}} \to 0$ singularity present in 
the massless case cannot be reached with on-shell quarks.
This suggests that the ``optimal'' trial function to use for
gluon splitting processes (massless and massive) is:
\begin{equation}
\label{eq:trialsplit}
	\ant_\text{trial-split} = 
\frac{\alphaStrial}{4 \pi} n_f \TRhat\frac{1}{m^2_{q\bar{q}}}~,
\end{equation}
where $n_f$ denotes the number of active flavours and we recall that
$\TRhat = 1$ in our conventions. 

While different choices for the evolution variable for gluon emissions 
are implemented in \Vc, the evolution variable for gluon splitting has been 
fixed to gluon virtuality, $m_{q\bar{q}}^2$, as also advocated in \cite{Seymour:1994ca}. 
This is
based on the fact that,  while gluon emission involves a sum
over terms that have different soft and collinear limits, with
different evolution variables assigning different ``times'' to each
region, here there is only one singular structure, in
$m_{q\bar{q}}^2$, and hence we consider this choice to be relatively
unambiguous. 
Further motivation is provided by a comparison between mass- and 
\pT-ordering given in \secRef{sec:masslesssplittings}.

What remains ambiguous is then the details of how to ``interleave''
gluon emissions and gluon splittings in the parton shower evolution. 
The \Vc algorithm sketched in this section is based on generating a
branching scale for every possible $2 \to 3$ branching 
and choosing the highest of those scales to determine which branching
occurs next. We therefore need a way to determine whether a possible gluon 
splitting is at a higher scale than a possible gluon emission.
Once we have defined how we compare a gluon splitting scale with a gluon 
emission scale, the interleaving of gluon emission and gluon splitting 
works exactly the same way as the interleaving of gluon emissions
from different dipole-antennae.
In order to compare the scales, we use the fact that all gluon
emission 
evolution variables as well as the evolution variable for gluon
splitting, $m_{q \qbar}$, 
range from $0$ for strictly soft/collinear branchings to $\mant$ 
for the hardest branching process kinematically allowed.
This suggests that we can compare $m_{q \bar{q}}$ directly 
with the gluon emission variable, for example $2 \pT$.
There is an ambiguity however in how we define this comparison. 
One could for example 
equally well compare $ \mant (m_{q \bar{q}}/\mant)^a$ ($a>0$) with the gluon 
emission variable. In VINCIA, we have so far chosen to maintain the nominal
evolution variable, with $a=1$.

Let us here consider the case where parton $K$ is the splitting gluon (as in
$qg\to q\bar{q}'q'$ or $gg\to g\bar{q}q$), parton
$I$ can be either massless or massive. 
The direct equivalent of \eqRef{eq:trialint}, the integrated trial 
antenna function for gluon splittings, is then 
\begin{equation}\label{eq:dipIntSplit}
\aint_\text{trial-split}( \mant, \qstart, \qemit ) = 
 n_f \TRhat
 \frac{1}{ \mant^2 - m_I^2 }
 \int_{\qemit^2}^{\qstart^2} \d{s_{g\bar{q}}}\d{s_{q\bar{q}}}
\frac{\alphaStrial}{4\pi }\frac{1}{m^2_{q\bar{q}}}  
\end{equation}
where $\mant^2 - m_I^2$ originates from the two-particle 
phase space volume (we have inserted $m_K = m_g = 0$) and 
$1/m^2_{q\bar{q}} $ corresponds to the trial function for the
gluon splitting branching. Since we fix $Q_E^2=m^2_{q\bar{q}}$, 
the most convenient definition for $\zeta=\zeta_\text{split}$ is simply 
the other phase space invariant  in the $3$-particle
phase space ${\rm d}\Phi_3$, normalized by the mass of the mother 
dipole-antenna, 
\begin{equation}
\zeta_\text{split} = \frac{m^2_{g\bar{q}}}{\mant^2}~,
\end{equation}
with the massless phase-space boundaries
\begin{equation}
\zeta_{\mrm{min}}(y_E) = 0 ~~~~~~~,~~~~~~~~~ \zeta_{\mrm{max}}(y_E) = 1-\sqrt{y_{E}}~.
\end{equation}
with, $y_E=Q^2_{E}/m^2_{IK}=m^2_{q\bar{q}}/m^2_{IK}$ for our specific
choice of evolution variable.
We recall that the definition of $\zeta$ has no physical 
significance in our formalism (neither for gluon emission nor for the
gluon splitting process).

As in the procedure for trial gluon emissions described in detail in 
\cite{Giele:2011cb} and adapted for the massive case in \secRef{sec:trials},
we replace the upper limit on $\zeta$ by an overestimate during trial generation,  
\begin{equation}
\hat{\zeta}_{\mrm{max}} = \left. \zeta_{\mrm{max}}( \qemin^2 ) \right|_{m_q = 0}~,
\end{equation}
such that the $\zeta$ integral in a given ``evolution window'' (with
lower boundary \qemin) becomes simply a constant 
\begin{equation}
I_\zeta (Q_E^2) = \int_0^{\hat{\zeta}_{\mrm{max}}} \d{\zeta} =
\hat{\zeta}_\mrm{max}~=~ 1-\sqrt{\frac{\qemin^2}{m_{IK}^2}}
\end{equation} 
The evolution integral, \eqRef{eq:dipIntSplit}, then acquires the form
\begin{equation}
\label{eq:trialsplitting}
\aint_\text{trial-split}( \mant, \qstart, \qemit )
 = n_f \TRhat \frac{1}{1 - \mu_I^2 } 
 	\hat{\zeta}_\mrm{max} \int_{\qstart^2}^{ \qemit^2 } \frac{\d{Q_{E}^2}}{Q_{E}^2}
\frac{\alphaStrial}{4\pi} ~
\end{equation}
where $\mu_I=\frac{m_I}{m_{IK}}$. 
If $\hat{\alpha}_s$ does not depend on $Q_{E}$, then the 
integrated trial function in \eqRef{eq:trialsplitting} simplifies to
\begin{equation}
\label{eq:trialsplitconstant}
\aint_\text{trial-split}( \mant, \qstart, \qemit ) = 
	n_f \TRhat \frac{1}{ 1 - \mu_I^2 } 
	\frac{\alphaStrial}{4\pi} 
\hat{\zeta}_\mrm{max} \ln \left( \frac{\qemit^2 }{ \qstart^2 }\right) ~
\end{equation}
while if we consider a first-order running $\alpha_s$ 
as a function of  $\mu_R^2 = k_\mu Q^2_E = k_\mu m^2_{q\bar{q}}$,
then the integral becomes
\begin{eqnarray}
\aint_\text{trial-split}( \mant, \qstart, \qemit )
& = & \frac{n_f \TRhat}{4\pi} \frac{1}{ 1 - \mu_I^2 }  \hat{\zeta}_{\mrm{max}}
\frac{1}{ b_0 }
	\ln \left( \frac{ \ln \left( k_\mu^2 \qstart^2/\Lambda^2 \right) }{ \ln \left( k_\mu^2 \qemit^2/ \Lambda^2 \right) } \right)~
\label{eq:IArun}
\end{eqnarray}
where the $[\ln(\ln())]$ structure seen
in \eqRef{eq:IArun}  reflects the single logarithms
generated by the antenna-function singularities folded with the
logarithm coming from the running of $\alpha_s$. 

In our treatment of flavour thresholds, a heavy flavour is treated as active, 
i.e.\ it contributes to the running of $\alpha_S$ and is allowed to be created 
in gluon splittings as long as we have
$m_{q \qbar} \geq m_{Q}$ in gluon splittings and as long as
$Q_E \geq m_{Q}$ -- which is $2 \pT \geq m_Q$ in the default settings -- in gluon emissions.
The threshold $m_{q \qbar} = m_{Q}$ chosen here instead of the
kinematical threshold $m_{q \qbar} = 2 m_Q$ 
is a consequence of our interleaving of gluon emissions and gluon splittings discussed above 
combined with the gluon emission threshold choice made in \cite{Giele:2011cb}.
Of course the kinematical conditions for the secondary production 
of the heavy flavour are always enforced. Therefore the only consequence 
of the fact that our flavour threshold is below the kinematical
threshold is a slight loss of efficiency of the algorithm
for $m_Q \leq m_{q \qbar} \leq 2 m_Q$ due to vetoed gluon splittings 
into the kinematically disallowed heavy flavour region.
The alternative of adjusting the gluon emission flavour thresholds such that 
the gluon splitting thresholds are at $m_{q \qbar} = 2 m_Q$ -- which would be equally valid 
at leading logarithmic accuracy -- is not
implemented at present in \Vc.

With this definition of $\aint$ given in \eqRef{eq:trialsplitting}, 
the generation of trial branchings can be carried over from the formalism presented in
\cite{Giele:2011cb}. 

\subsection{Matrix-Element Corrections (Matching)}
\label{sec:matching}

The procedure for matching \Vc to leading-order matrix
elements \cite{Giele:2011cb} is not affected by the presence of mass
terms and can be adapted to the massive case by just upgrading the
massless matrix elements and antenna functions to massive ones. 
In accordance with the antenna-factorization formalism, all particles
are treated as being on shell, both in the antenna functions and in
the matrix elements.

Briefly summarized, the strategy is as follows. Similarly to the
\Py~\cite{Bengtsson:1986hr} and \texttt{GENEVA}~\cite{Bauer:2008qj}
approaches, the \Vc matching formalism 
relies on the antenna shower itself to provide an all-orders phase-space
generator that  captures the leading
behaviour of full QCD by construction. At each trial branching in the shower, the
accept/reject probability can then be augmented by a multiplicative factor
that goes to unity in the collinear and soft limit, but which 
modifies the branching probability outside those limits. The
modification factor is constructed precisely such that the 
full-colour leading-order matrix element squared is obtained
after summing over shower histories. The approach relies heavily on
 unitarity and is 
qualitatively different from other multi-leg approaches in the literature, 
such as the MLM~(see \cite{Alwall:2007fs} for a description) and
CKKW~\cite{Catani:2001cc} ones. An important technical
difference is that \Vc only requires a Born-level phase-space
generator, with all higher multiplicities being generated by
the shower. There is therefore no need for separate phase-space
generators for the higher-multiplicity matrix elements, which 
can result in significant speed gains, both in terms of initialization
time (virtually zero in \Vc), and in terms of running speed. The reader is referred
to \cite{Giele:2011cb} for further details.

\subsection{Uncertainty Estimates \label{sec:uncertainties}}
Another crucial point concerns how to estimate reliably the accuracy
of the resulting calculation. Arguing that variations only of 
the renormalization scale is insufficient at best (and misleading at
worst), a more comprehensive approach for all-orders (matched-shower)
perturbative calculations was proposed in \cite{Giele:2011cb} and
implemented in VINCIA. As with the prescription for matrix-element
matching, this approach can again be adapted to the massive case
straightforwardly.  

Briefly summarized, \Vc is able to compute a number of weights
corresponding to alternative shower settings along with each
event. The central weight, corresponding to the current user settings,
is unity, while each of the alternative weights represents the
relative probability that the event would have been produced when
running with the corresponding alternative  
setting. The uncertainties are evaluated in a  way that explicitly
preserves unitarity, and hence the weights for a particular 
alternative setting  average to $1$ over a  large number of events.

The uncertainties accounted for in the present version of \Vc do
not differ from those presented for the original formulation
\cite{Giele:2011cb}. Those include, for each branching: 
variation of the renormalization scale by a factor of 2 in either
direction, variation of the non-singular terms in the antenna
functions from a ``MIN'' setting to a ``MAX'' setting (see
\secRef{sec:massiveantennae}), variation of
the shower evolution variable between $p_\perp$-like and 
mass-like choices, and variations proportional to $1/N_C^2$. In the
context of matrix-element corrections (see above), variations of the numerical
value of any ``matching scale'' applied can also be included. See
\cite{Giele:2007di,Giele:2011cb} for further details.

\section{Comparison to Fixed-Order Matrix Elements \label{sec:fixed-order}}
\label{sec:comparison}

By construction, the massive dipole-antenna shower formalism presented in the
preceding sections reproduces the (quasi-)collinear
and soft limits of the amplitude squared for a single shower branching.
In this section we present some examinations of 
its behaviour across multiple (combinations of) gluon emissions and/or
gluon splittings.
Specifically, we compare tree-level expansions of the 
shower to fixed-order matrix elements for $Z\to4,\,5$ and $6$ partons,
treating each (leading-)colour structure
separately\footnote{Subleading-colour properties were studied for the massless case in \cite{Giele:2011cb} and are not repeated here.}.

We consider three possible evolution orderings: no ordering, strong
ordering in transverse momentum and smooth
ordering in transverse momentum, as defined in
\secRef{sec:ordering}. 
 A  comparison with other orderings can be found in
\cite{Giele:2011cb} for massless partons.
For gluon splitting, we also consider the
difference between ordering in transverse momentum and ordering in
gluon virtuality. 
The dipole-antenna functions are the default ones given in
\secRef{sec:massiveantennae}, with the phase space mapping defined by 
\eqRef{eq:massivemapping}. The matrix elements are obtained from
\Mg~4.4.26~\cite{Alwall:2007st}. 

For each parton multiplicity, we make a flat (uniform) scan over 
the relevant $n$-parton phase space using an implementation of
the RAMBO algorithm \cite{Kleiss:1985gy} provided in \Vc. In
each phase space point, the tree-level expansion of the 
shower weight, $w_\text{PS}$, is given by a sum over 
nested antenna functions, subjected to the selected ordering
criterion. E.g., for $Z\to qgg \bar{q}$, the tree-level expansion of
the shower weight is 
\begin{equation}
\begin{split}
w_\text{PS} = 
\bigg(&
	\bar{a}_{g/qg} \left( q, g_1, g_2 \right) 
		\bar{a}_{g/q\bar{q}} \left( \widetilde{q g_1}, \widetilde{g_1 g_2}, \qbar \right)
		\Theta \left( 
			Q_E \left( \widetilde{q g_1}, \widetilde{g_1 g_2}, \qbar \right) - 
			Q_E \left( q, g_1, g_2 \right)  
		\right)
	+
 \\ &
	\bar{a}_{g/qg} \left( \qbar, g_2, g_1 \right) 
		\bar{a}_{g/q\bar{q}} \left( q, \widetilde{g_1 g_2}, \widetilde{g_2 \qbar} \right)
		\Theta \left( 
			Q_E  \left( q, \widetilde{g_1 g_2}, \widetilde{g_2 \qbar} \right) - 
			Q_E  \left( \qbar, g_2, g_1 \right)  
		\right)
\bigg) |\me^{(0)}_2|^2
\end{split}
\end{equation}
where $Q_E$ denotes the evolution variable and tilded variables are
obtained by reclustering the final-state momenta to intermediate
3-parton states, using the inverse of the shower kinematics map
described in  \secRef{sec:phasespacemap}.  
The $\Theta$ functions express the strong-ordering condition for each
of the two possible clustering histories that lead from 2 to 4 partons
in the shower. For an unordered shower,
they would be absent (i.e., unity), whereas for a smoothly ordered shower, they
would be replaced by the $P_\mrm{imp}$ factor defined in
\secRef{sec:ordering}. For higher numbers of partons, more terms are
generated, for which we use an iterative code structure to compute the
relevant sums.  

The tree-level expansion of the shower weight  $w_\text{PS}$ can then
be compared to the norm squared of the
appropriate fixed-order colour-ordered sub-amplitude squared,
$|\me^{(0)}_n|^2$, forming the ratio
\begin{equation}
R_n = \frac{w_\text{PS}}{|\me^{(0)}_n|^2}~ \label{eq:Rn}~.
\end{equation}
This ratio thus represents an estimate of the relative accuracy of the
shower (or, rather, its tree-level expansion) phase-space point by
phase-space point. For simplicity, all couplings and colour factors
are set to unity in this comparison.

By studying how the distribution of $R_n$ evolves with $n$, we obtain
a useful indication of the accuracy of the shower, and how this
accuracy evolves with parton multiplicity. We consider these
comparisons to be fairly conservative,  since the soft- and
collinear-enhanced regions only occupy a relatively 
small corner of phase space in a flat scan. 
Similar comparisons to tree-level matrix
elements were carried out for massless showers
in \cite{Andersson:1991he,Skands:2009tb,Giele:2011cb}.
Here, we focus in particular on the modifications to 
these comparisons caused by non-vanishing masses. 

For comparison purposes,
we will plot the logarithm of such ratios,
$\log_{10}(R_n)$, for different shower approximations, final state
multiplicities, and parton masses. 
This logarithm gives a way of quantifying 
the amount of over or undercounting by
the shower approximation.
For phase space points for which the shower
approximation reproduces the matrix element exactly, this logarithm is
zero. For those points for which the logarithm is positive, the shower
overestimates the matrix elements while it underestimate the
matrix elements for negative values of this logarithm.

\subsection{$\mathbf{Z \to Q\bar{Q}}$ + gluons
\label{sec:comparisonemit}}

For massless quarks, the default $q\bar{q}\to qg\bar{q}$ dipole-antenna 
function, which describes the first $2 \to 3$ branching in the shower
evolution, coincides with the matrix element for $Z \to q g \qbar$,
and therefore the shower is ``automatically'' matched to the $Z \to 3$
matrix element.  
For massive quarks, however, the shower with default antennae differs slightly from 
the exact matrix element for $Z$ decay already at this order. This was
discussed in \secsRef{sec:massiveantennae} and
\ref{sec:trials}, with an illustration provided in \figRef{fig:trial}. 

\begin{figure}[tp]
\centering
\subfloat[no ordering]{
	\includegraphics[width = 0.95\textwidth]{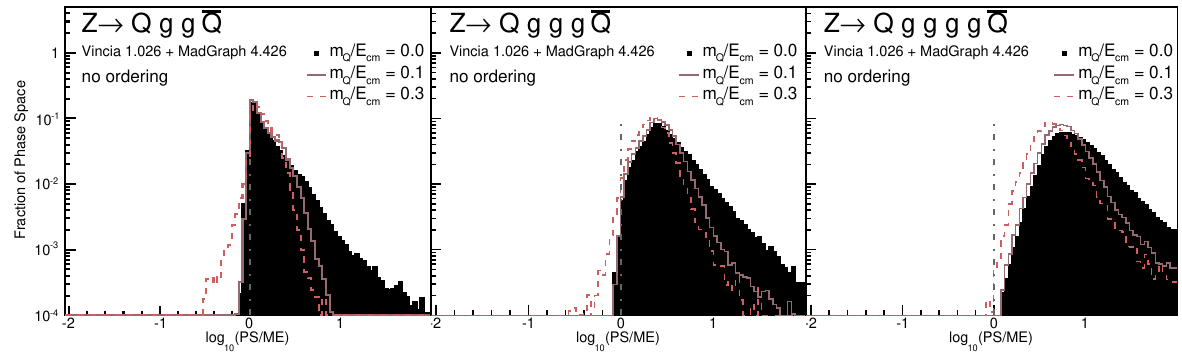}
}

\subfloat[strong ordering in $\pT{}$]{
	\includegraphics[width = 0.95\textwidth]{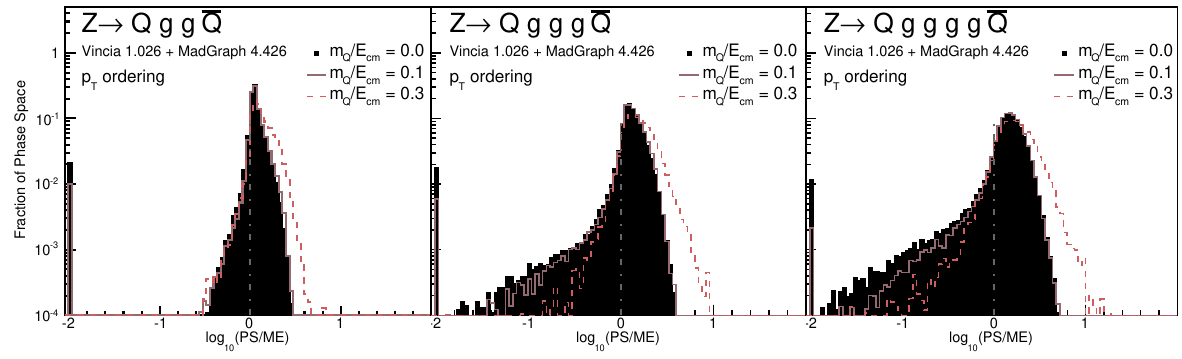}
}

\subfloat[smooth suppression of emissions not ordered in $\pT{}$ \label{fig:RpTsmooth}]{
	\includegraphics[width = 0.95\textwidth]{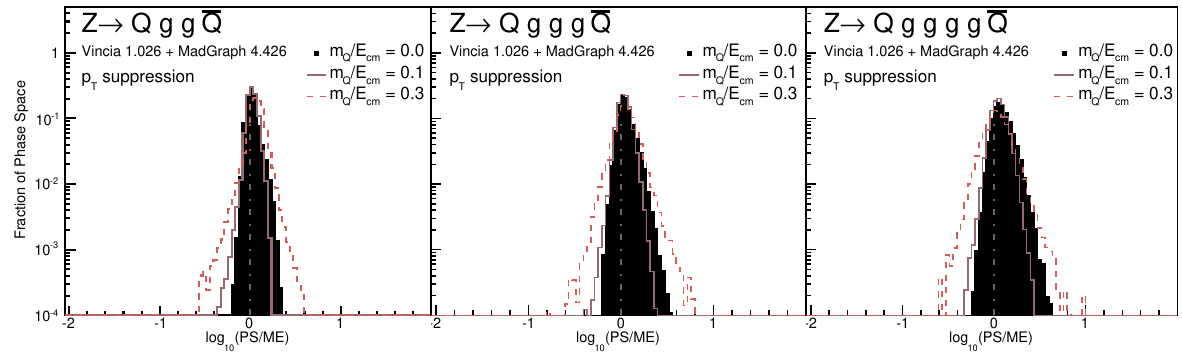}
}
\caption{Histograms of $\log_{10} (R_n)$, as defined in the text, in a
  flat phase space scan, for $n=4$ (left panes), $n=5$ (middle panes), 
and 
  $n=6$ (right panes), for not ordered (top row), strongly ordered
(middle row), and smoothly ordered (bottom row) shower
approximations. The parton shower uses the default
dipole-antenna set
defined in \secRef{sec:showerformalism}.}
\label{Fig:RpT}
\end{figure}

Turning to $Z \to 4$ partons and more, in figure \ref{Fig:RpT}, we
illustrate how quark masses affect the distribution of $R_n$,
for $Z\to q\bar{q}$ + 2~gluons (left), +~3~gluons (middle), and
+~4~gluons (right), for three different evolution criteria: 
no ordering (top), strong ordering in $p_\perp$
(middle), and smooth ordering in $p_\perp$ (bottom).

With no ordering (top row), the massless shower (solid black)
has a large tail to the right, i.e., it substantially
overcounts the matrix elements, as was also discussed in \cite{Giele:2011cb}. 
Towards the left of zero, it falls off extremely sharply (notice the
logarithmic $y$ axis), implying that the unordered shower
approximation is an almost strict overestimate of the matrix
elements. 
The introduction of masses changes this picture drastically, with a
moderate ratio $m_Q/E_\mrm{cm}=0.1$ shown as a thin solid histogram
and the larger $m_Q/E_\mrm{cm}=0.3$ shown with dashes: for both mass
values, 
a lot of the overcounting on the right of zero is removed, and for the
larger ratio the regions below zero, where the shower underestimates the
matrix-elements, is populated. 

In the second row of \figRef{Fig:RpT}, the introduction of
strong ordering in \pT improves systematically on the
non-ordered approximation and the introduction of quark masses
does not spoil this improvement.
Even for quite large quark masses,
the distributions remain almost centred around $\log_{\mrm{10}}(R) =
0$. However, as discussed in \secRef{sec:ordering}, the
price for strong ordering is the
introduction of a dead zone, which we illustrate by plotting the
underflow bin at $\log_{\mrm{10}}(R) =-2$.
Its size corresponds to a few percent of the phase space volume. 

In the last row, the change to a smooth ordering condition is
illustrated. The dead zone is removed and, at least in the massless
case, this smooth ordering condition further improves 
the agreement with the matrix elements relative
to the strong-ordering case. It eliminates the tail of large
undercounting that was present for $Z\to 5$ and 6 partons  in the
strong-ordering case and sharpens the peak around  $\log_{\mrm{10}}(R) =
0$ also for $Z\to 4$. This improvement is much 
less significant in the massive case, but since the dead zone is still
removed, we use the smooth ordering option as our default choice
for massive partons as well.

\begin{figure}[t]
\centering
\subfloat{\includegraphics[width = 0.45 \textwidth]{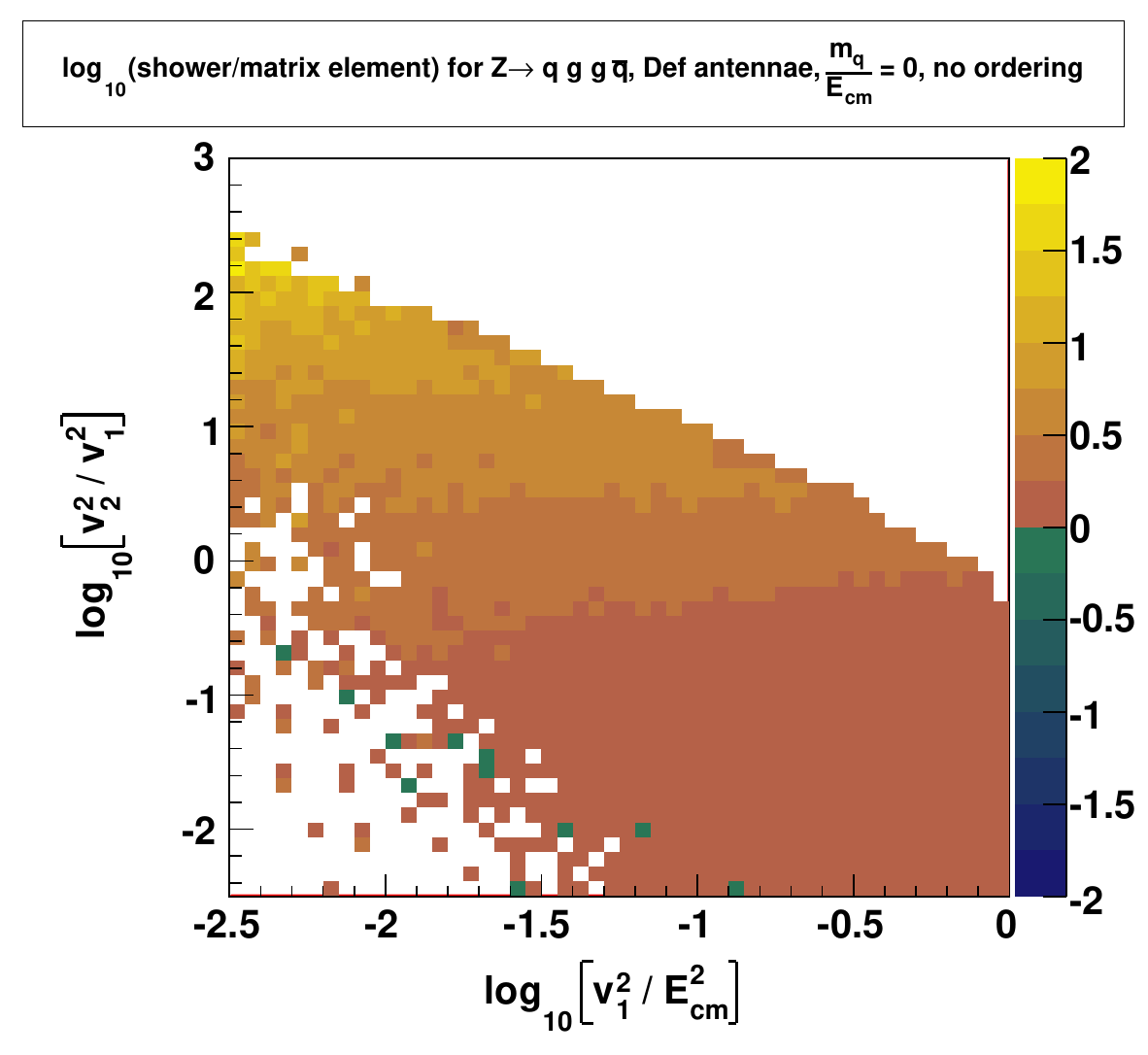}}
\subfloat{\includegraphics[width = 0.45 \textwidth]{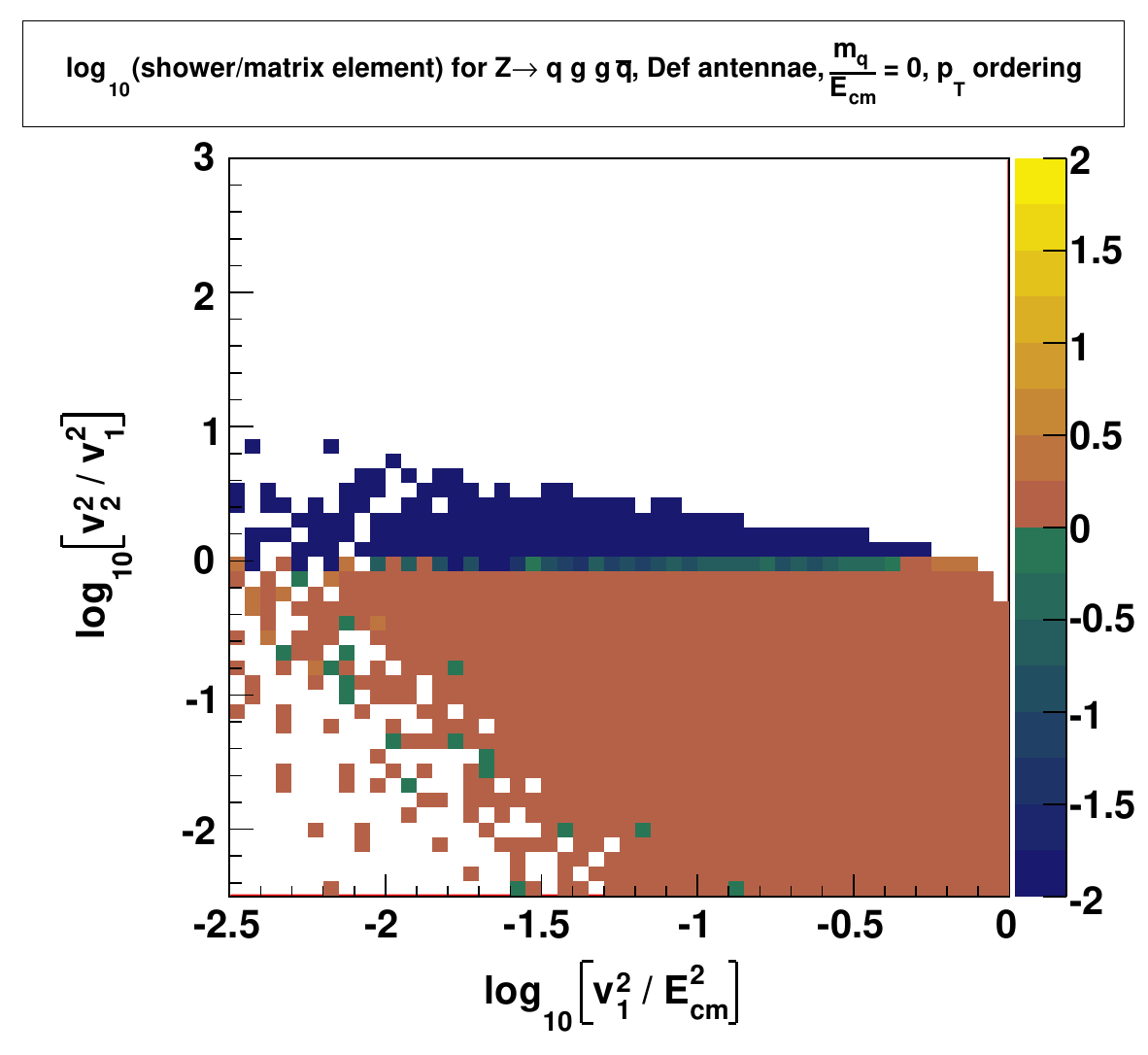}}
\caption{
$\log_{10} (R_4)$ for $Z \to q g g \qbar$, for $m_q=0$, 
as a function of $x=\log_\mrm{10}(p_{\perp,1}^2/E^2_{\text{cm}})$
and $y=\log_\mrm{10}(p_{\perp,2}^2 / p_{\perp,1}^2)$. Left pane: 
no ordering. Right pane: strong ordering in \pT.
\label{Fig:qg1}
}
\end{figure}
\begin{figure}[t]
\centering
\subfloat{\includegraphics[width = 0.45 \textwidth]{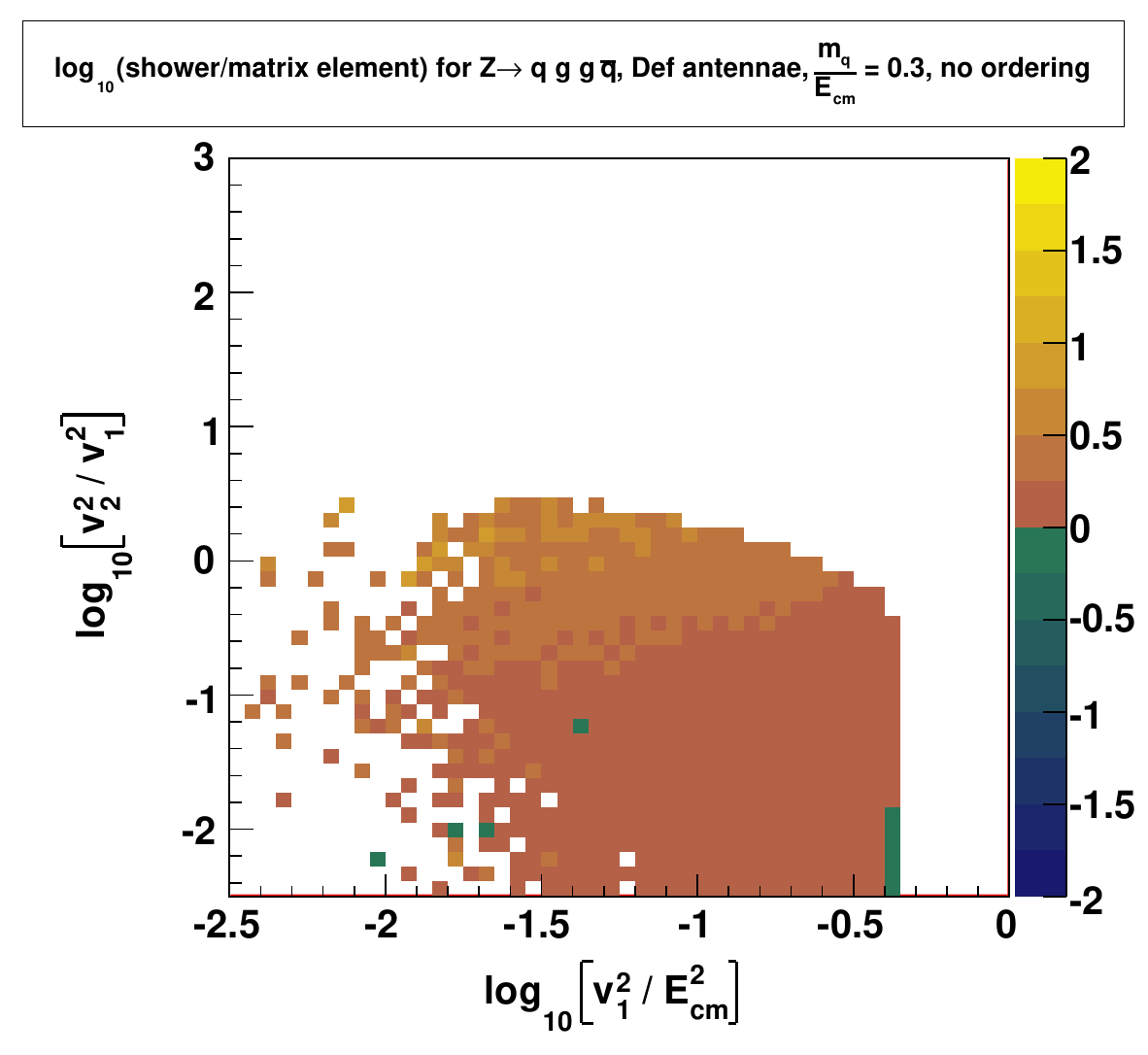}}
\subfloat{\includegraphics[width = 0.45 \textwidth]{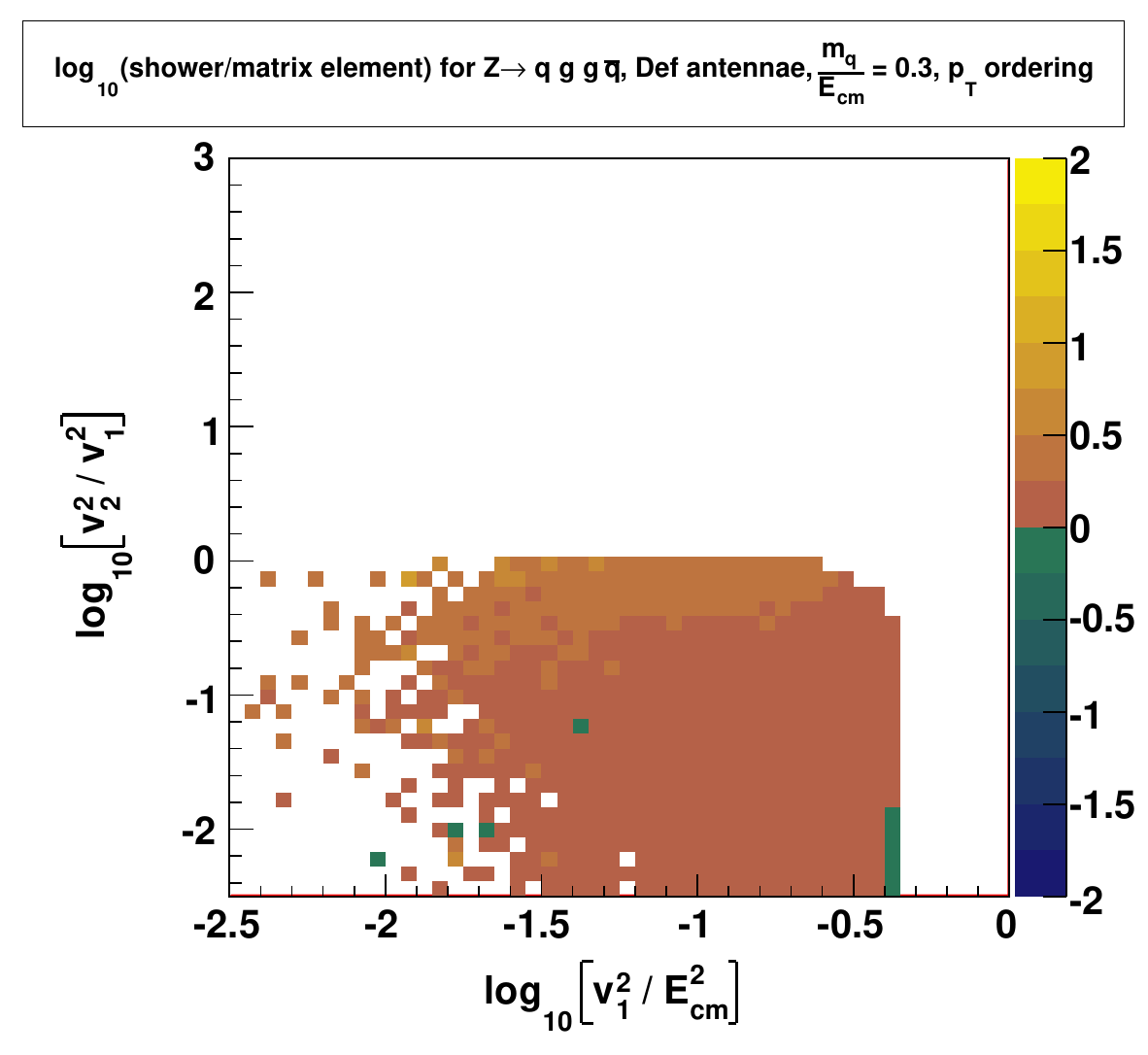}}
\caption{
$\log_{10} (R_4)$ for $Z \to q g g \qbar$, for $m_q/E_\mrm{cm}=0.3$, 
as a function of $x=\log_\mrm{10}(p_{\perp,1}^2/E^2_{\text{cm}})$
and $y=\log_\mrm{10}(p_{\perp,2}^2 / p_{\perp,1}^2)$. 
 Left pane: 
no ordering. Right pane: strong ordering in \pT.
\label{Fig:qg2}
}
\end{figure}

We note also that the effect of imposing strong ordering in $p_\perp$ 
is much more pronounced for massless quarks than for massive
ones. To see this, we compare for instance the change in the black (massless)
histogram between the top (unordered) and middle (strongly-ordered) 
left-hand panes of \figRef{Fig:RpT}. We investigate this further in
the 2D phase-space plots presented in 
\figsRef{Fig:qg1} and \ref{Fig:qg2}.

\FigRef{Fig:qg1} shows the case for massless quarks. In the left-hand
pane, no ordering condition is imposed; in the right-hand pane, strong
ordering in \pT. The axes of the figure have been chosen to be logarithmic 
in the two successive branching scales $p_{\perp, 1} / E_\mrm{cm} $ and 
$p_{\perp, 2} / p_{\perp, 1}$, respectively, with $p_{\perp, 1}$ the 
emission scale of the first branching and $p_{\perp, 2}$ the emission 
scale of the second branching. Among the two shower histories for  
$Z \to q g g \bar{q}$, we show the \pT values of 
the larger contribution on the plot. An average over the phase space 
points compatible with the corresponding values on the $x$- and $y$-axes 
is shown, obtained using the same flat scans of the phase 
space as for the one-dimensional phase space 
plots above.

The effect of the strong-ordering condition is clearly visible 
in the right-hand pane of \figRef{Fig:qg1}, removing the shower contributions in the 
upper half of the plot, corresponding to unordered branchings.

Comparing the left-hand pane in \figRef{Fig:qg1} (massless case, no
ordering) to that in \figRef{Fig:qg2} (massive case, with
$m/E_\mrm{cm}=0.3$, no ordering), 
one notices that the part of the phase space in which the 
overestimate of the shower is biggest in the massless case
corresponding to the region where 
the second emission is at a higher scale than the first one, is almost 
inaccessible in the massive case due to kinematic restrictions.

In accordance with this, the effect of imposing strong ordering in
$\pT{}$ is comparably small for
heavy quarks, as shown in the right-hand pane of 
\figRef{Fig:qg2}. We conclude that to impose an ordering condition is much more
important for massless quarks than it is for massive ones.

\subsection{Including massless $\mathbf{g\to \bar{q}'q'}$ splittings}
\label{sec:masslesssplittings}

The gluon-splitting $g\to q\bar{q}$ dipole-antenna functions 
($a_{\bar{q}'/qg}$ and $a_{\bar{q}/gg}$ in the notation adopted here) 
only contain single poles in the region where the secondary quark-antiquark 
pair becomes collinear.
Those antennae are therefore less singular than their gluon-emission 
counterparts, and hence there are intrinsically
fewer $g\to q\bar{q}$ splittings than gluon emissions occurring 
(independently of the difference $C_A$ vs $n_f \TRhat$ in colour/flavour
factors). However, for those gluon splittings that \emph{do} occur,  
the smaller relative size of the (universal) singular terms, on which
the shower approximations are based, 
as compared to possible non-singular (and non-universal) terms, 
imply that one can expect an overall worsening of the shower
approximation for processes involving $g\to
q\bar{q}$, as compared to ones involving only gluon emission. An immediate
consequence of this is, for instance, that the amount of strange and
heavier quarks produced in perturbative $g\to q\bar{q}$ splittings is associated
with substantial uncertainties in all current shower models. 

\begin{figure}[t]
\centering
\includegraphics[width = 0.95\textwidth]{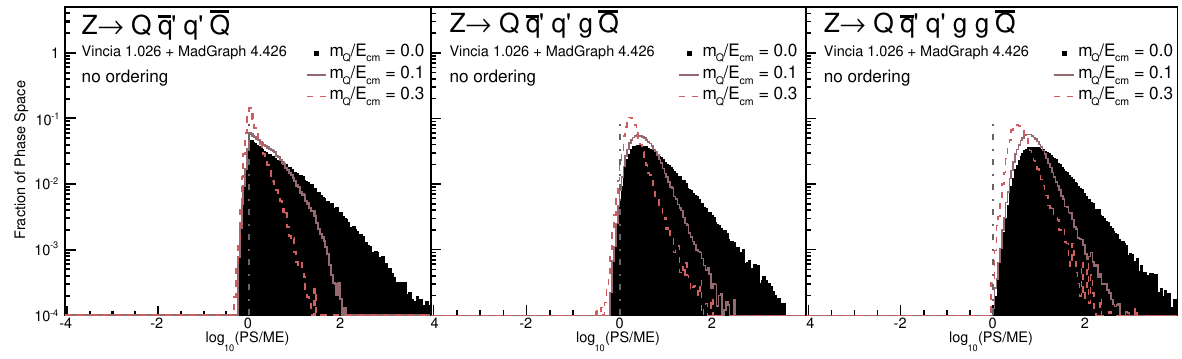}
\caption{
Histograms of $\log_{10} (R_n)$, as defined in the text, 
for $Z \to Q \qbar' q' (g g) \bar{Q}$ in a
  flat phase space scan, for $n=4$ (left pane), $n=5$ (middle pane), 
and 
  $n=6$ (right pane). No ordering 
condition is imposed.
}
\label{Fig:RQuarkPair}
\end{figure}

A first illustration of this feature is given by \figRef{Fig:RQuarkPair} 
which shows the ratio of the expanded parton shower (with no ordering
condition imposed) to the tree-level 
matrix element for $Z \rightarrow Q \bar{q}' q' \bar{Q}$. 
The $x$ axis now ranges from -4 to 4, rather than -2 to 2, allowing
for a much larger range of shower-to-matrix-element ratios. This
accommodates the most important feature  in figure
\ref{Fig:RQuarkPair}: the tail of high overestimates of the unordered 
shower approximation for a
massless primary quark-antiquark pair has $R$ values extending up  
to approximately $10^4$ as opposed to approximately $10^2$ 
for gluon emission, for the same $y$ range (i.e., same fraction of
flat phase space).  

\begin{figure}[t]
\centering
\includegraphics[width = 0.95\textwidth]{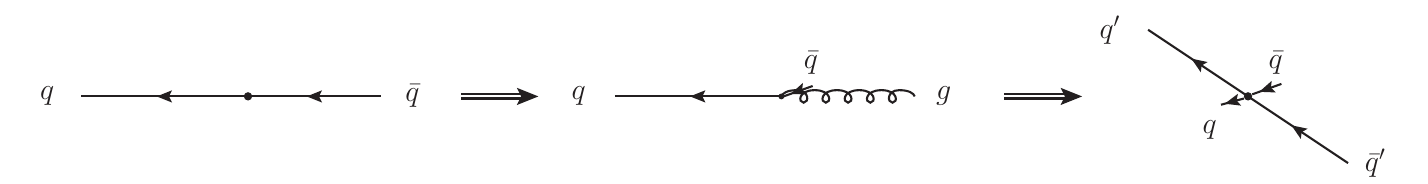}~.
\caption{A high-$z$ gluon emission followed by a hard $g\to q\bar{q}$
  splitting.
\label{Fig:history}}
\end{figure}
A phase-space scan similar to the ones shown in \figsRef{Fig:qg1} and
\ref{Fig:qg2}, revealed that most of the very high overestimates occur for 
configurations where the secondary quark-antiquark pair takes up almost all
of the energy, which in turn forces the primary quarks to be soft. 
One possible shower history leading to such a particular configuration
is obtained by having  
a collinear high-$z$ gluon emission followed by a very  hard  
$g\to q\bar{q}$ splitting, or represented pictorially illustrated in
\figRef{Fig:history}. 
Such occurrences are apparently all too frequent, in the unordered
shower. Physically, we interpret this as a screening effect which is
missing in the unordered approximation. By independently adding
the splitting probabilities in each of the $qg$ and $g\bar{q}$
antennae, we are not taking into account any screening effects
produced by the collective $qg\bar{q}$ system, which become
particularly relevant when two or more of those partons are collinear
with respect to each other and hence \emph{should} maximally screen
each other. 

As illustrated by the histograms for light primary quarks (thin solid line) 
and for heavy primary quarks (dashed) in \figRef{Fig:RQuarkPair}, the
introduction of non-zero masses for 
the primary quarks improves this situation, since 
the configurations where the shower overestimates 
are the largest in the massless case simply cannot be reached for
heavy primary quarks due to kinematical constraints.
In terms of coherence, the strong dampening of
the collinear singularity for massive emitters leads to an absence of
the subsequent very strong coherence dampening that is present in the
massless case. 

We now turn to ordered showers. 
By analogy with the case for gluon emission,  \figRef{Fig:RpT}, 
we expect that we can get rid of a significant part of the high 
shower overestimates for highly energetic secondary quark-antiquark pairs  
by imposing a strong ordering condition in the secondary 
quark-antiquark mass $m_{q \bar{q}}$ for gluon
splittings. An alternative choice that cannot a priori be excluded
would be to use \pT-ordering for gluon splitting as well.
\begin{figure}[t]
\centering
\subfloat[Evolution variable {\mqqbar} for gluon splitting.]
{\includegraphics[width = 0.45 \textwidth]{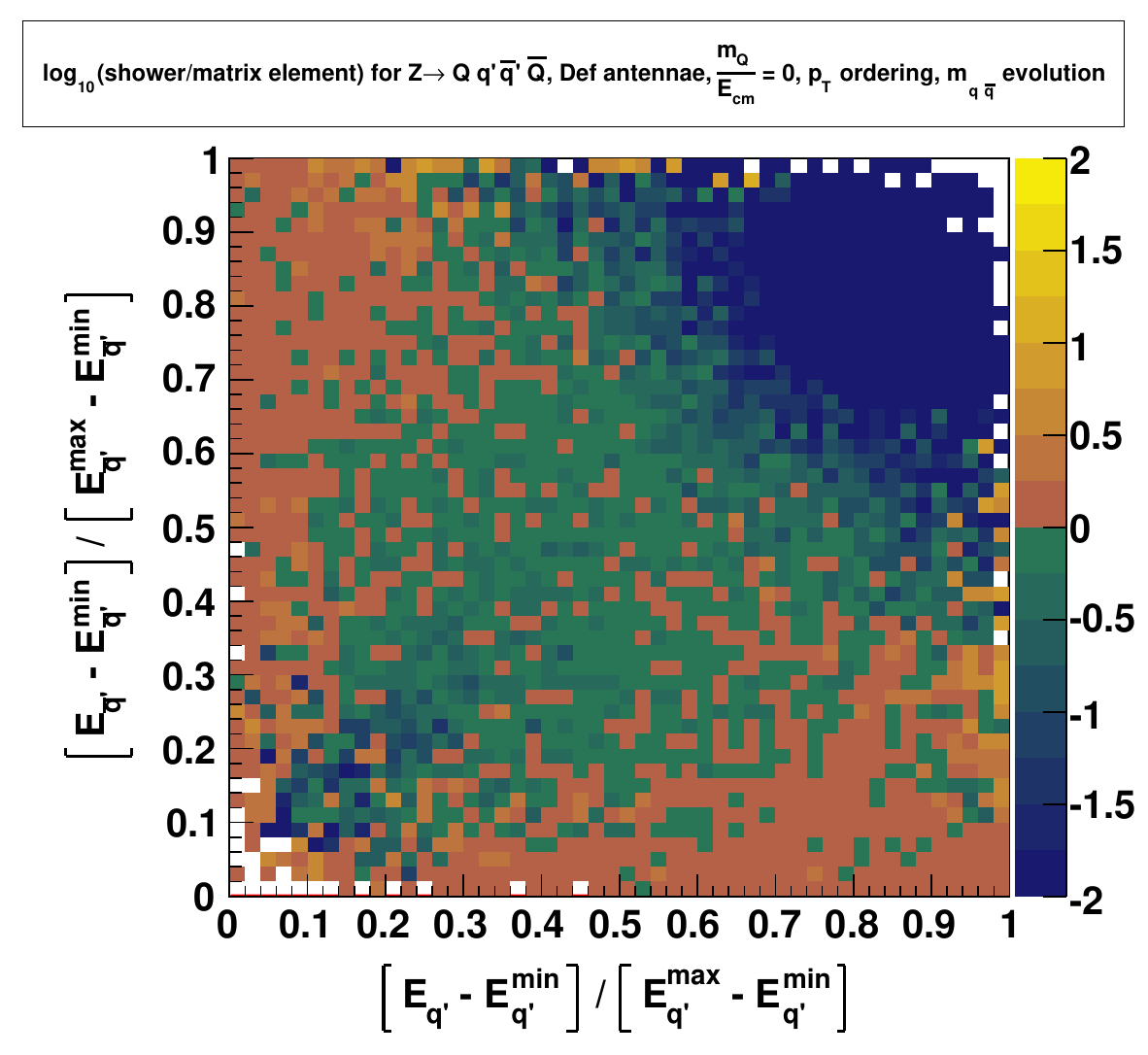}}
\subfloat[Evolution variable {\pT{}} for gluon splitting]
{\includegraphics[width = 0.45 \textwidth]{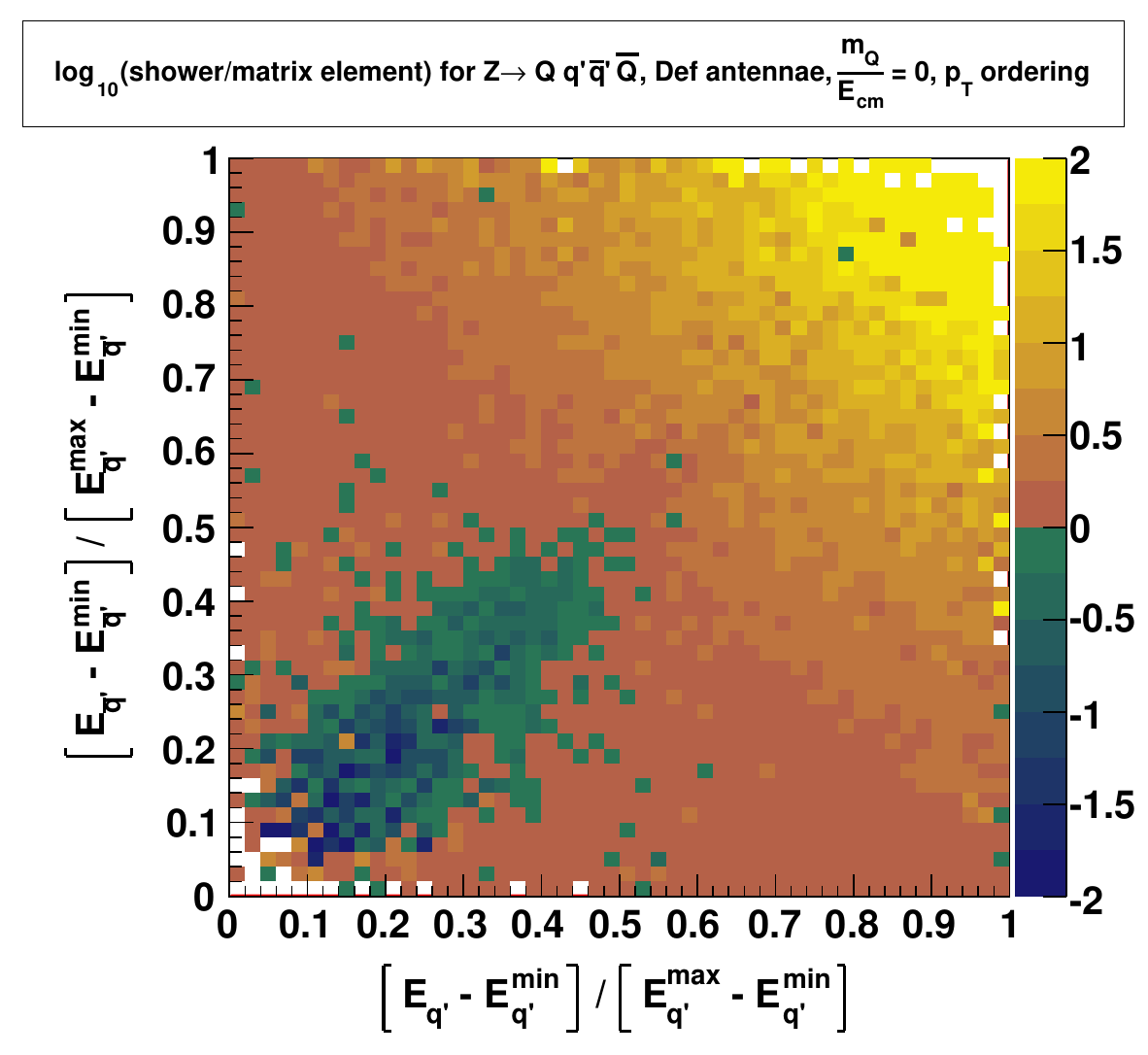}}
\caption{
$\log_{10} (w_\text{PS}/| \me |^2)$ for $Z \to q \qbar' q' \qbar$ as a function of the secondary 
quark energies, rescaled to range from $0$ to $1$.
Left: Evolution variable {\mqqbar} for gluon splitting, 
resulting in the strong ordering condition 
$m_{q'\bar{q}'} < 2 \, p_\perp^{q \qbar \to q g \qbar}$. Right:
 Evolution variable {\pT{}} for gluon splittings, resulting in the strong ordering
 condition $p_\perp^{q g \to q \qbar' q'} < p_\perp^{q \qbar \to q g \qbar}$.
 All quarks are massless.
 }
\label{Fig:RQuarkPairStrongOrdering}
\end{figure}
In \figRef{Fig:RQuarkPairStrongOrdering}, we make a first comparison
of these two
possibilities, for processes with
primary massless $q\bar{q}$ parent partons and 
involving gluon splittings in a second branching step. 
Either we use strong ordering in $Q_E=2\pT$ for gluon emission and
ordering in $Q_E=m_{q\bar{q}}$ for 
gluon splitting (shown in the left-hand pane of the figure), 
or we use strong ordering in $Q_E=2\pT$ for both branching processes 
(shown in the right-hand pane). Specifically, 
in the left-hand pane, the subsequent gluon splitting 
is vetoed if $m_{q'\bar{q}'} > 2 \pT(q, g ,\bar{q})$, while on the right-hand pane,
the gluon splitting is vetoed if  $\pT(q, \bar{q}', q') > \pT(q, g, \bar{q})$ or 
$\pT(\bar{q}, q', \bar{q}') > \pT(q, g, \bar{q})$ respectively.
A substantial over-counting for highly energetic secondary
quark-antiquark pairs remains in the \pT-ordered case (right-hand pane, top right
corner), while an
undercounting results when the evolution variable is changed to
$m_{q\bar{q}}$ in the second branching step (left).
We conclude that the strong ordering in $m_{q \qbar}$ does a better 
job of suppressing the high overestimates for the regions of the phase 
space where there is no leading-log contribution (top right of the
plots). 

\begin{figure}[t]
\centering
\subfloat[Evolution variable {\mqqbar} for gluon splitting.]
{\includegraphics[width = 0.45 \textwidth]{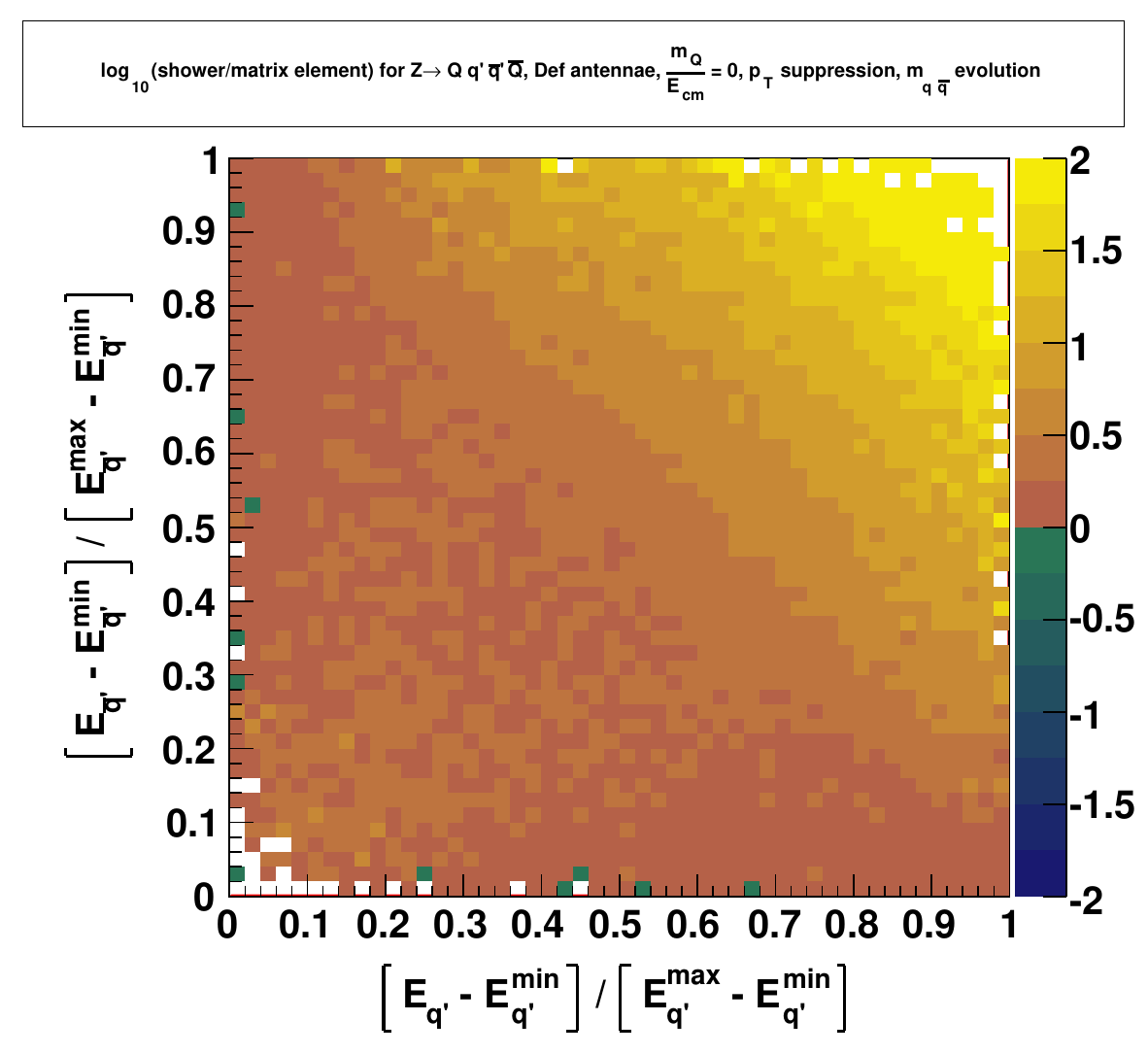}}
\subfloat[Evolution variable {\pT{}} for gluon splitting]
{\includegraphics[width = 0.45 \textwidth]{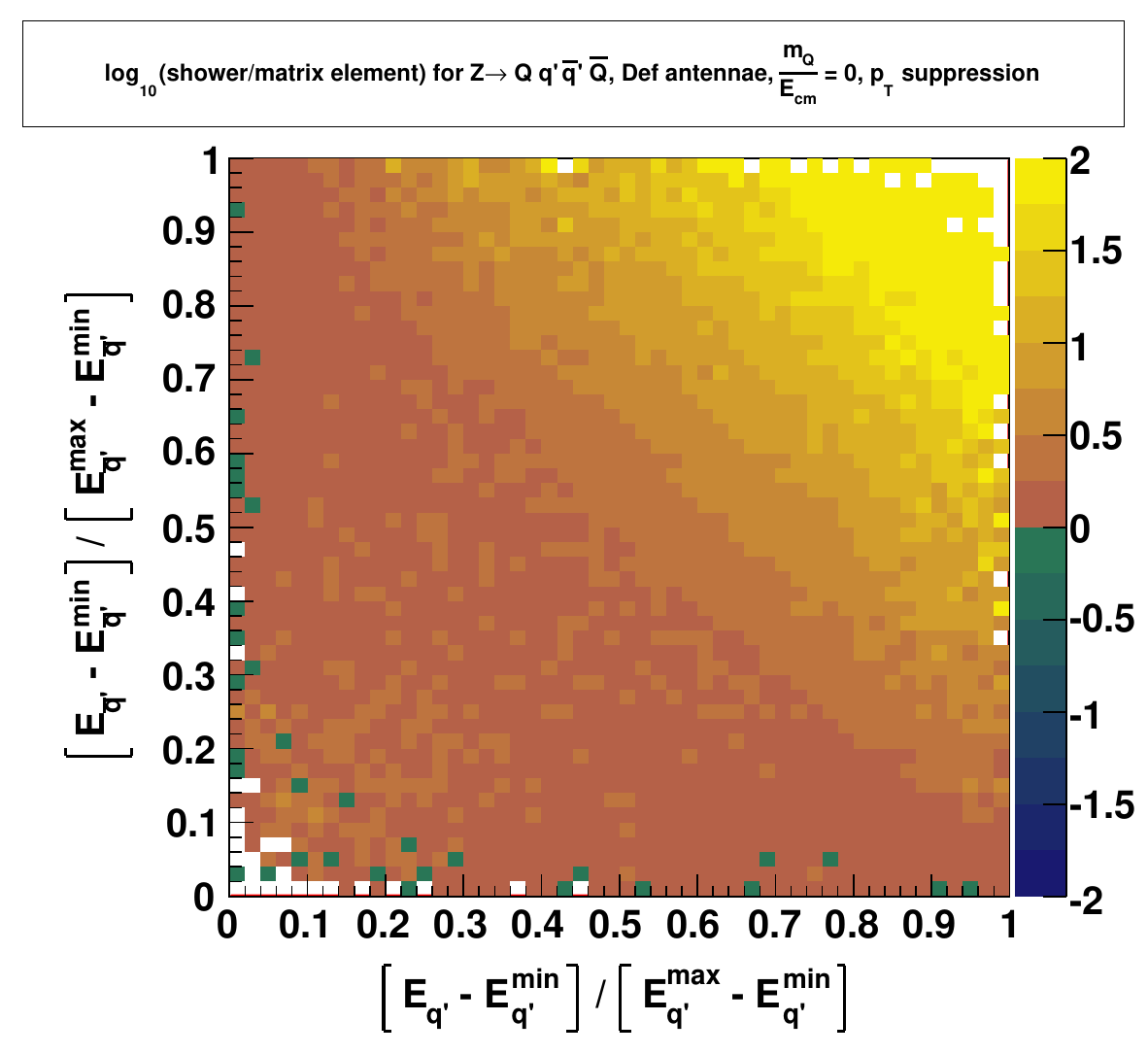}}
\caption{
$\log_{10} (w_\text{PS}/|\me|^2)$ for $Z \to q \qbar' q' \qbar$ as a function of the secondary 
quark energies, rescaled to range from $0$ to $1$. All quarks are massless, smooth ordering 
is imposed, \pT is used as the evolution variable for gluon emission.
Left: Use $m_{q \bar{q}}$ for gluon splitting. 
Right: Use \pT for gluon splitting. 
  }
\label{Fig:RQuarkPairUnOrdering}
\end{figure}

As in the case of gluon emission, we wish to avoid dead zones 
by switching to a smooth suppression of  gluon splittings,
using the suppression factor  $P_{\mrm{imp}}$, defined in \eqRef{eq:pimp}. 
However, in \figRef{Fig:RQuarkPairUnOrdering}, we illustrate that 
in this gluon splitting case,  a naive application of this
suppression factor  $P_{\mrm{imp}}$,  results in an overcounting 
for \emph{both} choices of gluon-splitting variables. This figure shows the same
distributions as in \figRef{Fig:RQuarkPairStrongOrdering}, but with the
strong-ordering condition replaced by a smooth one. 

In a dipole-antenna shower that employs ordering in \pT for all
branchings, or for our smooth-ordering shower variant, 
an additional suppression mechanism is therefore
needed to remove this overcounting and get reasonable agreement
between approximated shower vs matrix elements for processes involving
gluon splittings. 
The Lund dipole cascade implemented in the
\Ar\ program \cite{Lonnblad:1992tz} uses the following factor to modify its gluon splitting
antenna functions,
\begin{equation}
\pAri = \frac{2m^2_N}{\mant^2 + m^2_N}~,
\end{equation}
where $\mant^2$ is the invariant mass squared of the parent
antenna-dipole and $m^2_N$ is that of the
\emph{neighbouring} dipole-antenna. Thus, if the preceding branching
was collinear, with $m_N^2\to 0$, this factor
produces a very strong suppression, while if the two dipole-antennae
that share the splitting gluon have exactly equal sizes it goes to
unity.\footnote{If the neighbouring 
dipole-antenna is much larger than the parent antenna, it even produces a
slight enhancement, by up to a factor of 2, but this is more than
compensated for by the reduction in the splitting probability of the
neighbour itself.} 
\begin{figure}[t!p]
\centering
\subfloat[Evolution variable {\mqqbar} for gluon splitting.]{\includegraphics[width = 0.45 \textwidth]{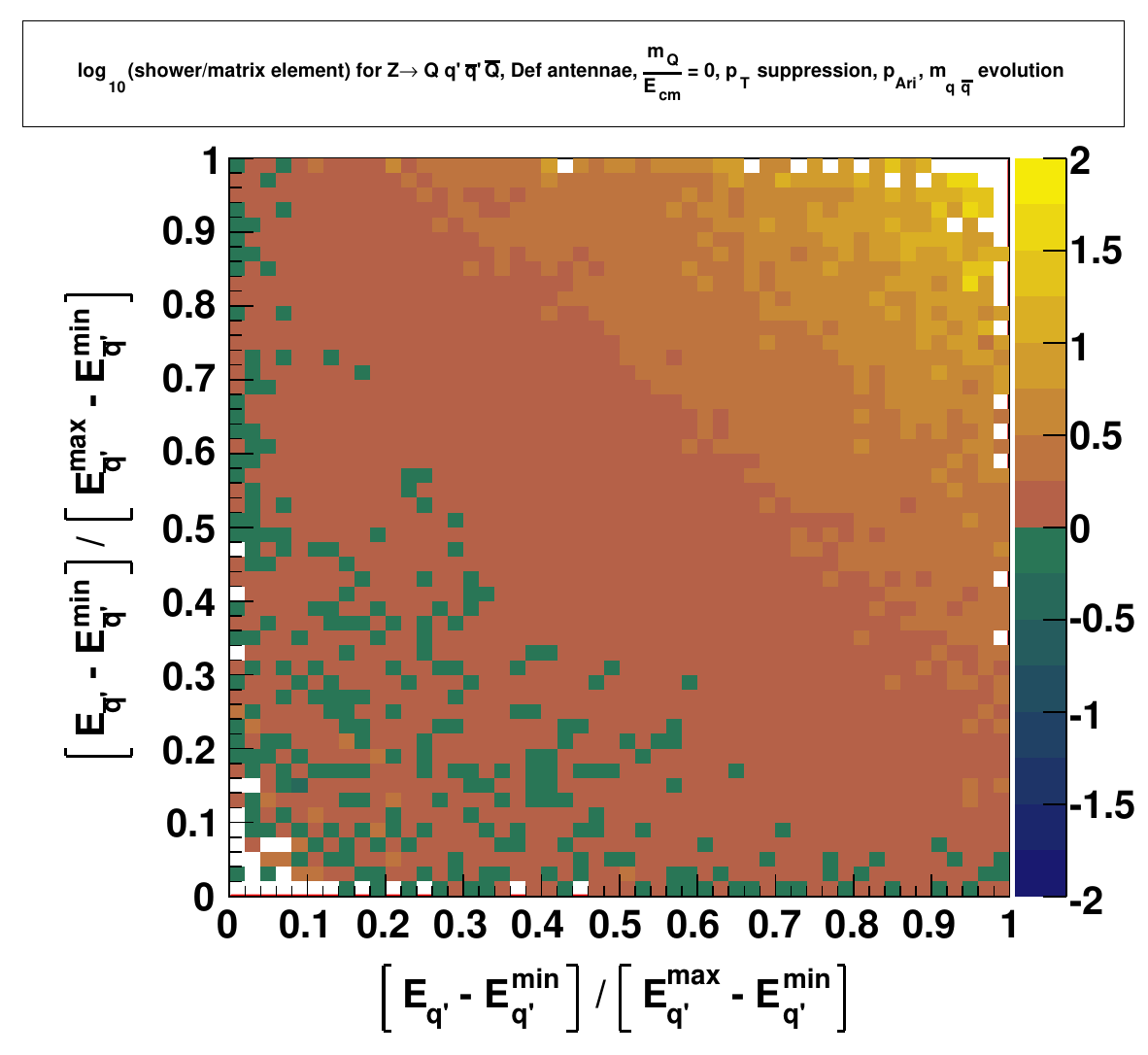}}
\subfloat[Evolution variable {\pT{}} for gluon splitting]{\includegraphics[width = 0.45 \textwidth]{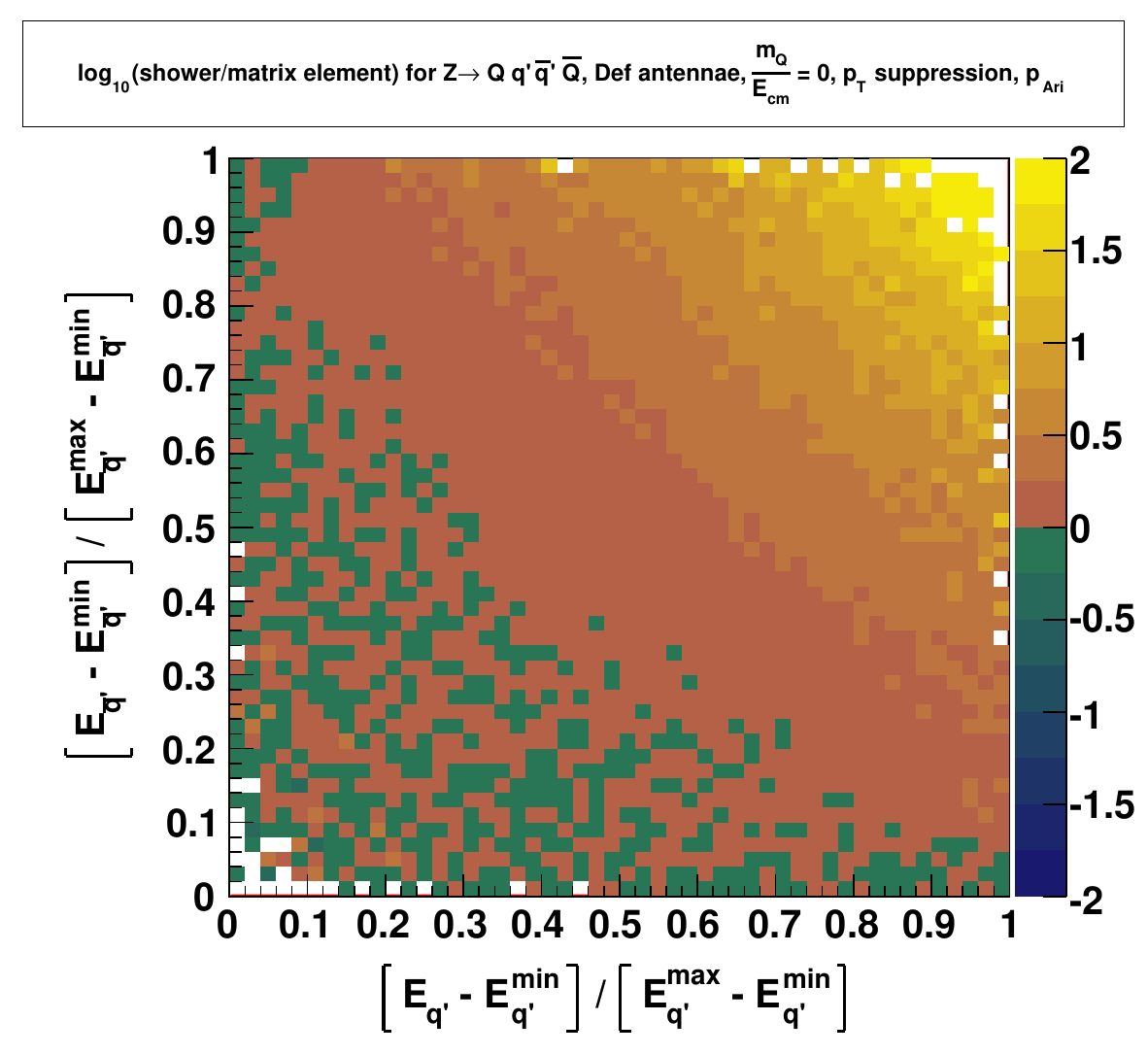}}
\caption{
$\log_{10} (R_4)$ for $Z \to q \qbar' q' \qbar$ as a function of the secondary 
quark energies, rescaled to range from $0$ to $1$. Smooth suppression of unordered
  emissions, with \pAri{} factor.
  }
\label{Fig:RQuarkPairUnOrderingpAri}
\end{figure}
In \figRef{Fig:RQuarkPairUnOrderingpAri}, we show that 
the use of the {\pAri} factor suppresses the overestimates visible in
\figRef{Fig:RQuarkPairUnOrdering} to a large degree, with a slightly
better agreement obtained in the left-hand pane (for interleaved
\pT and $m_{q\bar{q}}$ evolution) than in the right-hand one (with all
processes ordered in \pT).

\begin{figure}[t]
\centering
\subfloat[Using {\pT{}} for gluon emissions and $m_{q \bar{q}}$ for secondary $q \bar{q}$ production.]{
\label{Fig:RQuarkPairpTUnopAriQuarkPairMassEv}
\includegraphics[width = 0.95\textwidth]{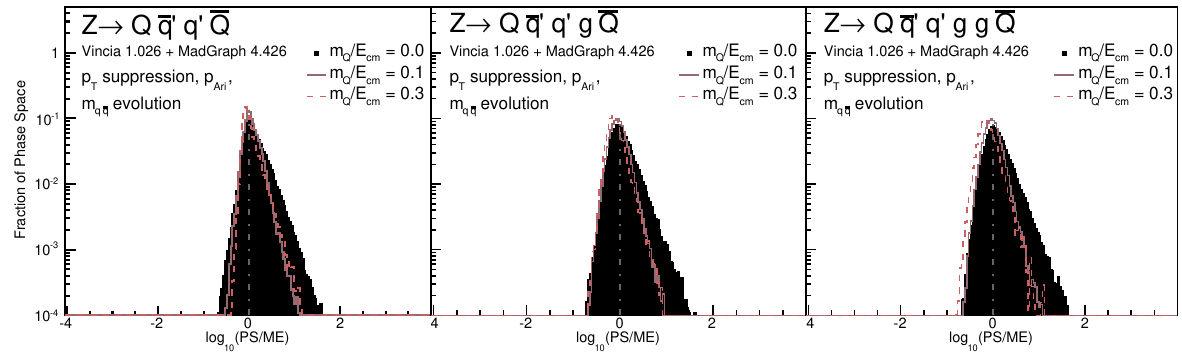}
}

\subfloat[Using {\pT{}} for both gluon emissions and for secondary $q
  \bar{q}$ production.]{
\label{Fig:RQuarkPairpTUnopAri}
\includegraphics[width = 0.95\textwidth]{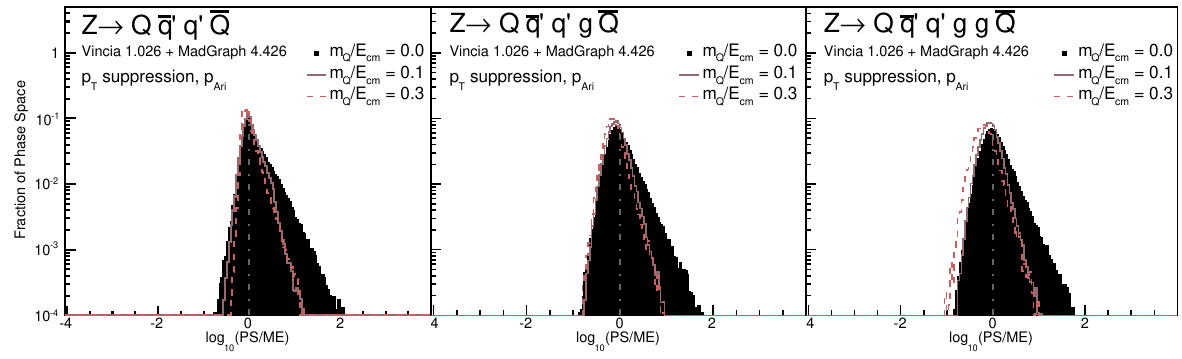}
}
\caption{Histograms of $\log_{10} (R_n)$,  for 
$Z$ decay into a massive primary quark pair, a massless secondary quark 
pair and up to two gluons in a
  flat phase space scan, for $n=4$ (left panes), $n=5$ (middle panes), 
and 
  $n=6$ (right panes). The {\pAri} factor is used for gluon splittings.
  Top row: smooth ordering using 
  \pT as the scale for gluon emission and $m_{q \qbar}$ 
  for gluon splittings. Bottom row: smooth ordering  
  using \pT for both gluon emissions and 
  gluon splittings.}
\end{figure}
The expansion of the resulting weights for  $Z\to4$, $5$, and $6$
partons with smooth ordering in
$2 \pT{}$ for gluon emissions and in $m_{q\bar{q}}$ for gluon splittings,
are shown in \figRef{Fig:RQuarkPairpTUnopAriQuarkPairMassEv} and for
smooth ordering in \pT{} for all branching processes in
\figRef{Fig:RQuarkPairpTUnopAri}. 

Although still far from the very good agreement obtained in the pure
gluon-emission case (as compared with \figRef{fig:RpTsmooth}), 
both the centre and the width of the weight distributions shown in
\figRef{Fig:RQuarkPairpTUnopAri} where smooth ordering including
this additional suppression factor $\pAri$ is imposed for secondary
massless $q \bar{q}$ production,
are now in tolerable agreement with the leading order (LO) 
matrix elements over a substantial fraction of phase space.
Matching to the LO matrix elements can obviously be used to improve 
this agreement further, up to the orders for which matrix elements 
are available (see \secRef{sec:matching}).  

It should be emphasized, however, that the centre position of the distribution 
$w_\text{PS}/\abs{\me}^2$ is highly sensitive to the finite parts of 
$\ant_{\bar{q}'/qg}$.
Since these pieces of $\ant_{\bar{q}'/qg}$ are not universal, the fact 
that the most frequent ratio of $w_\text{PS}/\abs{\me}^2$ is almost 
unity as
demonstrated in \figRef{Fig:RQuarkPairpTUnopAriQuarkPairMassEv}
cannot be expected to be universal for all processes involving gluon 
splitting either. This is an unavoidable consequence of the less
pronounced singular behaviour for these antennae, as compared to the
gluon-emission ones.

\subsection{Including massive $\mathbf{g\to
    \bar{Q}'Q'}$ splittings}

As a final set of comparisons we include massive $g\to \bar{Q}Q$
splittings in the shower expansion. Since the corresponding dipole-antenna
functions do not contain any poles at all (though one does appear for
$m_Q\to 0$), we should expect the shower approximation to be at its
worst for this case, translating to very large uncertainties on,
e.g., the amount of $g\to b\bar{b}$ splittings produced by it.

\begin{figure}[t]
\centering
\subfloat[Using {\pT{}} for gluon emissions and $m_{q \bar{q}}$ for gluon splittings.]{
\includegraphics[width = 0.95\textwidth]{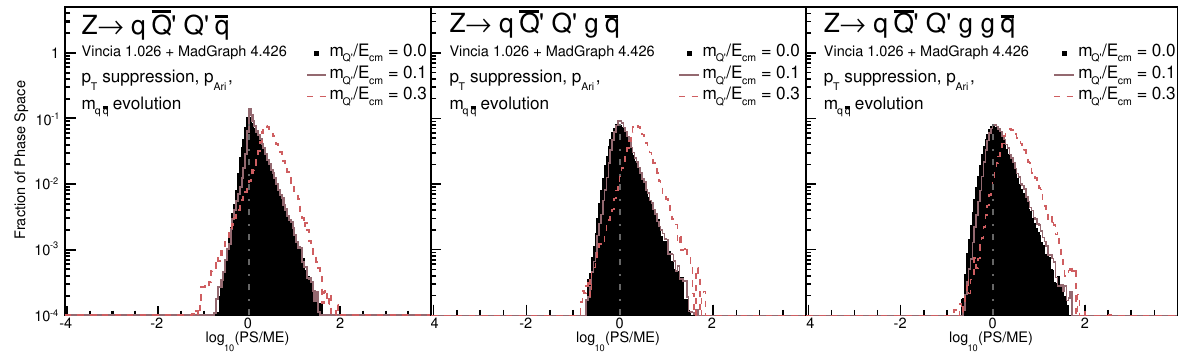}
\label{Fig:Rprimary0QuarkPairMassEv}
}

\subfloat[Using {\pT{}} for gluon emissions and gluon splittings.]{
\label{Fig:Rprimary0AllpT}
\includegraphics[width = 0.95\textwidth]{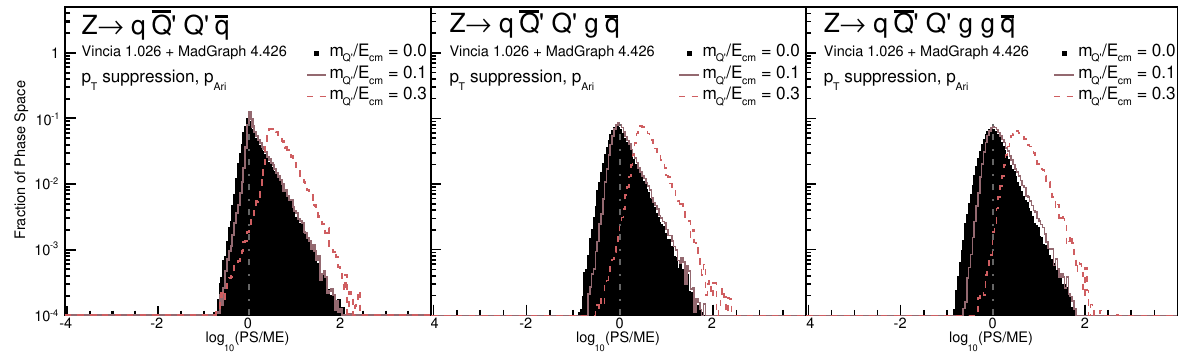}
}
\caption{Histograms of $\log_{10} (R_n)$, as defined in the text, for 
$Z$ decay into a massless primary quark-antiquark pair, a massive secondary quark-antiquark 
pair and up to two gluons in a
  flat phase space scan, for $n=4$ (left panes), $n=5$ (middle panes), 
and 
  $n=6$ (right panes). Unordered emissions are suppressed 
  smoothly, for gluon splittings {\pAri} is used.
  In the top row, the suppression factor for unordered branchings 
  is calculated using 
  \pT as the scale for gluon emission and $m_{q \qbar}$ as the 
  scale for gluon splittings. In the bottom row, the suppression 
  factor is calculated using \pT for both gluon emissions and 
  gluon splittings.
  Note: the parton shower uses the default antenna set.}
\label{Fig:Rprimary0}
\end{figure}

However, as illustrated by the plots in 
\figRef{Fig:Rprimary0}, the agreement is in fact at the same level as that
obtained for massive parents in the previous subsection, 
except for a slight tilt of the distribution for rather heavy secondary quarks. We note that this 
is especially true for the  ``interleaved evolution'' choice of using
$Q_E=2\pT{}$ for gluon emissions and $Q_E=m_{q\bar{q}}$ for gluon splittings,
cf.~\figRef{Fig:Rprimary0QuarkPairMassEv}, as compared to using
$\pT{}$ for all branchings as illustrated in \figRef{Fig:Rprimary0AllpT}.

This strengthens our motivation for using the interleaved \pT{}- and
mass-ordered evolution as the default in \Vc. 
Note also that  we have checked that this conclusion appears to be 
robust against at least moderate variations of the antenna function
finite terms. We conclude that there is still significant
uncertainties surrounding massive $g\to Q\bar{Q}$ splittings, but that
the default choices made in \Vc\ can at least be considered a sensible
starting point.
Of course, matching to matrix elements can still improve the situation, 
in particular for secondary quark-antiquark pairs of high invariant mass, by 
increasing  the multiplicity at which the arbitrary finite parts of the antenna 
functions start to matter.

\section{Comparison to Analytic Resummation \label{sec:resummation}}

Observables involving massive particles, like heavy-quark fragmentation 
processes, can be considered as being collinear-safe since
collinear divergencies are regulated by the finite value of
the heavy quark mass $m$. Thus, such processes 
 can be computed order by order in perturbation theory. 
Nevertheless, as mentioned in \secRef{sec:polestructure}, 
mass-dependent logarithms of the form $\ln (Q^2/m^2)$, 
where $Q$ is the typical scale of the hard 
scattering process, appear at each order in perturbation theory.
When the hard scale $Q$ is much larger than $m$, 
these quasi-collinear logarithmic 
contributions can be large and have to be resummed to all orders 
to obtain reliable predictions for these observables.

This resummation can be performed analytically, in a process-independent
way, by using the perturbative fragmentation formalism 
\cite{Mele:1990cw, Cacciari:1993mq}, which is summarized briefly 
below. Alternatively it can be performed numerically, 
using a parton-shower, such as \Vc.
 
In this section, our aim is to verify on a particular example, that 
the \Vc implementation developed in this paper, based on the exponentiation 
of massive dipole-antenna functions which reproduce 
the soft and the quasi-collinear limit of tree-level matrix-elements, 
performs the resummation of quasi-collinear logarithms 
correctly. We do this by comparing the predictions of \Vc 
with those obtained from the analytic calculation 
of \cite{Cacciari:2001cw} for the inclusive production of a single
heavy meson $H$ in $e^+e^-$ collisions.

The single inclusive production of a heavy meson $H$ 
as obtained by the process $e^+e^- \to V \to H(p) +X$  
can be described by the production of a heavy quark pair $Q-\bar{Q}$  
which subsequently hadronize to yield a heavy meson $H$.
This hadronization process can be
described by non-perturbative (heavy quark)-to-hadron fragmentation
function (such as for example in the model of Peterson ~\cite{Peterson:1982ak}) 
whose free parameters have to be extracted from the data. 

Since  we are mainly interested in the perturbative contributions to
the inclusive cross section, we shall here ignore this
non-perturbative contribution. Consequently, we   
compare the predictions obtained from the calculation 
of the inclusive production of a heavy quark pair
 with our predictions obtained with \Vc without hadronization.   

We consider the inclusive production of a heavy quark pair in $e^+e^-$
collisions in the kinematical region where 
the centre-of mass energy $E_\text{cm}$ of the collisions 
is much larger than the heavy quark mass $m$, i.e.\ $E_\text{cm} \gg m$.
At the same time the heavy quarks are produced in the
perturbative regime with a mass $m$ which is large enough so that 
($m \gg \Lambda_\text{QCD}$) .
We consider the following distribution, 
\begin{equation}
	\mathcal{D}(x, E_\text{cm}^2, m^2) \equiv \frac{1}{ \sigma_\text{tot}} \frac{ \d{\sigma} }{ \d{x} }
\end{equation}
with $\sigma_\text{tot}$ the total hadronic cross section in $e^+ e^-$.
The energy fraction $x$ of the heavy quark system with momentum $p$ is given by
$x=2 \, p\cdot (p_{e^+} + p_{e^-}) /E_\text{cm}^2$.

Since we are mainly interested in the large $x$ behaviour of the cross
section, we concentrate on its so-called flavour-non-singlet
contribution. In this case, the ingredients to the cross section are
dependent on the difference of two flavour-non-identical parton species.

In the following we shall first recall the main ingredients of the
analytic calculation of \cite{Cacciari:2001cw}
before presenting our comparison.
 
\subsection{The heavy quark fragmentation formalism}

The inclusive production of heavy quarks in $e^+e^-$ annihilation is a
purely perturbative process which can be described by the perturbative
fragmentation formalism in which collinearly-enhanced
 contributions are resummed.
In the large $x$ region,
logarithmically enhanced contributions due to soft radiation can also
occur. A reliable theoretical prediction for this heavy quark
production process can therefore only be obtained
if both class of logarithms are appropriately resummed. 
This combined resummation has been performed in \cite{Cacciari:2001cw}
up to  the next-to-leading logarithmic (NLL) accuracy.
In the following we shall first describe 
how the collinear-enhanced logarithms are resummed 
before including soft-gluon resummation effects as well. 

Thanks to the factorization theorem of mass singularities 
the cross section for the production of a hadron $H$ in $e^+e^-$
collisions can be written as a convolution  
of a process-dependent coefficient function $C^{e^+e^-}_{(a)}$
and a parton-to-hadron fragmentation function denoted by $D_{a/H}$, 
for each parton $a=(q,\bar{q},g)$ involved, treating 
all flavours as massless.
Performing a power series expansion in $\alpha_s$, one can compute 
each of the coefficient functions $C^{e^+e^-}_{(a)}$ 
as a massless QCD partonic cross section
and the $\mu_F^2$ evolution of $D_{a/H}$, within perturbation theory.
The $\mu_F^2$ evolution of  $D_{a/H}$ is performed using the 
well-known 
Altarelli-Parisi evolution equations \cite{Altarelli:1977zs, Owens:1978qz} which
enable the resummation of logarithmic contributions of collinear
origin in $D_{a/H}$.
Furthermore, provided $D_{a/H}$ 
is known at some initial factorization scale called $\mu_{0F,H}$, 
its form at any other higher factorization scale $\mu_{F,H}$  
can be determined from this evolution equation.
Its form at the initial scale $\mu_{0F,H}$, chosen such that $\mu_{0F,H}\sim \Lambda_\text{QCD}$, 
is a purely non-perturbative contribution which has either to be
modelled phenomenologically and/or determined from data.

Within the framework of the perturbative fragmentation formalism, the
cross section  $\sigma_Q$ for the inclusive production in $e^+e^-$
collisions, of a heavy quark pair $Q-\bar{Q}$ is given as a
generalization of the fragmentation formalism for hadrons described
above, as follows:
For each parton $a=q,\bar{q},g$, (still treating all flavours as
massless), and for a given type of heavy quark $Q$ (or $\bar{Q}$), the
heavy quark production cross section $\sigma_Q$  is given as a convolution 
of the previously defined coefficient function $C^{e^+e^-}_{(a)}$ 
with $D_{a/Q}$, the perturbative fragmentation function 
of the massless parton $a$ into  $Q$.
In the most general case, all massless partonic contributions
(including gluonic ones) to coefficient functions and fragmentation 
functions have to be taken into account.
As we restrict ourselves in this section to the non-singlet part of
the cross section, only the primary production of a quark-antiquark 
pair of the given type will contribute, secondary production will be neglected. 

In calculations involving all-order resummation, 
it is usual to consider the equivalent expression in Mellin 
moment space 
where one uses the customary definition of the Mellin transform
\begin{equation}
\label{eq:Mellin}
f_N = \int_0^1 \d{x} x^{N-1} f(x) ~
\end{equation}
which transforms the convolution in  $x$-space  into a product in Mellin space.
Using the perturbative fragmentation formalism as
described above, the $N$ moments of the inclusive distribution 
$\mathcal{D}$  may be written as,
\begin{equation}
\label{eq:D}
	\mathcal{D} \left( E_\text{cm}^2, m^2 \right) = 
		\frac{ \sigma^{(LO)} }{ \sigma_\text{tot} } 
		C_N^{(e^+ e^-)} \left( \alpha_S(\mu^2), E_\text{cm}^2, \mu^2, \mu_F^2 \right)
		D_N \left( \mu_F^2, m^2 \right)~
\end{equation}
where $ \sigma_\text{tot} /\sigma^{(LO)} $ is given at order
$\alpha_s$ by $(1+ \alpha_s / \pi)$. 
$C_N^{(e^+ e^-)}$ denotes the Mellin transform of the $e^+ e^-$ coefficient 
function while $D_N$ stands for the Mellin moments of the 
flavour non-singlet component of the 
perturbative fragmentation 
function $D(x)$, as defined in \eqRef{eq:Mellin}.
Both of these contributions depend on the factorization scale $\mu_F$ 
which was introduced by factorizing the cross section.

As mentioned before,
we  concentrate on the large-$x$ behaviour, i.e $ x
\to 1$ (in momentum space) or equivalently,
in Mellin space on the large-$N$ limit ($N \gg 1$) of this cross section.
Note that, in \cite{Cacciari:2001cw} the inversion of the results 
to $x$ space is performed using the 
Minimal Prescription of \cite{Catani:1996dj,Catani:1996yz}, 
but we shall not use those results in what follows.
We shall in fact compare the results obtained with \Vc to the
predictions of the analytic calculation obtained only in Mellin space,
using \eqRef{eq:D}.    

The resummation of collinear-enhanced logarithms of $E_\text{cm}^2/m^2$
is achieved in \eqRef{eq:D} 
by writing the N moments of the perturbative fragmentation
function, denoted by $D_N$ in this equation, as the product of an
evolution operator $E_N$ and $D^\text{ini}_N$, the perturbative initial
condition for the heavy quark fragmentation function. 
Both of these functions depend on the factorization scale $\mu_{F}$
and on the starting point of the evolution called  $\mu_{0F}$.
The perturbative fragmentation function in Mellin space denoted by $D_N$ reads,
\begin{equation}
	D_N(\mu_F^2, m^2) = E_N(\mu_F^2, \mu_{0F}^2) 
		D_N^\text{ini} (\alpha_S(\mu_0^2), \mu_0^2, \mu_{0F}^2, m^2).
\end{equation}
The evolution operator $E_N$ is the solution of the 
Altarelli-Parisi evolution equations written in Mellin space as,
\begin{equation}
	\frac{ \d{E_N} (\mu_F^2, \mu_{0F}^2 ) }{\d{ \ln \mu_F^2 } } = 
		\gamma_N \left( \alpha_S( \mu_F^2 ) \right) E_N( \mu_F^2, \mu_{0F}^2 )
\end{equation}
with the boundary condition $E_N( \mu_{0F}^2, \mu_{0F}^2) = 1$.
In this equation, the anomalous dimension $\gamma$ is 
related to the well-known time-like Altarelli-Parisi splitting
functions in Mellin space $P_N$ 
\cite{Curci:1980uw,Furmanski:1981cw,Kalinowski:1980wea,Floratos:1980hk,Antoniadis:1981zv,Mitov:2006wy},  
(up to the second order in $\alpha_{s}$)  by 
\begin{equation}
\label{eq:gamma}
	\gamma_N (\alpha_S) = \frac{ \alpha_S}{ 2 \pi } P_N^{(0)} 
		+ \left( \frac{ \alpha_S }{ 2 \pi } \right)^2 P_N^{(1)} + \order{ \alpha_S^3 }~.
\end{equation}
For heavy quark fragmentation, the starting point of 
the perturbative evolution $\mu_{0F}$ is chosen to be of the same order as
$m$, the mass of the heavy quark. 
As such it is a perturbative scale (i.e.\ $m \gg \Lambda_\text{QCD}$).
As a consequence, unlike the parton-to-hadron fragmentation
function defined at the initial scale $\mu_{0F,H}$ close to $\Lambda_\text{QCD}$ , 
the initial condition $D_N^\text{ini}$ 
for the parton-to-(heavy quark) fragmentation function 
which depends on this perturbative starting scale $\mu_{0F}$, can be 
computed in perturbation theory as a power series in $\alpha_s$.
At leading order, we have $D^\text{ini}(x) = \delta(1-x)$ 
(and therefore $D_N^\text{ini} = 1$), 
which expresses nothing but the fact that at leading order, 
a $b$-flavoured jet is a $b$ quark.
As such the LO initial condition is trivial and 
independent of the mass $m$ of the heavy-quark.  
 
In \eqRef{eq:D}, the process dependence of the inclusive cross
section is entirely contained in the 
coefficient function $C_N^{(e^+ e^-)}$. It denotes the Mellin 
transform of the $e^+ e^-$ coefficient function 
in the \msbar scheme . It is given in \cite{Altarelli:1979kv}.
For conciseness it is not presented here, its
behaviour for large Mellin moment $N$ is given by, 
\begin{multline}
\label{eq:coefficientfunctionN}
	C_N^{(e^+ e^-)} = 1 + \frac{ \alpha_S( \mu^2 ) }{ \pi } C_F
	\left( 
		- \left( \ln N + \gamma_E - \frac{3}{4} \right) \ln \left( \frac{ E_\text{cm}^2 }{ \mu_F^2 } \right)
		+ \frac{1}{2} \ln^2 N
		+ \left( \frac{3}{4} + \gamma_E \right) \ln N
		\right. \\ \left.
		+ \left( \frac{5}{12} \pi^2 - \frac{9}{4} + \frac{1}{2} \gamma_E^2 + \frac{3}{4} \gamma_E \right)
		+ \order{ \frac{1}{N} }
	\right)~.
\end{multline}

The perturbative fragmentation function
$D_N$, as well as its components the evolution operator $E_{N}$ 
and the initial fragmentation function $D_\text{ini}$ are instead universal.
Beyond leading order, those depend on the factorization scales 
$\mu_F$ and $\mu_{0F}$ as well as on the renormalization scales 
$\mu$ and $\mu_{0}$. 
We start by describing the required evolution operator $E_N$ 
needed for our comparison.
 
The resummation of large collinear logarithms is performed by
solving the AP equations perturbatively and by setting factorization
and renormalization scales equal. 
This is realized  by setting 
$\mu_F \approx \mu \approx E_\text{cm}$ and $\mu_0 \approx \mu_{0F} \approx m$.

Using in addition the second-order expansion 
of the anomalous dimensions as given in \eqRef{eq:gamma}
the evolution operator for the non-singlet channel
reads,
\begin{equation}
\label{eq:evolutionoperator}
	E_N( \mu_F^2, \mu_{0F}^2 ) = 
		\left( \frac{ \alpha_S(\mu_{0F}^2) }{ \alpha_S(\mu_F^2) } \right)^\frac{P_N^{(0)}}{2 \pi b_0 }
		\exp \left(
			\frac{\alpha_S(\mu_{0F}^2)  - \alpha_S(\mu_F^2) }{4 \pi^2 b_0 }
			\left( P_N^{(1)} - \frac{ 2 \pi b_1 }{b_0} P_N^{(0)} \right)
		\right)~.
\end{equation}
Using this equation, one sees that the leading collinear 
logarithms of the form $( \alpha_s \ln(E_\text{cm}^2/m^2) )^n$ of \eqRef{eq:D} 
are resummed by combining the LO expression of the evolution operator
 with the LO expressions of the initial condition $D_\text{ini}$ 
and the LO expression of the coefficient function $C^{e^+e^-}$.
Since at this order, these three ingredients do not depend on the heavy quark mass $m$,  
it corresponds to using a massless calculation 
to describe the heavy quark fragmentation process, 
which is a very crude approximation.
The resummation of the 
next-to-leading collinear terms of the form  
$\alpha_{s} (\alpha_s \ln(E_\text{cm}^2/m^2) )^n$
in the evolution operator given 
in \eqRef{eq:evolutionoperator} above requires the equivalent
NLO expressions for these three components which are now mass-dependent.
As a consequence, although the collinear-enhanced logarithms are
formally resummed in \Vc only at the LL level, to compare our predictions, 
we will use the full evolution operator $E_N$ defined above in
\eqRef{eq:evolutionoperator} which includes subleading logarithmic effects.
The comparison can in fact only be made with the analytic result which 
shows a clear mass-dependent behaviour.

To come to $D_N^\text{ini}$,
as mentioned before, the initial scale $\mu_{0F}$ is perturbative as it is chosen 
of the order of the heavy-quark mass $m$ and by this choice, 
the appearance of large logarithms of the ratio 
$\mu_{0F}/m$ are avoided in the perturbative initial condition $D_N^\text{ini}$.
The perturbative initial condition up to order $\alpha_{S}$ which
only resums the collinear-enhanced logarithms of the form $E_\text{cm}^2/m^2$  
is given below as, 
\label{eq:Dinifull}
\begin{multline}
	D_N^\text{ini} ( \alpha_S( \mu_0^2), \mu_0^2, \mu_{0F}^2, m^2 ) = \\
		1 + \frac{ \alpha_S( \mu_0^2)  }{ 2 \pi } C_F 
		\int_0^1 \d{z} (z^{N-1}-1) \left( 
			\frac{ 1 + z^2 }{1-z} \left( \ln \frac{ \mu_{0F}^2 }{(1-z)^2 m^2 } -1 \right) 
		\right)+ \order{ \alpha_S^2}~.
\end{multline}
The corresponding expression in the large-$N$ limit is given by,
\begin{multline}
\label{eq:initialconditionN}
	D_N^\text{ini} ( \alpha_S( \mu_0^2), \mu_0^2, \mu_{0F}^2, m^2 ) = \\
		1 + \frac{ \alpha_S ( \mu_0^2 ) }{ \pi } C_F \left(
			- \ln^2 N + \left( \ln \frac{m^2}{\mu_{0F}^2} - 2 \gamma_E + 1 \right) \ln N 
			+ \order{1} \right) + \order{ \alpha_S^2 }~.
\end{multline}
As can be seen from \eqRef{eq:coefficientfunctionN} for $C_N$ and 
\eqRef{eq:initialconditionN} for $D_{N}^\text{ini}$ given 
in the large $N$ limit, 
for $1-x \ll 1$ (corresponding to $N \gg 1$), both the 
perturbative initial condition and the $e^+ e^-$ 
coefficient function cannot be computed in fixed-order 
perturbation theory as they contain
logarithmic contributions proportional to $\ln N$ and $\ln^2 N$ respectively.
Those logarithmic contributions, arising through the radiation of soft
gluons from the heavy quarks at the renormalization scale $\mu_0
\approx m$,  
spoil the convergence of the fixed order perturbative expansion at
large $N$ (or equivalently at large $x$) and have therefore to be
resummed.

Note that we expect the soft-gluon effect to be quantitatively 
more important for the initial condition of the heavy-quark
fragmentation function $D_\text{ini}$ 
than for the coefficient function $C^{e^+e^-}$
since in the first case this soft-gluon or so-called Sudakov effect is controlled by the coupling
$\alpha_s(\mu_{0}^2) $ which is larger than $\alpha_s(\mu_F^2)$ 
present in the partonic cross section $C^{e^+e^-}$ instead. 
 
In \cite{Cacciari:2001cw}, this soft-gluon resummation was performed to 
NLL accuracy for both the coefficient function 
and the perturbative initial condition. 
The expression of the coefficient function including soft-gluon
resummation effects given in Mellin space, which we shall call 
soft-gluon-resummed part of the coefficient function is denoted by $C^S_N$ reads,
\begin{equation}
\label{eq:CS}
	C_N^\text{S} ( \alpha_S (\mu^2 ), E_\text{cm}^2, \mu^2, \mu_F^2 ) = 
		\exp \left( 
			\ln N g^{(1)} (\lambda) + g^{(2)}
				\left( \lambda, \frac{E_\text{cm}^2}{\mu^2}, \frac{E_\text{cm}^2}{\mu_F^2} \right)
			\right) 
\end{equation}
with $\lambda = b_0 \alpha_S( \mu^2 ) \ln N$ and 
\begin{equation}
\begin{split}
	g^{(1)} ( \lambda ) &= \frac{ A^{(1)} }{ \pi b_0 \lambda } 
		\left( \lambda + (1-\lambda) \ln ( 1 - \lambda) \right),
	\\
	g^{(2)} \left( \lambda, \frac{E_\text{cm}^2}{\mu^2}, \frac{ E_\text{cm}^2 }{ \mu_F^2 }  \right) &=
		\frac{ A^{(1)} b_1 }{ 2 \pi b_0^3 } 
			\left( 2 \lambda + 2 \ln (1-\lambda) + \ln^2 (1-\lambda) \right)
		+ \frac{ B^{(1)} - 2 A^{(1)} \gamma_E }{ 2 \pi b_0 } \ln (1-\lambda)
		\\ & \quad
		- \frac{1}{ \pi b_0 } \left( \lambda + \ln (1-\lambda) \right)
			\left( \frac{ A^{(2) } }{ \pi b_0 } - A^{(1)} \ln \frac{ E_\text{cm}^2 }{ \mu^2 } \right)
		- \frac{ A^{(1)} }{ \pi b_0 } \lambda \ln \frac{ E_\text{cm}^2 }{ \mu_F^2 } ~.
\end{split}
\end{equation}
The coefficients $A^{(1)}$, $A^{(2)}$ and $B^{(1)}$ are given in the \msbar scheme by
\begin{equation}
\begin{split}
	A^{(1)} &= C_F, A^{(2)} = 
		\frac{1}{2} C_F \left( 
			C_A \left( \frac{ 67 }{18} - \frac{ \pi^2 }{6 } - \frac{5}{9} N_f \right)
		\right),
	\\
	B^{(1)} &= - \frac{3}{2} C_F ~.
\end{split}
\end{equation}

The initial condition $D_\text{ini}$ including soft-gluon resummation 
which we shall call soft-gluon-resummed part of the initial condition
is denoted by $D^\text{ini,S}_N$ and reads, 
\begin{equation}
\label{eq:DiniS}
	D_N^\text{ini, S} ( \alpha_S(\mu_0^2), \mu_0^2, \mu_{0F}^2, m^2 ) =
		\exp \left( \ln N g_\text{ini}^{(1)} ( \lambda_0 ) + g_\text{ini}^{(2)}  
		\left( \lambda_0, \frac{m^2}{ \mu_0^2}, \frac{ m^2 }{ \mu_{0F}^2 } \right) \right)	
\end{equation}
with $\lambda_0 = b_0 \alpha_S( \mu_0^2 ) \ln N$ and
\begin{equation}
\begin{split}
	g_{\text{ini}}^{(1)} ( \lambda_0 ) &= - \frac{ A^{(1)} }{ 2 \pi b_0 \lambda_0 }
		\left( 2 \lambda_0 + (1 - 2 \lambda_0 ) \ln (1 - 2 \lambda_0 ) \right), \\
	g_\text{ini}^{(2)} \left( \lambda_0, \frac{m^2}{\mu_0^2}, \frac{ m^2 }{\mu_{0F}^2 } \right) &= 
		\frac{ A^{(1)} }{ 2 \pi b_0 } \left( \ln \frac{ \mu_{0F}^2 }{ m^2 } + 2 \gamma_E \right)
			\ln (1 - 2 \lambda_0 )
		\\ & \quad
		- \frac{ A^{(1) b_1 } }{4 \pi b_0^3 } \left( 4 \lambda_0 + 2 \ln ( 1- 2 \lambda_0 ) + 
			\ln^2 (1 - 2 \lambda_0 ) \right)
		\\ & \quad
		+ \frac{1}{2 \pi b_0 } \left( 2 \lambda_0 + \ln(1-2 \lambda_0) \right)
			\left( \frac{ A^{(2)} }{ \pi b_0 } + A^{(1)} \ln \frac{ \mu_0^2 }{ \mu_{0F}^2 } \right)
		\\ & \quad
		+ \frac{ H^{(1)} }{ 2 \pi b_0 } \ln (1 - 2 \lambda_0 ) ~.		 
\end{split}
\end{equation}

From the formulae above, one can see that the soft-gluon 
resummed parts of the coefficient function and of the initial
condition are not valid for arbitrarily large $N$ values.  
Those have branch cuts in the complex $N$-plane, the former starting at
$N = \exp ( 1/(b_0 \alpha_S(\mu^2)) )$, the latter 
at $N = \exp ( 1/(2 b_0 \alpha_S(\mu_0^2)) )$.
This behaviour signals the onset of non-perturbative physics at 
values of $x$ very close to $1$.
Keeping this fact in mind, we will restrict our comparisons 
of the \Vc results with the soft-gluon 
resummed calculation to moderate ($N \leq 25$) Mellin moments.

The final step is to match the
 soft-gluon resummed part of the coefficient function, \eqsRef{eq:CS}, 
and  initial condition, \eqsRef{eq:DiniS}, 
to the fixed-order result 
in such a way that the truncation of the matched result reproduces 
the fixed-order result and the logarithmic accuracy of the resummed
part is preserved.
This requirement does not fix the matched result uniquely, however. In 
\cite{Cacciari:2001cw}, the matching was performed additively.\footnote{
The numerical difference of this matching scheme to the so-called 
"log-R matching scheme" used for example in \cite{Cacciari:2005uk} 
for the same observable is  small for the parameter values we consider.}

The final NLO +NLL resummed expression for the N moments 
of the (non-singlet) initial condition denoted by $D_N^\text{ini, matched}$
is given by,  
\begin{equation}
\label{eq:Dmatched}
	D_N^\text{ini, matched} = D_N^\text{ini}  
		+ \left( D_N^\text{ini, S} - \left. D_N^\text{ini, S} \right|_{\alpha_S} \right)~.
\end{equation}
In this equation, $D_N^\text{ini}$ is given above in \eqRef{eq:Dinifull},
$D_N^\text{ini, S}$ is the Sudakov-resummed part given in \eqRef{eq:DiniS} while 
$D_N^\text{ini, S} |_{\alpha_S}$ denotes the expansion of this expression 
$D_N^\text{ini, S}$ given in \eqRef{eq:DiniS} to the first order in $\alpha_S$.
The definition of the matched $e^+ e^-$ coefficient function is analogous to 
\eqRef{eq:Dmatched}, i.e.\ we have,
\begin{equation}
\label{eq:Cmatched}
	C_N^\text{matched} = C_N
		+ \left( C_N^\text{S} - \left. C_N^\text{S} \right|_{\alpha_S} \right)~.
\end{equation}
In this equation, $C_N$ is the full $e^+e^-$ coefficient function 
at $\order{\alpha_s}$ given in \cite{Altarelli:1979kv}, $C_{N}^\text{S}$ 
is the resummed part as given in \eqRef{eq:CS} above 
and $ C_N^\text{S} |_{\alpha_S}$ corresponds to its expansion at order $\alpha_s$.

\subsection{Comparison with VINCIA}

We now compare the analytical calculation of the fragmentation
function $\mathcal{D}(E_\text{cm}^2, m^2)$, 
\eqRef{eq:D}, described in the previous subsection, to VINCIA results
transformed to Mellin space. For the analytical calculation, we use 
the evolution operator defined by \eqRef{eq:evolutionoperator} and consider 
two possibilities for the 
choice of coefficient function $C_N$ and fragmentation-function
initial condition $D_N^\text{ini}$. The simplest is to take 
the analytic results obtained using the fixed-order coefficient function given in  
\cite{Altarelli:1979kv}, and the form for $D_N^\text{ini}$ which only resums
the collinear-enhanced logarithms given in \eqRef{eq:D}. 
Alternatively, we use the equivalent matched expressions defined 
above in \eqsRef{eq:Dmatched} and \eqRef{eq:Cmatched}, 
which include soft-gluon resummation effects up to the NLL accuracy. 

In order to do the comparison appropriately, we need to fix 
a kinematical range where both the analytical results and 
the prediction from \Vc are reliable.
The range of validity of the analytic resummation calculation 
was discussed in the previous subsection. We expect it to work 
for a centre-of-mass energy $E_\text{cm}$ such that $E_\text{cm} \gg m$  and for 
$m \gg \Lambda_\text{QCD}$.
For \Vc, the requirements for the mass of the heavy quark $m$ are given by,
\begin{equation}
\qstop \ll m_Q \ll E_\text{cm}~,
\end{equation}
where \qstop denotes the emission scale at which \Vc 
would make the transition to a hadronization model.
The first hierarchy $\qstop \ll m_Q$ is necessary to ensure that 
gluons which are soft compared to the quark mass can be emitted, 
whereas the second hierarchy $m_Q \ll E_\text{cm}$ ensures that 
the quasi-collinear logarithms play an important role.

Furthermore, since the accuracy of \Vc is formally LL, there still
remain considerable uncertainties. 
The uncertainty due to the choice of finite terms, estimated by 
using the ``MIN'' and ``MAX'' antenna sets defined in
\secRef{sec:massiveantennae}, was found to be 
small in comparison to the uncertainty coming from the choice of
renormalization scale. Hence we only consider the default antennae set 
for our results here. The renormalization-scale choice was then varied between 
 $\mu_R=2 \pT$ and $\mu_R=\pT/2$, to define the uncertainty range.
(Note that, in the parton shower context, we do not have an explicit 
factorization scale, the evolution variable $Q_E$ 
has the role of a factorization scale instead.)

The predictions obtained with the analytic resummation calculation 
will be taken at factorization and renormalization scale 
respectively given by 
$\mu_F = \mu = Q$ and $\mu_0 = \mu_{0F} = m$.

\begin{figure}
\centering
\subfloat{
\begin{tabular}{c}
  \includegraphics[width = 0.45
    \textwidth]{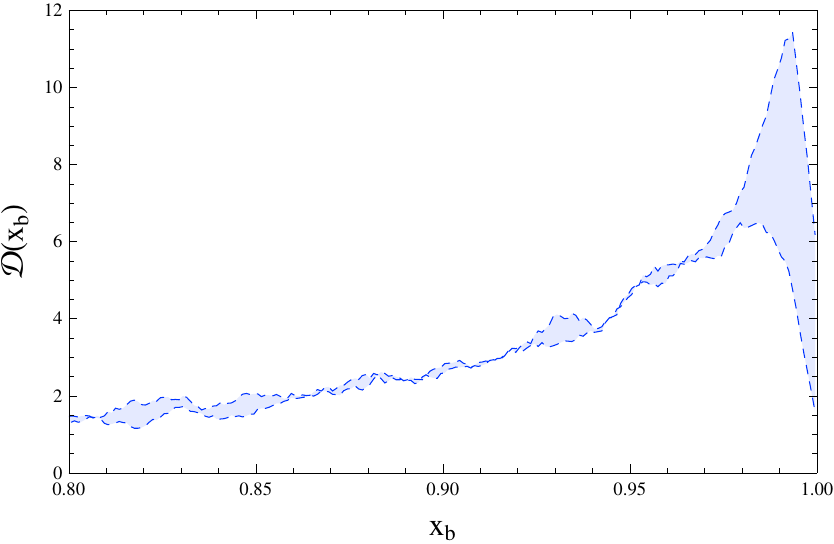}\\[-4.5cm]
\tiny \Vc~1.026 (no hadr.)\\
\tiny $\mu_R = \frac12 \pT\ , \ 2\pT$
\\[3.5cm]
\end{tabular}
}
\subfloat{
\begin{tabular}{r}
	\includegraphics[width = 0.45
          \textwidth]{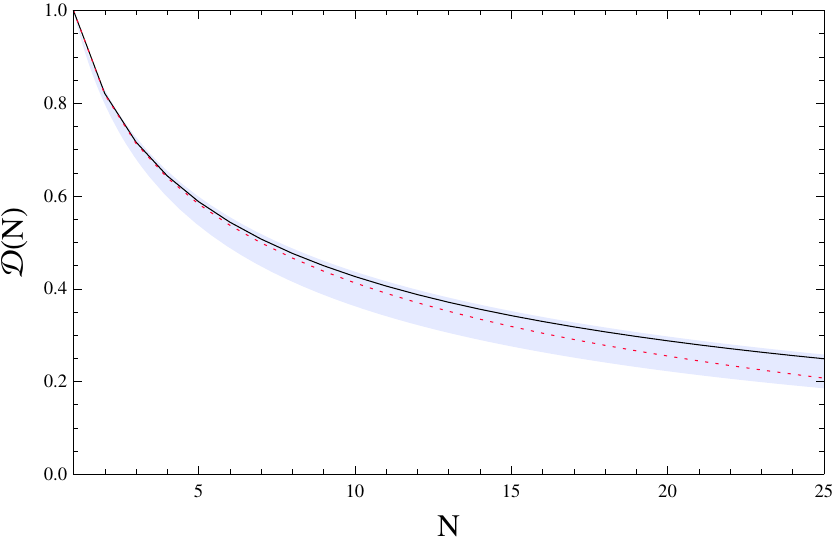}\\[-4.5cm] 
\tiny Band : \Vc~1.026 ($\mu_R = \frac12 \pT,2\pT$, no hadr.)\hspace*{5mm}\\[-1mm]
\tiny Solid Line : Analytic  pQCD (incl NLL soft)\hspace*{5mm}\\[-1mm]
\tiny Dotted Line : Analytic pQCD (without soft)\hspace*{5mm}\\[3.5cm]
\end{tabular}
}
\caption{
Left: the $\mathcal{D}(x_b)$ distribution obtained with \Vc with
hadronization switched off, i.e., at $\qstop = 2p_\mrm{\perp stop} = 1\,$GeV. 
The shaded band shows the uncertainty obtained by 
varying the renormalization scale in $\alpha_S$ by a factor of $2$ and $1/2$
respectively (with scale-stabilization switched off, see
\cite{Giele:2011cb}).
Right: the Mellin transform, $\mathcal{D}(N)$,
including comparisons to  analytic
 resummation  with (solid) and without (dotted) 
 soft-gluon resummation  at NLL.  
$E_\text{cm} = 10^4 \, \text{GeV}$, $m_Q = 0.02 E_\text{cm}$,
$\alpha_S(m_Z^2) = 0.139$. 
The \Vc predictions use the default dipole-antenna set, strong ordering of the emissions in \pT, no matching, and no secondary quark-antiquark production.
}
\label{fig:rimathiasuggestionforpaper}
\end{figure}

The $x$-space fragmentation function obtained with \Vc for a 
centre-of-mass energy of $10^4 \GeV$, quark mass $m_Q/E_\text{cm} =
0.02$ and $\alpha_S ( m_Z^2 ) = 0.139$ (using one-loop runing) is shown in the left-hand pane of
\figRef{fig:rimathiasuggestionforpaper}, in the range 
 $0.8 \leq x \leq 1$. The right-hand pane shows the 
comparison to the analytical results in Mellin space, in the range 
$1<N<25$, using the same parameters as for the \Vc result.
We find reasonable agreement between the two predictions over this
range. The \Vc prediction reproduces 
the main feature of the analytic resummation calculation. 
Due to its large uncertainty, it cannot however distinguish clearly between 
the different analytic predictions.\footnote{
There are parameter values for which the resummation of soft-gluon 
logarithms has a much more pronounced effect than
in \figRef{fig:rimathiasuggestionforpaper}. It turned out, however, that the 
dependence of the \Vc results on the cutoff \qstop is too large to 
allow for a comparison in these cases.}
We find comparable agreement for light ($m_Q\lesssim 0.04 \, E_\text{cm}$) 
quarks, under the condition that the dependence of the \Vc result on the cutoff \qstop is small.

\section{Comparison to $\mathbf{b}$-tagged experimental data \label{sec:lep}}

As a final cross-check, we 
include three basic comparisons to 
data published by the  SLD \cite{Abe:2002iq}, DELPHI \cite{Abdallah:2005cv},
and L3  \cite{Achard:2004sv} experiments. 
We do not intend this to represent a full-fledged
phenomenological study of $b$ fragmentation in VINCIA. Rather, we 
wish to demonstrate that our implementation of
mass effects in \Vc yields sensible numbers also when compared
directly to experimental data, including the effects of hadronization.

We also include comparisons to default
\Py~8.150, on the same distributions. (This is trivial for us to do
--- we merely switch \Vc off.) The resulting  
comparison represents a further validation and cross check (of both
models) since the shower formalisms are quite
different between \Py and \Vc, especially for massive particles. We
also obtain a concrete check that the events generated by \Vc
are being treated by \Py's string model of hadronization in a consistent
manner.

For \Py, we use the default
parameters of \Py~8.150 \cite{Corke:2010yf}. We also note that default
\Py includes matching through $Z\to 3$ partons.
For \Vc, we use the  \Vc-specific tune of the light-flavour
parameters reported in \cite{Giele:2011cb} and include matching
through $Z\to 5$ partons.
Since tuning is not our main purpose here, we have not
attempted to retune \Py's $b$-specific non-perturbative parameters to
(re)optimize them for use with \Vc. It is therefore 
possible that some further improvements could be made in the
non-perturbative description. 

\begin{figure}[t]
\centering
\subfloat[Hadronization OFF]{\includegraphics*[scale=0.36]{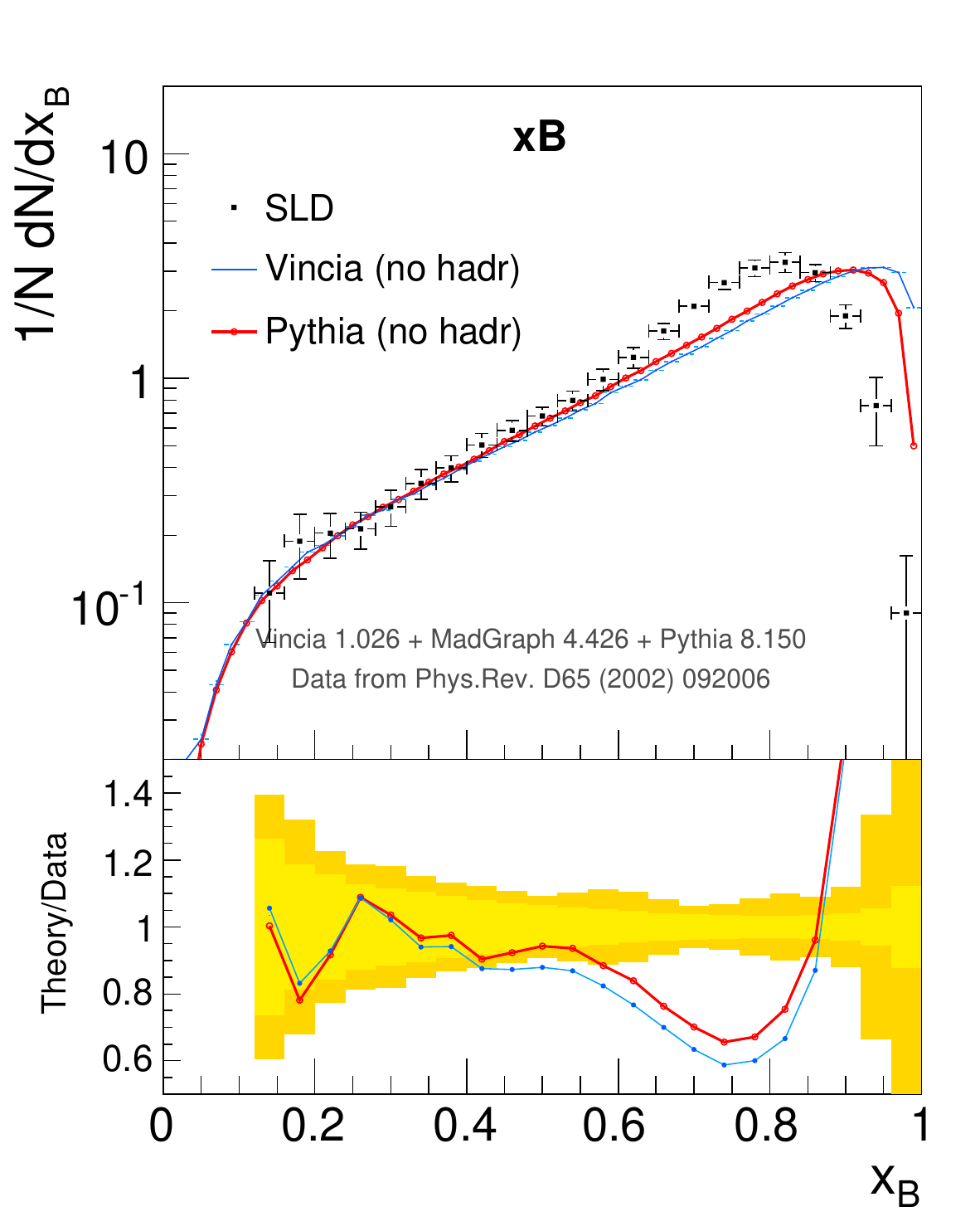}}
\subfloat[Hadronization ON]{\includegraphics*[scale=0.36]{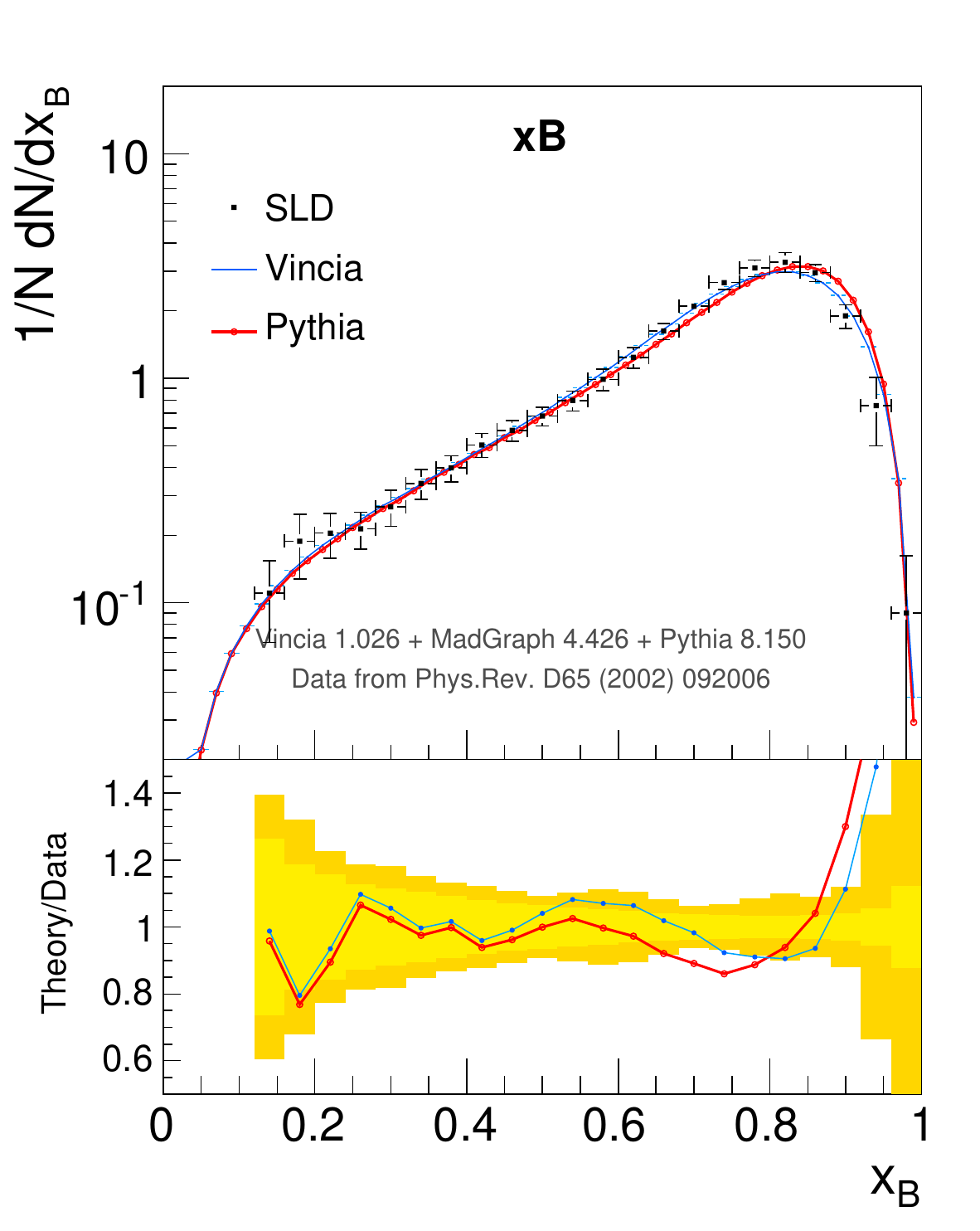}}
\caption{\Vc (thin lines) and \Py~8 (thick lines), before (left) and
  after (right) hadronization,  compared to SLD \cite{Abe:2002iq} 
data on the fragmentation function for weakly
decaying $B$ mesons\label{fig:SLDxB}} 
\end{figure} 
In \figRef{fig:SLDxB}, we show the fragmentation function (a) for $b$
quarks at the parton level\footnote{Obtained by switching off PYTHIA's
  hadronization model, i.e., the $b$ quarks are evolved perturbatively
  to a scale $\qstop = 2p_{\perp\mrm{had}} = 1\,$GeV
  (or $2p_{\perp\mrm{evol}}=0.8\,$GeV in standalone PYTHIA). We ignore the
  slight difference in the exact value and 
  definition of $p_\perp$ between \Py
  and \Vc, see \cite{Sjostrand:2004ef} for a discussion.} 
and (b) at the hadron level, compared to hadron-level SLD data for weakly
decaying $B$ mesons \cite{Abe:2002iq}. The top panes show
the distributions normalized to unity, and the bottom panes show the
ratios of theory to data. In the bottom panes, the inner (lighter) 
yellow bands indicate the statistical uncertainty, and the outer
(darker) bands indicate the combined statistical plus systematical
uncertainty, added linearly. Note that the 
vertical error bars on the Monte Carlo predictions 
correspond to $\pm 1.645 \sigma$ statistical uncertainty from the
number of generated MC points, equivalent to 90\%
confidence. This is the
default for how statistical MC uncertainties are displayed by \Vc's
plotting tool.  

Comparing the two panes of \figRef{fig:SLDxB}, we conclude that 
the non-perturbative corrections 
are significant in the region above $x_B \sim 0.5$, 
while the spectrum at lower $x_B$ is dominated by the
perturbative prediction, for which the two codes are in good
agreement, both with each other and with the data. In the high-$x_B$
region, VINCIA generates a slightly harder parton-level spectrum (peaked at
higher $x_B$) than PYTHIA. That is, VINCIA generates slightly
\emph{less} perturbative radiation. After hadronization, the VINCIA
spectrum is somewhat softer than the PYTHIA one. The total
non-perturbative component of the ``parton-to-hadron'' correction
would therefore be evaluated as being slightly  
larger for VINCIA than for PYTHIA. The main properties of this
correction are driven by the tuning to light flavours. As mentioned
above, a dedicated $b$-specific tuning has not yet been carried
out. Nonetheless, we note that the fragmentation spectrum obtained
with the default tuning already appears to be in reasonable 
agreement with the data, cf.\ the right-hand pane of \figRef{fig:SLDxB}.  

\begin{figure}[t]
\centering
\subfloat[Hadronization OFF]{\includegraphics*[scale=0.36]{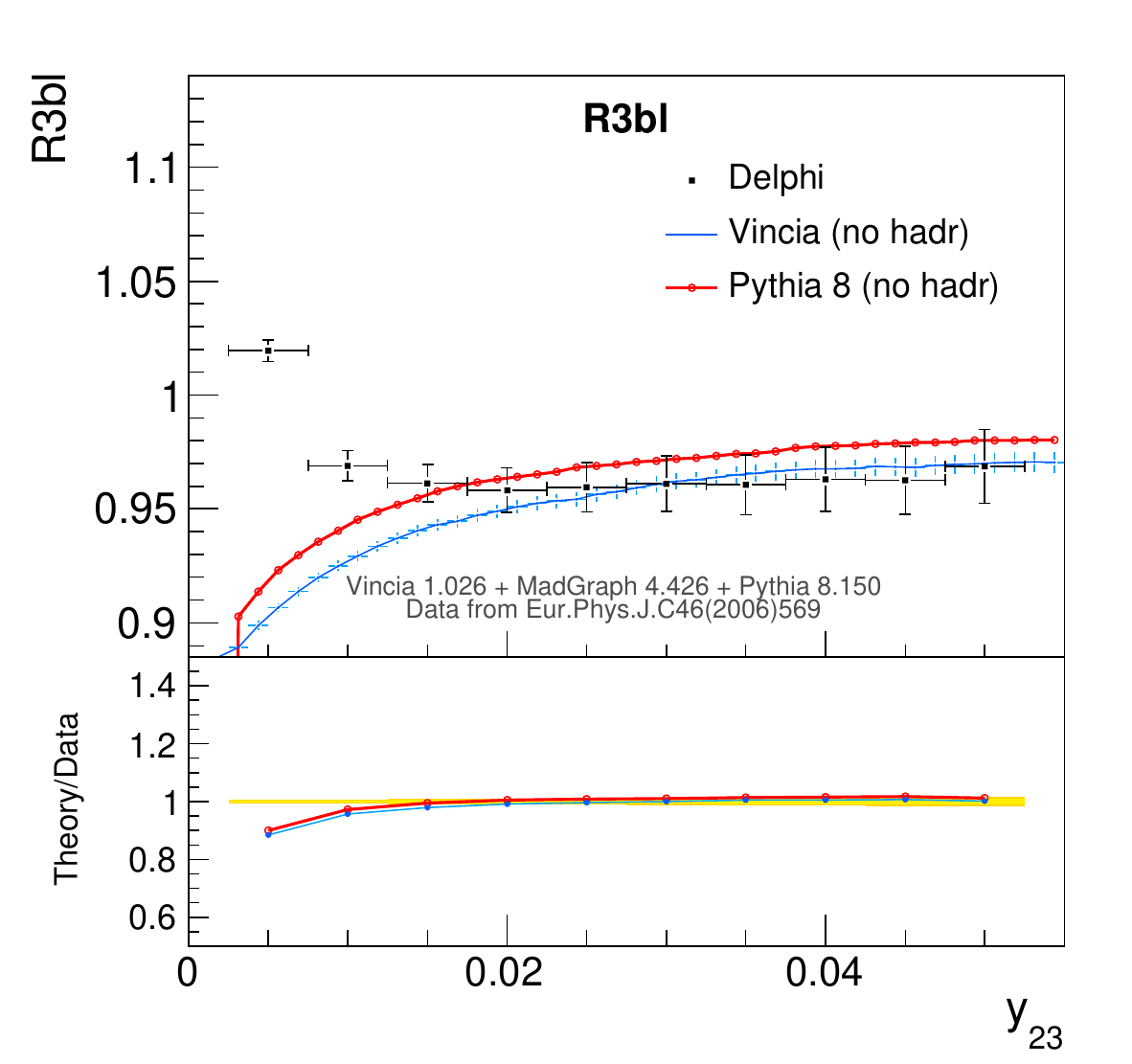}}
\subfloat[Hadronization ON]{\includegraphics*[scale=0.36]{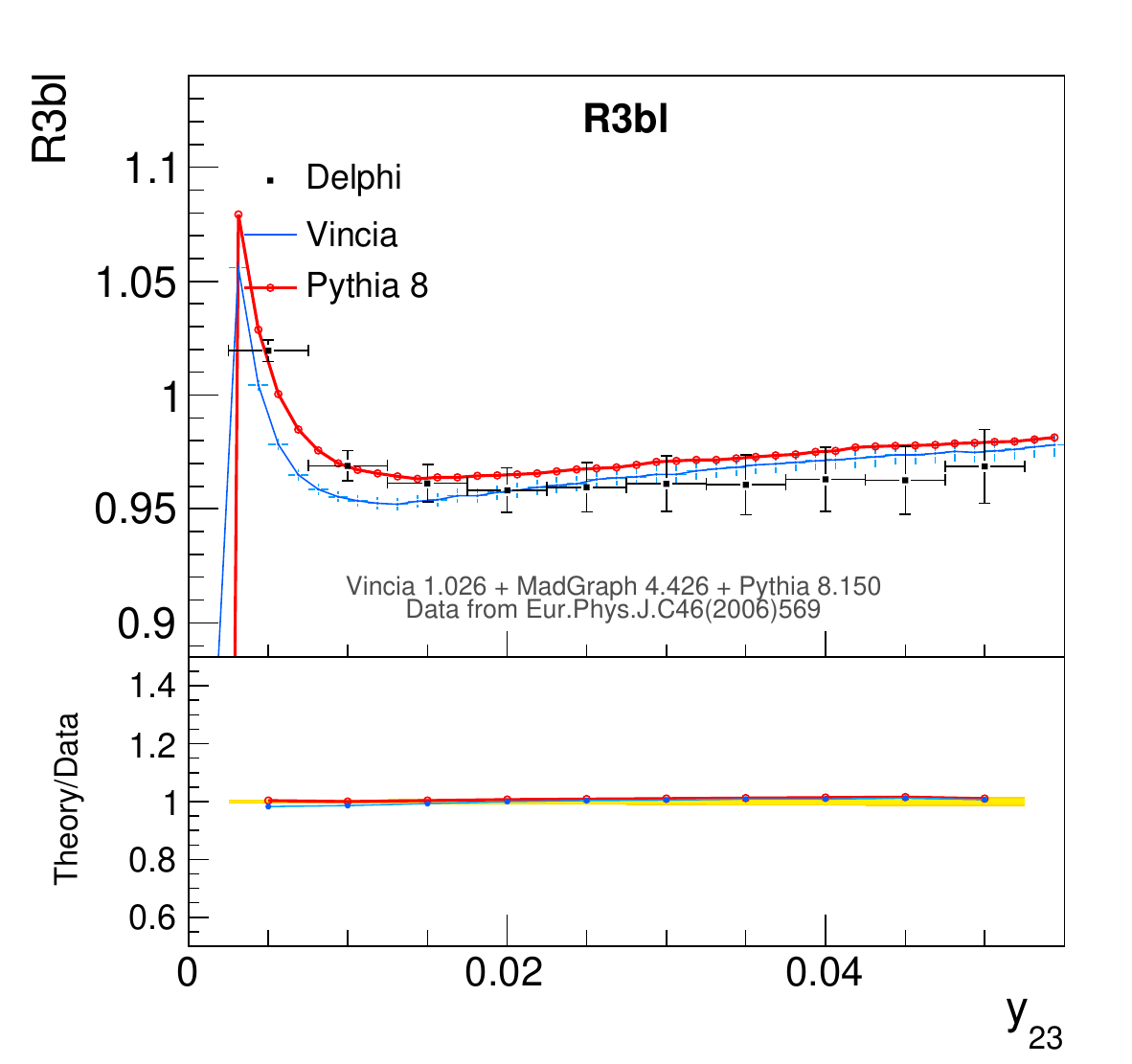}}
\caption{\Vc (thin lines) and \Py~8 (thick lines), before (left) and
  after (right) hadronization,   compared to DELPHI \cite{Abdallah:2005cv}
data on the ratio of $y_{23}$ in $b$- vs.\ light-flavour
events.\label{fig:R3bl}}  
\end{figure}
In \figRef{fig:R3bl}, we have used FASTJET \cite{Cacciari:2005hq} to  
compare VINCIA and PYTHIA to DELPHI data \cite{Abdallah:2005cv} 
on $R_3^{bl}$, the ratio of the $y_{23}$ distribution in $b$-tagged events to  
light-flavour events, with $y_{23}$ the dimensionless resolution 
scale at which the event goes from being a 2- to a 3-jet event,
according to the $k_T$ clustering algorithm (using the $E$
scheme). This distribution is sensitive to the value of the $b$-quark
mass. Comparing left- (without hadronization) to right-hand (with
hadronization) panes, we see that hadronization effects are large in
the region below, roughly, $y_{23}\sim 0.02$. Above that value, the deviation
from unity is generated mainly by perturbative quark mass
effects. Both models use the default PYTHIA value of
$m_b=4.8\,$GeV, which appears to give a reasonable agreement with the
measurement. 

\begin{figure}[t]
\centering
\subfloat[Hadronization OFF]{\includegraphics*[scale=0.36]{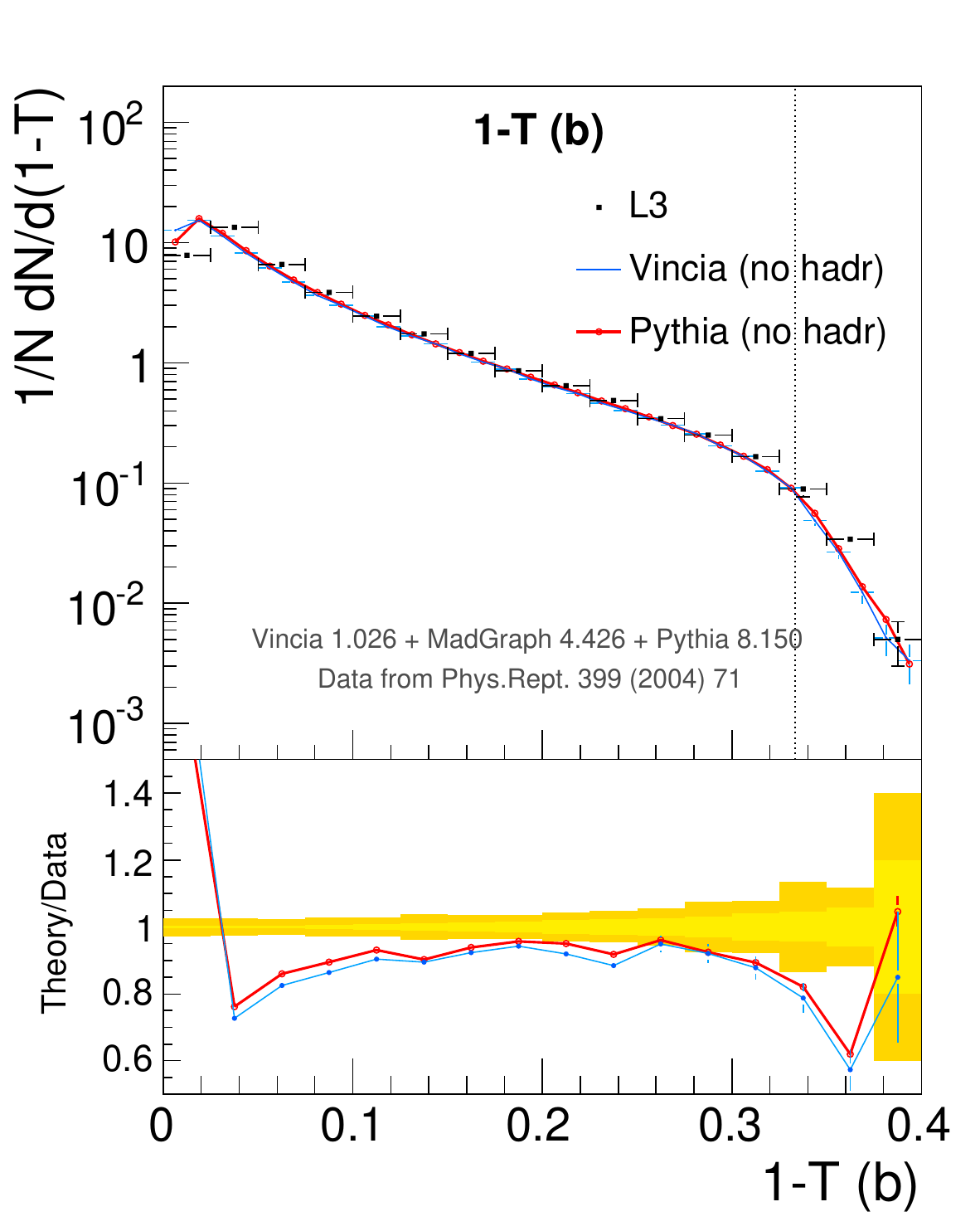}}
\subfloat[Hadronization ON]{\includegraphics*[scale=0.36]{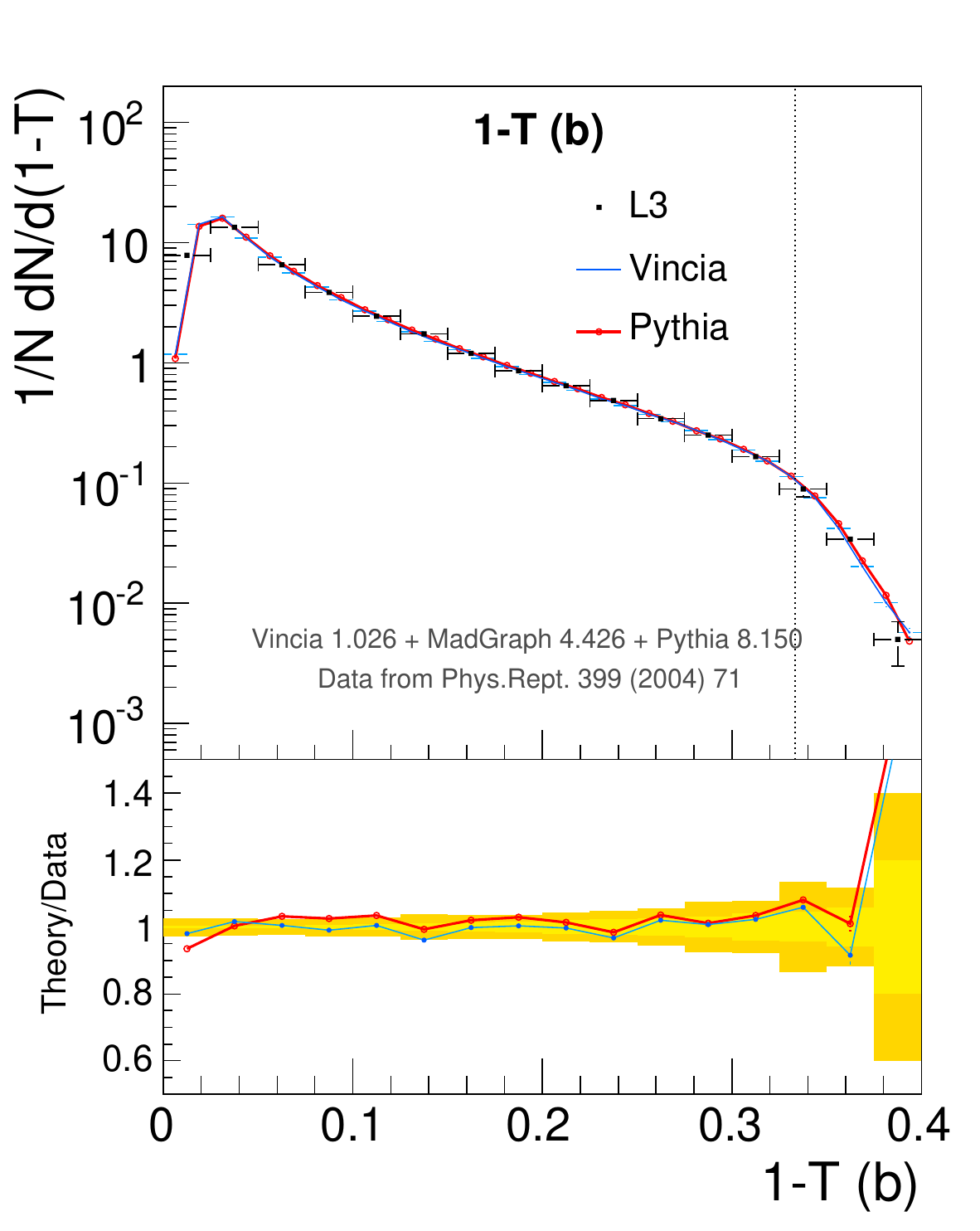}}
\caption{\Vc (thin lines) and \Py~8 (thick lines), before (left) and
  after (right) hadronization,   compared to L3 \cite{Achard:2004sv} 
data on Thrust in $b$-tagged events.\label{fig:L3T}} 
\end{figure} 
In \figRef{fig:L3T}, we show a similar comparison for the Thrust event
shape variable in $b$-tagged events, compared to a measurement
performed by the L3 collaboration \cite{Achard:2004sv}. Comparing (a)
parton- to (b) hadron-level results, an interesting pattern can be
seen, for both PYTHIA and VINCIA. The non-perturbative corrections are
significant not only at low values of $\tau = 1-T < 0.1$, but also around
$\tau = 1/3$, shown with a vertical dotted line in the upper panes of
the plots. At this point, the distribution changes slope\footnote{Due
  to the underlying change from a
3-parton to a 4-parton quantity that the Thrust variable undergoes at
that point.}; thus, even though hadronization effects are
parametrically strongly suppressed in that region, the small
``smearing'' they provide of the underlying perturbative prediction
becomes relatively more important at exactly that point.
With hadronization effects
included, both models describe the data acceptably well. 
Combined with the good agreement found with the light-flavour 
variables included in \cite{Giele:2011cb}, we conclude that the 
\Vc code can at this point be considered validated as a 
Monte Carlo model for final-state showering and hadronization effects.


\section{Conclusions and Outlook \label{sec:conclusion}}

A precise description of processes involving heavy coloured particles
is of prime importance to the physics programme at current and future 
high-energy collider facilities. 
In this article, we have extended the timelike dipole-antenna shower
formalism of \cite{Giele:2007di,Giele:2011cb} to include massive
fermions. Advantages of this treatment include the exact on-shell 
phase-space factorization which is inherent to the antenna formalism
\cite{GehrmannDeRidder:2005cm,GehrmannDeRidder:2009fz,Abelof:2011jv},
the ability to vary the remaining ambiguous parts of the calculation
in a similar manner as was already done for the 
massless case but now including explicit mass-dependent terms, and the
smooth merging with the ``GKS'' matching formalism \cite{Giele:2011cb}.  

In this paper, we have shown evidence, through extensive comparisons to
other calculational methods, that the algorithm is physically sensible
and that it can be expected to yield reasonably precise results 
over large parts of the phase space. Furthermore, using the GKS formalism,
it can be matched to leading-order matrix elements over all of phase
space up to any given fixed order 
(in practice, the limit is currently Born + 4 partons), which should help
to give a systematic improvement even for fairly soft and/or collinear
emissions. 

Although we have not attempted a dedicated tuning of the
non-perturbative fragmentation parameters for heavy quarks, the
preliminary tuning of the massless parameters reported in
\cite{Giele:2011cb}, combined with the \Py~8 defaults for the
$b$-specific ones, appears to give a reasonably good description of
the $B$ fragmentation function and of $b$-tagged event shapes and jet
rates at the $Z$ pole.  

For a full-fledged application to physics at hadron
colliders, two further ingredients remain to be developed: an
extension of the formalism to spacelike (initial-state) showers, and
the inclusion of finite-width decays, which can be important in the
description of (chains of) resonance decays. Nonetheless, 
the step taken here is a necessary prerequisite, and can already be
used to study questions involving final-state $c$ and $b$ quark
fragmentation.  

\subsection*{Acknowledgments}
We thank W. Giele and D. Kosower for many useful discussions 
and comments on the manuscript. 
This work was carried out partly at, and with the support of, the 
Institute of Theoretical Physics at ETH Zurich 
and the CERN Theory Division. The authors acknowledge both of 
these institutions for their kind hospitality.  

This work was supported in part by the Marie Curie research training
network ``MCnet'' (contract number MRTN-CT-2006-035606), by the 
Swiss National Science Foundation
(SNF) under contract PP0022-118864 and by the European 
Commission through the 'LHCPhenoNet' Initial Training 
Network PITN-GA-2010-264564', which are hereby acknowledged. 

\bibliography{bibliography}

\end{document}